\documentclass[11pt,a4paper]{article}
\usepackage{jheppub}

\newcommand{\valpha}{v_\alpha}

\newcommand{\vmu}{v_\mu}

\newcommand{\vTbare}{v_{T,0}}

\newcommand{\MWsq}{M_W^2}
\newcommand{\MHsq}{M_h^2}
\newcommand{\MZsq}{M_Z^2}
\newcommand{\MTsq}{m_t^2}
\newcommand{\lag}{\mathcal{L}}

\usepackage{multirow}
\usepackage{caption}
\usepackage{subcaption}
\usepackage{placeins}

\newcommand{\mwhat}{\hat{M}_W}

\def\msbar{$\overline{\hbox{MS}}$}

\title{Electroweak input schemes and universal corrections in SMEFT}
\preprint{IPPP/23/23, MITP-23-019, MPP-2023-91}

\author[a]{Anke Biek\"otter,}
\author[b]{Benjamin D.~Pecjak,}
\author[c]{Darren J.~Scott,}
\author[b]{Tommy Smith}
\affiliation[a]{PRISMA$^+$ Cluster of Excellence \& MITP, Johannes Gutenberg University, 55099 Mainz, Germany}
\affiliation[b]{Institute for Particle Physics Phenomenology, 
Durham University,  Durham DH1 3LE, UK}
\affiliation[c]{Max-Planck-Institut f\"ur Physik, F\"ohringer Ring 6,  80805 M\"unchen, Germany}

\abstract{The choice of an electroweak (EW) input scheme is an important component of perturbative calculations in
Standard Model Effective Field Theory (SMEFT).  In this paper we perform a systematic study of three different EW input schemes
in SMEFT, in particular those using the parameter sets $\{M_W, \, M_Z, \, G_F\}$, $\{M_W, \, M_Z, \, \alpha\}$, or
$\{\alpha, \, M_Z, \, G_F\}$. We discuss general features and calculate decay rates of $Z$ and $W$ bosons to leptons and Higgs decays to bottom quarks in these three  schemes up to next-to-leading order (NLO) in dimension-six SMEFT.  
We explore the sensitivity to Wilson coefficients and perturbative convergence in the different schemes, 
and show that while the latter point is more involved  than in the Standard Model, the dominant scheme-dependent 
NLO corrections are universal and can be taken into account by a simple set of substitutions on the leading-order results.   
Residual NLO corrections are then of similar size between the different input schemes, and performing 
calculations in multiple schemes can give a useful handle on theory uncertainties in SMEFT predictions and fits to data.
}

\begin{document}
\maketitle

\section{Introduction}
\label{sec:intro}
Standard Model Effective Field Theory (SMEFT) is an important tool for investigating small deviations from Standard Model (SM) predictions. Such indirect descriptions of new physics can be made more robust by including quantum corrections not only in the SM, but also in SMEFT.
Indeed, the study of next-to-leading order (NLO)
 corrections (and in a few instances next-to-next-to-leading order (NNLO) corrections) in dimension-six SMEFT has received much attention in recent years, either calculated on a case-by-case
 basis for specific processes,  \cite{Zhang:2013xya,Crivellin:2013hpa,Zhang:2014rja,Pruna:2014asa,Grober:2015cwa,Hartmann:2015oia,Ghezzi:2015vva,Hartmann:2015aia,Zhang:2016omx,BessidskaiaBylund:2016jvp,Maltoni:2016yxb,Degrande:2016dqg,Hartmann:2016pil,Grazzini:2016paz,deFlorian:2017qfk,Deutschmann:2017qum,Baglio:2017bfe,Dawson:2018pyl,Degrande:2018fog,Vryonidou:2018eyv,Dedes:2018seb,Grazzini:2018eyk,Dawson:2018liq,Dawson:2018jlg,Dawson:2018dxp,Neumann:2019kvk,Dedes:2019bew,Boughezal:2019xpp,Dawson:2019clf,Baglio:2019uty,Haisch:2020ahr,David:2020pzt,Dittmaier:2021fls,Dawson:2021ofa,Boughezal:2021tih,Battaglia:2021nys,Kley:2021yhn,Faham:2021zet,Haisch:2022nwz,Heinrich:2022idm,Bhardwaj:2022qtk,Asteriadis:2022ras,Bellafronte:2023amz}
or moving towards full automation as in the case of QCD corrections~\cite{Degrande:2020evl}. 

An important consideration for SMEFT predictions and fits is the choice of the electroweak (EW) input scheme. Ideally, the input parameters should be measured with very high accuracy such that their effect on SMEFT fits is subdominant or even negligible. 
However, even beyond that, the choice of the input parameters influences perturbative convergence as well as 
the pattern of Wilson coefficients appearing in leading-order (LO) and NLO predictions.
Typical choices of the input parameters include the Fermi constant~$G_F$, the mass of the $W$ and $Z$ bosons, $M_W$ and $M_Z$, as well as the electromagnetic coupling constant $\alpha$.  Invariably, the NLO SMEFT calculations described above have been performed in one of three different schemes, which use either  $\{M_W, \, M_Z, \, G_F\}$ ($\alpha_\mu$ scheme), $\{M_W, \, M_Z, \, \alpha\}$ ($\alpha$ scheme) or $\{\alpha, \, M_Z, \, G_F\}$ (LEP scheme) as inputs. Some discussions of
these input schemes can be found in~\cite{ Brivio:2017bnu, Brivio:2021yjb}. However, there has been no systematic study which elucidates general features of these EW input schemes beyond LO in SMEFT, much less a numerical exploration of benchmark results at NLO in the different schemes. The aim of this paper is to fill this gap.  

We structure the discussion as follows.  First, in Section~\ref{sec:three_schemes}, we describe the ingredients
needed to construct UV renormalised amplitudes in the three schemes, introducing a notation that makes the connections
between them transparent.  In Section~\ref{sec:salient} we identify salient features of the different schemes,
including patterns of perturbative convergence and Wilson coefficients associated with finite parts of counterterms 
for typical weak or electromagnetic vertices.  We give a first set of NLO results at the level of derived parameters such
as $M_W$ in the LEP scheme or $G_F$ in the $\alpha$ scheme in Section~\ref{sec:Derived}, also laying out 
our method for estimating perturbative uncertainties from scale variations in the SM  and SMEFT.  In 
Section~\ref{sec:HeavyBosonDecays} we perform a thorough numerical analysis of heavy boson decays at NLO in SMEFT
in the three schemes, covering $W$ and $Z$ decay into leptons, and Higgs decay into bottom quarks.  Finally, drawing
on the insights from the aforementioned sections, we propose in Section~\ref{sec:universal_corrections}
a simple procedure which can be used to deduce a set of universal and numerically dominant input-scheme dependent
NLO corrections in SMEFT.  Concluding remarks are given in 
Section~\ref{sec:conclusions}.

While the main focus of the paper is to elucidate the role of EW input schemes in SMEFT, as a by-product we have 
produced quite a few NLO results which were not available in the literature so far.  These have been obtained using an in-house \texttt{FeynRules}~\cite{Alloul:2013bka} implementation of the dimension-six SMEFT Lagrangian, 
and cross checked with \texttt{SMEFTsim}~\cite{Brivio:2017btx,Brivio:2020onw}. Matrix elements were computed using~\texttt{FeynArts} and~\texttt{FormCalc}~\cite{Hahn:2016ebn,Hahn:1998yk,Hahn:2000kx}, 
analytic results for Feynman integrals were extracted from \texttt{PackageX}~\cite{Patel:2015tea}, and 
numerical results were obtained with \texttt{LoopTools}~\cite{Hahn:1998yk}. 
Phase-space integrals arising from the real emission of photons and gluons 
were calculated analytically using standard methods. The results have been further cross checked by performing 
calculations in both unitary and Feynman gauge. We include the most important NLO SMEFT results, 
namely the heavy boson decay rates,  $\Delta r$ as defined in Eq.~(\ref{eq:DeltaR}), and the $W$-boson mass in the LEP 
scheme, as \texttt{Mathematica} files in the electronic submission of this work.


\section{Three EW input schemes}
\label{sec:three_schemes}

The dimension-six SMEFT Lagrangian can be written as
\begin{align}
\lag = \lag^{(4)} + \lag^{(6)}  ;  \quad \lag^{(6)} =  \sum_i C_i \, \mathcal{O}_i \, ,
\end{align} 
where $ \lag^{(4)}$ denotes the SM Lagrangian and $ \lag^{(6)}$ is the 
dimension-six Lagrangian with operators $\mathcal{O}_i$ and the corresponding 
Wilson coefficients $C_i = c_i/\Lambda^2$ which are inherently suppressed by the new 
physics scale $\Lambda$.  For the dimension-six operators we adopt the Warsaw 
basis~\cite{Grzadkowski:2010es} -- the  59 independent operators in this basis (which in general
carry flavour indices) are listed and grouped into eight classes in Table~\ref{op59}.  
Throughout this work, the SMEFT expansion of a given quantity is truncated to linear order in the Wilson coefficients 
and thus treated consistently at dimension six.\footnote{Power corrections appearing at dimension eight and beyond come in two distinct types: those which scale as the vacuum expectation value of the theory and those which scale as some kinematic factor $p^2$. A powerful formalism where the distinction between these two becomes important is the so-called Geometric SMEFT~\cite{Helset:2020yio}, where corrections of the former kind can in some cases be computed to all orders in the power series.}

Predictions in SMEFT depend on a number of input parameters and the renormalisation schemes 
in which they are defined  (see for example~\cite{Denner:2019vbn} for an excellent discussion
of renormalisation and input schemes in  the SM).  A number of these are rather standard and are adopted 
throughout this work. The Wilson coefficients 
$C_i \equiv C_i(\mu)$  are renormalised in the \msbar~scheme, and are thus 
functions of the renormalisation scale $\mu$. 
Moreover, we use on-shell renormalisation for the top-quark mass $m_t$ and the 
Higgs-boson mass $M_h$ and set masses of fermions lighter than the top quark equal to zero, 
with the exception of $h\to b\bar{b}$ decay where we keep a non-zero $m_b$ renormalised in
the \msbar~scheme in a five-flavour version of QCD$\times$QED as described 
in Section 5.2 of~\cite{Cullen:2019nnr}.  Furthermore, we approximate the CKM matrix as the unit matrix.
 
The difference in EW input schemes used in the literature is related to how the U(1) and SU(2) gauge couplings, denoted by $g_1$ and $g_2$ respectively, as well as the vacuum expectation value $v_T$ of the Higgs doublet field $H$, defined by 
\begin{align}
\langle H^\dagger H \rangle = \frac{v_T^2}{2} \,,
\end{align} 
are eliminated in favour of three physical input parameters.  In this paper, we consider the following three
schemes:\footnote{More input schemes have been proposed in the literature for specific processes.  For instance, it has been shown that using the sine of the Weinberg angle, $s_w$, as an input parameter leads to good convergence 
for the prediction of the forward-backward asymmetry at the $Z$~pole~\cite{Chiesa:2019nqb}. }
\begin{enumerate}
\item[(1)] The ``$\alpha$ scheme'', which uses as inputs  $\{\alpha, M_W, M_Z\}$, where $M_W$ and $M_Z$
are renormalised on-shell and $\alpha$ is the fine-structure constant renormalised in a given scheme. 
\item[(2)]  The ``$\alpha_\mu$ scheme'', which uses as inputs $\{G_F, M_W, M_Z\}$, where  $G_F$ is the Fermi 
constant as measured in muon decay. This scheme is sometimes called the ``$M_W$ scheme" in the SMEFT 
literature.\footnote{Unfortunately, the SMEFT and SM naming conventions for the schemes do not agree. We choose to use the SM naming conventions.}
\item[(3)] The ``LEP scheme", which uses as inputs $\{\alpha, G_F, M_Z\}$.  In contrast to the 
first two schemes, $M_W$ is not an input. This scheme is sometimes called the ``$\alpha$ scheme" in the SMEFT literature.
\end{enumerate}

In Sections~\ref{sec:AlphaScheme}--\ref{sec:The_LEP_scheme} we discuss the ingredients needed to implement these
three EW input schemes to NLO.  In order to treat them in a unified fashion, it is convenient to use as a starting 
point the tree-level Lagrangian written in terms of $v_T$, $M_W$, and $M_Z$. In practice, this is obtained by 
transforming to the gauge-boson mass-basis using the field rotations defined in~\cite{Alonso:2013hga} and making the substitutions
\begin{align}
\bar{g}_1 & = \frac{2M_W s_w}{c_w v_T}\left[1 - \frac{v_T^2}{4 s_w^2}\left(C_{HD} + 4 c_{w} s_{w} C_{HWB}\right) \right] \, , \\
\bar{g}_2 & = \frac{2M_W}{v_T} \, , 
\end{align}
which are valid up to linear order in the Wilson coefficients.  The sine and cosine of the Weinberg angle are defined as
\begin{align}
s_w = \sqrt{1-c_w^2} \, , \quad c_w = \frac{M_W}{M_Z}  \,.
\end{align}
The renormalised Lagrangian in a given scheme is then obtained by interpreting the tree-level parameters and fields 
as bare ones, denoted with a subscript $0$, and trading them for renormalised quantities through the addition of counterterms appropriate to that scheme.  For instance, all three schemes use  counterterms  $\delta C_i$ for the Wilson coefficients in the \msbar~scheme, which are defined as
\begin{align}
\label{eq:RenCi}
C_{i,0}= C_i + \delta C_i, \qquad \delta C_i \equiv \frac{1}{2\epsilon} \dot{C}_i \equiv \frac{1}{2\epsilon}\frac{d C_i}{d\ln \mu} \, , 
\end{align}
where $\epsilon$ is the dimensional regulator in $d=4-2\epsilon$ space-time dimensions.  Explicit results
for the $\delta C_i$ at one loop can be derived from~\cite{Jenkins:2013zja, Jenkins:2013wua,Alonso:2013hga}.

\subsection{The $\alpha$ scheme}
\label{sec:AlphaScheme}

The  $\alpha$ and  $\alpha_\mu$ schemes share as common inputs $M_W$ and $M_Z$, 
renormalised in the on-shell scheme.  They differ through the way the bare 
quantity $v_{T,0}$ is related to renormalised parameters and counterterms.  
In the $\alpha$ scheme we use
\begin{align}
\label{eq:vT_elim_vhat}
\frac{1}{v_{T,0}^2} =  \frac{1}{\valpha^2} \left[ 1  - \valpha^2  \Delta \valpha^{(6,0,\alpha)}    -\frac{1}{\valpha^2} \Delta \valpha^{(4,1,\alpha)} -  \Delta \valpha^{(6,1,\alpha)}\right] \, .
\end{align}
We have introduced the derived parameter
 \begin{align}
 \label{eq:def_vhat}
 \valpha = \frac{2 M_W s_w}{\sqrt{4\pi \alpha}} \, , 
 \end{align}
where $\alpha=e^2/(4\pi)$ is the QED coupling constant defined in a given
renormalisation scheme.  The superscripts $i$ and $j$ in the counterterms $\Delta v_\alpha^{(i,j,\alpha)}$ label 
the operator dimension and the number of loops ($j=0$ for tree-level and $j=1$ for one-loop) respectively, 
while the superscript $\alpha$ refers to the fact that the expansion coefficients are multiplied by explicit factors of $\valpha$. 
The dependence on the perturbative expansion parameter $1/\valpha^2\sim \alpha$ is then explicit.\footnote{There are 
a handful of exceptions to this in $\Delta v_\alpha^{(6,1,\alpha)}$; all appear in tadpoles, with the exception of
the contribution from the Class-1 coefficient $C_W$. }

\begin{table}[t]
	\begin{center}
		\def\arraystretch{1.3}
		\begin{tabular}{|c|c||c|c|}
			\hline $M_h$ & $125.1$~GeV & $\overline{m}_b(M_h)$       & 3.0~GeV \\ 
			\hline $m_t$ & $172.9$~GeV & $\valpha (M_h)$             & $241.7$~GeV \\ 
			\hline $M_W$ & $80.38$~GeV & $v_\mu$                     & 246.2 GeV \\ 
			\hline $M_Z$ & $91.19$~GeV & $\alpha_s \left(M_h\right)$ & 0.1125 \\
			\hline 
		\end{tabular} 
		\caption{\label{tab:Inputs}Input parameters employed throughout the paper. Note that $v_\alpha$ is a derived parameter.}
	\end{center}
\end{table}

The expansion coefficients in Eq.~\eqref{eq:vT_elim_vhat} are 
determined by the counterterms for the input parameters $M_W$, $M_Z$,  and 
the electric charge $e$.   These are calculated from two-point functions as in \cite{Cullen:2019nnr}. 
In the $\alpha$ scheme, we relate the bare and renormalised quantities up to NLO as
\begin{align}
\label{eq:Xalpha}
X_0 & = X \left(1 +  \frac{1}{v_\alpha^2} \Delta X^{(4,1,\alpha)} + \Delta X^{(6,1,\alpha)}\right) \, , 
\end{align} 
where $X\in\{M_W,M_Z,e\}$ and $X_0$ are the corresponding bare parameters.  We use the same notation
for the expansion coefficients of the derived parameters $c_w$ and $s_w$, so that, for instance,
\begin{align}
\label{eq:SWren}
\Delta s_w^{(i,1,\alpha)} = -\frac{c_w^2}{s_w^2} 
\left(\Delta M_W^{(i,1,\alpha)}- \Delta M_Z^{(i,1,\alpha)} \right) \,.
\end{align}
At tree level the relation between $v_T$ and $\valpha$ is given by \cite{Cullen:2019nnr}
\begin{align}
\label{eq:bare_vhat}
\frac{1}{v_T^2} & = \frac{1}{\valpha^2}\left(1 + 2 \valpha^2  \frac{c_w}{s_w}
\left[C_{HWB} + \frac{c_w}{4s_w}C_{HD} \right] \right) \,.
\end{align}
Interpreting this as a relation between the bare parameters, renormalising
them as in  Eq.~\eqref{eq:Xalpha}, and matching with
Eq.~\eqref{eq:vT_elim_vhat} we find
\begin{align}
\Delta v_\alpha^{(6,0,\alpha)} =&\, -2 \frac{c_w}{s_w}
\left[C_{HWB} + \frac{c_w}{4 s_w}C_{HD} \right]   \,,
\label{eq:valpha60} \\
\Delta v_\alpha^{(4,1,\alpha)} =&\, 2\left(\Delta M_W^{(4,1,\alpha)} + \Delta s_w^{(4,1,\alpha)} - \Delta e^{(4,1,\alpha)}\right)\, ,
\label{eq:dV2_vhat} \\
\begin{split}
\Delta v_\alpha^{(6,1,\alpha)} =&\, 2\left(\Delta M_W^{(6,1,\alpha)} + \Delta s_w^{(6,1,\alpha)} - \Delta e^{(6,1,\alpha)}\right) \\
&  +\frac{ 2}{c_w s_w} \left[C_{HWB}+\frac{c_w}{2s_w}C_{HD} \right]  \Delta s_w^{(4,1,\alpha)} \\
&   -  2 \valpha^2 \frac{c_w}{s_w}\left[\delta C_{HWB}+\frac{c_w}{4s_w}\delta C_{HD} \right] \,.
\end{split}
\label{eq:dV2_vhat_NLO}
\end{align}

When calculating EW corrections to heavy boson decay rates, it is natural to use a renormalisation scheme for $\alpha$
that avoids sensitivity to light fermion masses in counterterms.
In the remainder of the paper, unless otherwise stated,
we will use the \msbar~definition of $\alpha$ in a five-flavour version of 
QED$\times$QCD, where the top quark and heavy electroweak bosons have been integrated out and thus contribute finite parts to the  $\Delta e$ through decoupling constants, see the discussion in Section 4.2 of \cite{Cullen:2019nnr}.\footnote{The running $\alpha$ defined in the five-flavour version of QED$\times$QCD is denoted as $\overline{\alpha}^{(\ell)}$ in that reference.} This running coupling, $\alpha(\mu)$, is related to the effective on-shell definition at $\sqrt{s}=M_Z$, $\alpha^{\rm O.S.}(M_Z)$, according to
\begin{align}
\label{eq:ToOnShell}
\alpha(\mu = M_Z) &= \alpha^{\rm O.S.}(M_Z)\left[ 1 +\frac{\alpha^{\rm O.S.}(0)}{\pi} \frac{100}{27}\right] \, .
\end{align}
Numerically, using the values $1/\alpha^{\rm O.S.}(0)=137.036$ and $1/\alpha^{\rm O.S.}(M_Z)=128.946$ (from \cite{ParticleDataGroup:2022pth} and~\cite{Keshavarzi:2019abf} respectively), the coupling constant and the derived quantity $\valpha$ evaluate to 
\begin{align}
1/\alpha(M_Z) = 127.85 \, ,  \, &\qquad \valpha(M_Z) = 242.16\, {\rm GeV} \, .
\end{align}
The (fixed-order) solution to the RG equation for the running of $\alpha$ to other scales necessary for this work is discussed in more detail in Section 5.2 of~\cite{Cullen:2019nnr} and is given as
\begin{align}
\label{eq:run_alpha}
\alpha(\mu) =\alpha(M_Z)\left(1+2 \gamma_e(M_Z) \ln\frac{\mu}{M_Z}\right) \, ,
\end{align}
where $\gamma_e(M_Z) = \frac{\alpha(M_Z)}{\pi}\times \frac{20}{9}$.
Values for these parameters at other scales considered in this work are given as
\begin{align}
1/\alpha(M_W) = 128.03\, ,  \, &\qquad \valpha(M_W) = 242.33\, {\rm GeV} \, , \nonumber \\
1/\alpha(M_h) = 127.40\, ,  \, &\qquad \valpha(M_h) = 241.74\, {\rm GeV} \, . \nonumber
\end{align}

\subsection{The $\alpha_\mu$ scheme}
\label{sec:AlphaMuScheme}

In contrast to the $\alpha$ scheme, the $\alpha_\mu$ scheme uses $G_F$ rather than $\alpha$ as an input parameter.  
This can be implemented by modifying the counterterm for $v_T$ in the $\alpha$ scheme, 
Eq.~\eqref{eq:vT_elim_vhat}, to read 
\begin{align}
\label{eq:vT_elim_vmu}
\frac{1}{v_{T,0}^2} =  \frac{1}{\vmu^2} \left[ 1  - \vmu^2  \Delta v_\mu^{(6,0,\mu)}   -
\frac{1}{\vmu^2} \Delta v_\mu^{(4,1,\mu)} -  \Delta v_\mu^{(6,1,\mu)}\right] \, .
\end{align}
The superscripts on $\Delta v_\mu$ have the same meaning as in 
Eq.~\eqref{eq:vT_elim_vhat}, so that in particular the superscript $\mu$ means that the  
expansion coefficients multiply distinct powers of $\vmu$ instead of $v_\alpha$, where 
\begin{align}
\label{eq:v_mu_def}
\vmu \equiv \left(\sqrt{2}G_F\right)^{-\frac{1}{2}}
\equiv \frac{2 M_W s_w }{\sqrt{4\pi \alpha_\mu}}  \, .
\end{align}
We have introduced the derived EW coupling $\alpha_\mu$ in the final equality of the above equation. 
Using the PDG value of $G_F = 1.166 \times 10^{-5}$ GeV$^{-2}$~\cite{ParticleDataGroup:2022pth} gives $\alpha_\mu \approx 1/132$, and the corresponding value of $\vmu$ is given in Table~\ref{tab:Inputs}.

The expansion coefficients in Eq.~(\ref{eq:vT_elim_vmu}) are obtained by a renormalisation condition relating
muon decay in SMEFT with that in Fermi theory. We give the technical details of the calculation, and 
results for the coefficients $\Delta v_\mu^{(i,j,\mu)}$, in Appendix~{\ref{sec:NLO_vmu}.
A previous result for these coefficients has been given in~\cite{Dawson:2018pyl}, using a simplified flavour 
structure for the SMEFT Wilson coefficients, and omitting tadpoles such that the results are gauge dependent 
and limited to $R_\xi$ gauge.  While we have made no flavour assumptions and included tadpole contributions in the FJ tadpole scheme~\cite{Fleischer:1980ub}, so that the coefficients are gauge invariant, we have checked that our results
are consistent with those in \cite{Dawson:2018pyl} when the same calculational set-up is used, 
thus providing a strong check on both sets of results.

We can convert results in the $\alpha_\mu$ scheme to the  $\alpha$ scheme
using the perturbative relation between $v_\mu$ and $v_\alpha$.  A useful quantity for this purpose is 
\begin{align}
\label{eq:DeltaR}
\frac{\valpha^2}{\vmu^2} &\equiv 1+ \Delta r \, .
\end{align}
Two equivalent SMEFT expansions of this quantity are
\begin{align}
\Delta r & = \valpha^2 \Delta r^{(6,0)} + \frac{1}{\valpha^2} \Delta r^{(4,1)} + \Delta r^{(6,1)} \, , \\
\label{eq:DeltaRvmu}
  & =  \vmu^2  \Delta r^{(6,0)} + \frac{1}{\vmu^2} \Delta r^{(4,1)} + \Delta r^{(6,1)}\, .
\end{align}
The expansion coefficients are the same whether expanded in $v_\mu$ or $v_\alpha$, so we use superscripts for operator dimension and loop order only.\footnote{It is understood that any implicit $v_T$ dependence in the $(6,1)$ term  is expressed in terms of $v_{\alpha}$ in the first line or $v_\mu$ in the second.} They are obtained by equating the two expressions for $v_{T,0}$ given in Eq.~(\ref{eq:vT_elim_vhat})  and Eq.~(\ref{eq:vT_elim_vmu}) and performing a SMEFT expansion, yielding the result
\begin{equation}
\begin{split}
\label{eq:dr_exp_coeffs}
\Delta r^{(6,0)} & = \Delta v_{\mu\alpha}^{(6,0)} \, , \\
\Delta r^{(4,1)} & = \Delta v_{\mu\alpha}^{(4,1)}  \, , \\
\Delta r^{(6,1)} & = \Delta v_{\mu\alpha}^{(6,1)}  + 2\Delta v_\mu^{(4,1,\mu)} \Delta v_{\mu\alpha}^{(6,0)} \, , 
\end{split}
\end{equation}
where we have defined 
\begin{align}
 \Delta v_{\mu\alpha}^{(i,j)} = \Delta v_\mu^{(i,j,\mu)} -  \Delta v_\alpha^{(i,j,\alpha)}  \, .
\end{align}
For two-body decays of heavy bosons, the SMEFT expansion coefficients in the $\alpha_\mu$ or $\alpha$ scheme take the form
\begin{align}
\Gamma& = \frac{F}{\vmu^2}\left[1+ \vmu^2 \Delta_\Gamma^{(6,0,\mu)}  +
\frac{1}{\vmu^2} \Delta_\Gamma^{(4,1,\mu)} +\Delta_\Gamma^{(6,1,\mu)}  \right] \nonumber \\
& =  \frac{F}{\valpha^2}\left[1+ \valpha^2 \Delta_\Gamma^{(6,0,\alpha)}  +
\frac{1}{\valpha^2} \Delta_\Gamma^{(4,1,\alpha)} +\Delta_\Gamma^{(6,1,\alpha)}  \right]  \, ,
\end{align}
where $F$ does not depend on $v_\mu$ in the first line or $v_\alpha$ in the second. The relation between the expansion coefficients in the two schemes is
\begin{align}
\label{eq:mu_to_alpha}
\Delta_\Gamma ^{(6,0,\alpha)} & =\Delta_\Gamma ^{(6,0,\mu)} + \Delta r^{(6,0)} \, , \nonumber \\
\Delta_\Gamma ^{(4,1,\alpha)} & =\Delta_\Gamma ^{(4,1,\mu)} + \Delta r^{(4,1)}  \, , \nonumber  \\
\Delta_\Gamma ^{(6,1,\alpha)}  & = \Delta_\Gamma ^{(6,1,\mu)} + \Delta r^{(6,1)}  + 2 \Delta_\Gamma ^{(4,1,\mu)}  \Delta r^{(6,0)}   \, .
\end{align}
Conversions from the $\alpha$ to the $\alpha_\mu$ scheme work in a similar manner.  As a simple example, the expansion of
counterterms $X$ in Eq.~(\ref{eq:Xalpha}) in the $\alpha_\mu$ scheme is obtained by replacing $\alpha\to \mu$ in that
equation, with expansion coefficients related through
\begin{align}
\Delta X^{(4,1,\mu)} = \Delta X^{(4,1,\alpha)} \, ,\qquad \Delta X^{(6,1,\mu)} = \Delta X^{(6,1,\alpha)} -\Delta r^{(6,0)}\Delta X^{(4,1,\alpha)} \,.
\end{align}
Note that although both the $\alpha$ and the $\alpha_\mu$ scheme use on-shell renormalisation for $M_W$ and $M_Z$, 
the perturbative expansions of the counterterms differ at one-loop in SMEFT.

\subsection{The LEP scheme}
\label{sec:The_LEP_scheme}
At LEP and in SMEFT analyses, one often considers the LEP scheme, where the on-shell $W$-boson mass is not used as an input, but is instead expressed as a SMEFT expansion in terms of the three independent input parameters $\{\alpha,G_F,M_Z\}$.
The SMEFT expansion of the on-shell $W$-boson mass in this  scheme is most easily obtained 
by re-arranging Eq.~(\ref{eq:DeltaR}) and then expanding in $\Delta r$ to find
\begin{align}
\label{eq:GenMW}
M_W^2 
 & = \mwhat^2\left[1-\frac{\hat{s}_w^2}{\hat{c}_{2w}}\Delta r - \frac{\hat{c}_w^2\hat{s}_w^4}{\hat{c}_{2w}^3} \Delta r^2\right] + {\cal O}\left( \Delta r^3 \right) \, ,
\end{align}
where 
\begin{align}
 \mwhat^2 = \frac{M_Z^2}{2}\left(1+\sqrt{1-\frac{4\pi \alpha \vmu^2}{M_Z^2}} \right)\,, 
 \quad \hat{c}^2_w = \frac{\mwhat^2}{M_Z^2} = 1-\hat{s}_w^2 \, , 
 \quad \hat{c}_{2w}= 2\hat{c}_w^2-1 \,.
\end{align} 
In the LEP scheme, the appropriate SMEFT expansion of $\Delta r$ depends only on the derived parameter $\hat{M}_W$.
We therefore define expansion coefficients 
\begin{align}
\Delta r = \vmu^2  \hat{\Delta} r^{(6,0)} + \frac{1}{\vmu^2} \hat{\Delta} r^{(4,1)} + \hat{\Delta} r^{(6,1)}\, ,
\end{align}
where the ``hat'' on the expansion coefficients $\hat{\Delta} r^{(i,j)}$ means that the dependence on the on-shell mass 
$M_W$ in the $\Delta r^{(i,j)}$ in Eq.~(\ref{eq:DeltaRvmu}) has been eliminated in favour of $\hat{M}_W$ through iterative use of Eq.~(\ref{eq:GenMW}).  A short calculation yields the following results:
\begin{align}
\label{eq:rhat_coeffs_ders}
\hat{\Delta} r^{(6,0)}& = \Delta r^{(6,0)} \big|_{M_W = \mwhat} \, , \nonumber \\
\hat{\Delta} r^{(4,1)} & =  \Delta r^{(4,1)} \big|_{M_W = \mwhat} \, ,\nonumber  \\
\hat{\Delta} r^{(6,1)} & =  \Delta r^{(6,1)} 
-  \frac{\hat{s}_w^2}{2\hat{c}_{2w}}\left[ \Delta r^{(6,0)} \partial_W \Delta r^{(4,1)} 
+  \Delta r^{(4,1)}\partial_W\Delta r^{(6,0)}\right]\Bigg|_{M_W = \mwhat} \,,
\end{align}
where the notation $\big|_{M_W=\hat{M}_W}$ means that $M_W$ is to be replaced by $\hat{M}_W$ and we have defined
\begin{align}
\partial_W\equiv M_W \frac{\partial}{\partial M_W} \, .
\end{align}
Notice that the term $\hat{\Delta} r^{(6,1)}$ involves derivatives of Passarino-Veltmann functions, 
which at one-loop level are simple to evaluate.

We can now write the SMEFT expansion of $M_W$ in the LEP scheme as
\begin{align}
\label{eq:MW_prediction}
M_W & = \mwhat\left[1+ \vmu^2 \hat{\Delta}_W^{(6,0,\mu)}   + \frac{1}{\vmu^2} \hat{\Delta}_W^{(4,1,\mu)}  
+ \hat{\Delta}_W^{(6,1,\mu)} \right]  \, , 
\end{align}
where
\begin{align}
\label{eq:MW_prediction_coeffs}
\hat{\Delta}_W^{(6,0,\mu)} &  =  -\frac{\hat{s}_w^2}{2\hat{c}_{2w}} \hat{\Delta} r^{(6,0)} ,  \nonumber \\
 \hat{\Delta}_W^{(4,1,\mu)} &  = -\frac{\hat{s}_w^2}{2\hat{c}_{2w}} \hat{\Delta} r^{(4,1)} \, , \nonumber \\
  \hat{\Delta}_W^{(6,1,\mu)} &  =   -\frac{\hat{s}_w^2}{2\hat{c}_{2w}} \hat{\Delta} r^{(6,1)} 
  -\frac{\hat{s}_w^4}{4\hat{c}_{2w}^2} \left(1+\frac{4\hat{c}_w^2}{\hat{c}_{2w}} \right)\hat{\Delta} r^{(6,0)} \hat{\Delta} r^{(4,1)} \,.
\end{align}
The above expressions allows for the conversion of the SMEFT expansion of any quantity from the $\alpha_\mu$ scheme to the LEP scheme. The conversion takes the form
\begin{equation}
\label{eq:LEP_convert_X}
\begin{split}
& X(M_W)\left[ 1+ \vmu^2 \Delta_X^{(6,0,\mu)}   + \frac{1}{\vmu^2} \Delta_X^{(4,1,\mu)}  
  + \Delta_X^{(6,1,\mu)} \right]  \\
  & =X(\mwhat)\left[ 1+ \vmu^2 \hat{\Delta}_X^{(6,0,\mu)}   + \frac{1}{\vmu^2} \hat{\Delta}_X^{(4,1,\mu)}  
  + \hat{\Delta}_X^{(6,1,\mu)} \right] \, ,
\end{split}
\end{equation}
where the expansion coefficients $\Delta_X$ ($\hat{\Delta}_X$) are functions of $M_W$ ($\mwhat$).  They are related through
\begin{align}
\label{eq:LEP_convert}
\hat{\Delta}_X^{(6,0,\mu)}& =  \Delta_X^{(6,0,\mu)}  +  \hat{\Delta}_W^{(6,0,\mu)} \frac{\partial_W X}{X} ,\nonumber \\ 
\hat{\Delta}_X^{(4,1,\mu)} &=  \Delta_X^{(4,1,\mu)}  +   \hat{\Delta}_W^{(4,1,\mu)} \frac{\partial_W X}{X} , \nonumber \\
\hat{\Delta}_X^{(6,1,\mu)} &=  \Delta_X^{(6,1,\mu)}  +  \hat{\Delta}_W^{(6,0,\mu)} \partial_W \Delta_X^{(4,1,\mu)} +    \hat{\Delta}_W^{(4,1,\mu)} \partial_W \Delta_X^{(6,0,\mu)}\\
 &+ \frac{1}{X}\bigg[ \bigg(\hat{\Delta}_W^{(6,1,\mu)}+  \Delta_X^{(4,1,\mu)}\hat{\Delta}_W^{(6,0,\mu)}+\Delta_X^{(6,0,\mu)}\hat{\Delta}_W^{(4,1,\mu)}\bigg)\partial_W X \nonumber \\
 &+  
\hat{\Delta}_W^{(6,0,\mu)}\hat{\Delta}_W^{(4,1,\mu)}\partial^2_W X \bigg]\nonumber \, ,
\end{align}
where $X=X(M_W)$,  in a slight abuse of notation we have defined
\begin{equation}
\partial^2_W\equiv M_W^2 \frac{\partial^2}{\partial M_W^2} \, ,
\end{equation}
and one is to set $M_W=\mwhat$ on the right-hand side of the relations in Eq.~(\ref{eq:LEP_convert}).

As a simple example, we can relate the counterterm for $M_W$  in the on-shell scheme to that in the LEP scheme. Setting $X=M_W$ in Eq.~(\ref{eq:LEP_convert_X}), and recalling that the on-shell definition of $M_W$ has no tree-level dimension-six contributions, we can write 
\begin{equation}
\begin{split}
M_{W,0}& = M_W\left(1+\frac{1}{\vmu^2}\Delta M_W^{(4,1,\mu)}+\Delta M_W^{(6,1,\mu)}\right) \\
& = \hat{M}_W \left(1+\vmu^2 \hat{\Delta}\hat{M}_W^{(6,0,\mu)} + \frac{1}{\vmu^2}\hat{\Delta}\hat{M}_W^{(4,1,\mu)}+\hat{\Delta}\hat{M}_W^{(6,1,\mu)}\right) \, .
\end{split}
\end{equation}
The terms on the second line as determined from Eq.~(\ref{eq:LEP_convert}) read
\begin{align}
\hat{\Delta} \hat{M}_W^{(6,0,\mu)} =& \, \hat{\Delta}_W^{(6,0,\mu)} \, , \\
\hat{\Delta} \hat{M}_W^{(4,1,\mu)} =& \, \hat{\Delta}_W^{(4,1,\mu)} + \Delta M_W^{(4,1,\mu)} \bigg|_{M_W=\hat{M}_W} \, , \\
\begin{split}
\hat{\Delta} \hat{M}_W^{(6,1,\mu)} =& \,   \hat{\Delta}_W^{(6,1,\mu)} + \Delta M_W^{(6,1,\mu)} + \hat{\Delta}_W^{(6,0,\mu)} \Delta M_W^{(4,1,\mu)} \\
& + \hat{\Delta}_W^{(6,0,\mu)} \partial_W \Delta M_W^{(4,1,\mu)}  \bigg|_{M_W=\hat{M}_W} \, .
\end{split}
\label{eq:DeltaHMWh61mu}
\end{align}
We emphasise, however, that the LEP scheme uses $\{\alpha,G_F, M_Z\}$ as input parameters, so the result is ultimately a function of these parameters and the associated counterterms $\{\hat{\Delta}e, \hat{\Delta}v_\mu, \hat{\Delta}M_Z\}$, which can be obtained from expansion coefficients in the $\alpha$ or $\alpha_\mu$ scheme similarly to $\hat{\Delta}\hat{M}_W$.

\section{Salient features of the EW input schemes}
\label{sec:salient}

We are mainly interested in two features of the EW input schemes: 
the number of Wilson coefficients they introduce into physical observables through renormalisation, 
and perturbative convergence. Ideally, one would like a small number of 
coefficients to appear, so that the finite parts of observables are dominated by  process-specific   
Wilson coefficients rather than those related to the EW renormalisation
scheme.  Furthermore, one would like to avoid large corrections between orders,
so that perturbation theory is well behaved and can safely be truncated 
at a low order.  We discuss these two issues in the following subsections.

\subsection{Number of Wilson coefficients}
\label{sec:Counting}

It is a simple matter to count  the number of Wilson coefficients appearing in the finite parts of counterterms for the bare 
parameters $M_{Z,0}$, $M_{W,0}$ and $v_{T,0}$ in the different input
schemes.  The results at LO and NLO are  listed in Table \ref{tab:NumWC}.  Here and below we exclude Wilson coefficients which contribute only through tadpoles and therefore drop out of observables. This includes $C_H$ and  $C_{\substack{uH\\ 33}}$ in each of the three counterterms considered here.   
Note that although all schemes use the on-shell renormalisation scheme for $M_Z$, its dimension-six counterterm still
differs between the schemes. To see this explicitly, we note that expansion coefficients in 
$\alpha_\mu$ and $\alpha$ schemes can be written in the form
\begin{align}
\label{eq:61diffs}
M_{Z,0}  = M_Z \left(1 +  \frac{1}{v_\sigma^2} \Delta M_Z^{(4,1)} + \Delta M_Z^{(6,1)}  - \Delta v_\sigma^{(6,0,\sigma)} \Delta M_Z^{(4,1)}  \right)  \, ,
\end{align}
where here and throughout the remainder of the paper the choice of $\sigma \in \{\mu,\alpha\}$ selects between the 
$\alpha$ and $\alpha_\mu$ schemes.  An analogous equation holds for the counterterms for $M_W$. The coefficients  $\Delta M_Z^{(4,1)}$ and $\Delta M_Z^{(6,1)}$ are the same in the two schemes, but differences in the dimension-six piece arise due to the renormalisation of $v_T$. In the LEP scheme one must use $\sigma=\mu$ and in addition apply Eq.~(\ref{eq:MW_prediction}) to trade $M_W$  for
$\hat{M}_W$, which gives an additional scheme-dependent contribution.  

\begin{table}[t]
\centering
\begin{tabular}{cc||ccc||c}
                              &     & $M_W$ & $M_Z$ & $v_T$ & Total \# unique WC \\ \hline \hline
\multirow{2}{*}{$\alpha$}     & LO  &   0    &         0                 &  2   &      2              \\
                              & NLO &   12    &     29                     &    29     &   29                 \\ \hline
\multirow{2}{*}{$\alpha_\mu$} & LO  &   0    & 0                       &    3     &     3               \\
                              & NLO &    13   &  30     &                 12         &         33          \\ \hline
\multirow{2}{*}{LEP}          & LO  &  5      &      0                    &  3       &            5        \\
                              & NLO &    33   &      30                  &    12     &       33            
\end{tabular}
\caption{Number of Wilson coefficients introduced in the dimension-six counterterms for the bare $M_W$, $M_Z$ and $v_T$ at LO and NLO, as well as the number of unique coefficients between them.}
\label{tab:NumWC}
\end{table}

The specific Wilson coefficients appearing in the various counterterms in the $\alpha$ scheme are determined by 
the two-point functions shown in Figure~\ref{fig:twopoint_diags}.  The counterterm for the $W$-boson mass
contains the following coefficients:
\begin{align}
\label{eq:MW_coeffs}
\Delta M_W^{(6,1,\alpha)}:  \quad \{C_W,C_{H\Box},C_{HD}, C_{HW},C_{HWB}, C_{\substack{Hl \\ ii}}^{(3)} ,C_{\substack{Hq \\ ii}}^{(3)},C_{\substack{uW \\ 33}}  \} \,, \quad i=1,2,3 \,.
\end{align}
$C_{H\Box}$ and $C_{HW}$ contribute to the two left-most topologies in Figure~\ref{fig:twopoint_diags}, while $C_{HWB}$ and $C_W$ contribute to topologies three and four, which involve vertices with at least three gauge bosons. $C_{HD}$ appears in all four purely bosonic diagrams. We see that 7 of the 12 coefficients appearing are due to flavour-specific $W$ couplings to fermions,  arising from the right-most graph in Figure~\ref{fig:twopoint_diags}.  Since in the SM
the $W$ boson couples only to left-handed fermions, the SMEFT operators must also be left-handed unless they 
contain a top-quark loop (in which case a chirality flip is associated with a power of $m_t$), which explains the relatively
small number appearing. For the $Z$-boson mass, on the other hand, both left and right-handed couplings are relevant even
for massless fermions, and operators containing the field-strength tensor for the hypercharge field $B_\mu$, namely 
$C_{HB}$ and $C_{uB}$, contribute as well. This leads to a much larger number of coefficients compared to $M_W$. The full set is:
\begin{align}
\label{eq:MZ_coeffs}
\Delta M_Z^{(6,1,\alpha)}: & \quad  \{C_W,C_{H\Box},C_{HD}, C_{HW}, C_{HB}, C_{HWB},  C_{\substack{Hl \\ ii}}^{(1)} ,C_{\substack{Hq \\ ii}}^{(1)}, C_{\substack{Hl \\ ii}}^{(3)} ,C_{\substack{Hq \\ ii}}^{(3)},C_{\substack{uW \\ 33}}, C_{\substack{uB \\ 33}} , \nonumber \\
 & \quad C_{\substack{He \\ ii}}, C_{\substack{Hd \\ ii}}, C_{\substack{Hu \\ ii}}  \} \,, \quad i=1,2,3 \,.
\end{align}
The counterterm $\Delta v_\alpha^{(6,1,\alpha)}$ requires also the counterterm $\Delta e$, as shown in Eq.~({\ref{eq:dV2_vhat_NLO}).  Only those Wilson coefficients appearing in $W$, top-quark or Higgs loops contribute to the finite parts of the counterterm for electric charge renormalisation (through decoupling constants, as explained in \cite{Cullen:2019nnr}),
which limits the result to the following 6 coefficients: 
\begin{align}
\label{eq:E_coeffs}
\Delta e^{(6,1,\alpha)}: \quad \{C_W, C_{HW}, C_{HB}, C_{HWB}, C_{\substack{uW \\ 33}}, C_{\substack{uB \\ 33}}  \} \, .
\end{align}
All of these are already contained in $\Delta M_Z^{(6,1,\alpha)}$, so the set of coefficients contributing to $\Delta v_\alpha^{(6,1,\alpha)}$ is the same as in Eq.~(\ref{eq:MZ_coeffs}).

In the $\alpha_\mu$ scheme, one needs the counterterms $\Delta v_\mu^{(6,j,\mu)}$, which are calculated from muon 
decay in Appendix~\ref{sec:NLO_vmu}. In SMEFT, two kinds of coefficients appear at NLO -- 
those that involve modified couplings of the external fermions, including 
four-fermion operators of the kind shown in Figure~\ref{fig:fermi_4ferm_diags_duplicate}, or those that contribute
to the $W$-boson two-point function at vanishing external momentum.  The latter condition eliminates some operators
compared to what is seen in $\Delta M_W$ itself (in the case of massless fermions or certain derivative couplings), while
the former increases it mainly due to four-fermion operators. The end result is that the following set appears:
\begin{align}
\label{eq:dVmu_coeffs}
\Delta v_\mu^{(6,1,\mu)}:  \quad \{C_{H\Box},C_{HD}, C_{HWB}, C_{\substack{Hl \\ jj}}^{(1)} ,C_{\substack{Hl \\ jj}}^{(3)},C_{\substack{Hq\\ 33}}^{(3)}, C_{\substack{ll\\ 1221}},C_{\substack{ll\\ 1122}},C_{\substack{lq\\ jj33}} \}\, , \quad j=1,2 \,.
\end{align}
The counterterms for $M_W$ and $M_Z$ are also modified compared to the $\alpha$ scheme, as follows 
from Eq.~(\ref{eq:61diffs});  one finds that the $\alpha_\mu$ scheme contains 
the four-fermion coefficient $C_{\substack{ll\\ 1221}}$ in addition to the $\alpha$-scheme coefficients listed in Eqs.~(\ref{eq:MW_coeffs}, \ref{eq:MZ_coeffs}). 

Finally, in the LEP scheme the counterterm $\hat\Delta \hat{M}_W^{(6,1,\mu)}$ (see Eq.~(\ref{eq:DeltaHMWh61mu}})) is a function of those for $e$, $M_Z$, and $v_T$ (renormalised in the $\alpha_\mu$ scheme), and thus contains the full set of 33 unique coefficients that also appear in the $\alpha_\mu$ scheme, while no additional coefficients appear in the counterterms for $M_Z$ or $v_T$  compared to the $\alpha_\mu$ scheme.

\begin{figure}[t]
	\centering
    \includegraphics[width=0.19\textwidth]{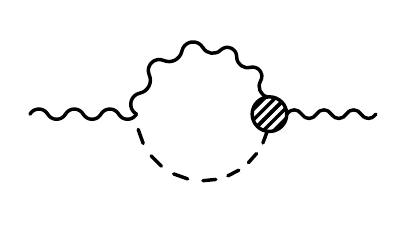}
    \includegraphics[width=0.19\textwidth]{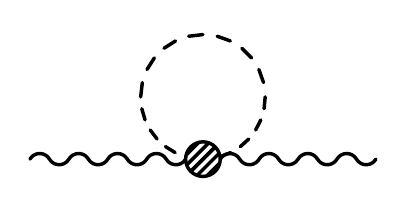} 
    \includegraphics[width=0.19\textwidth]{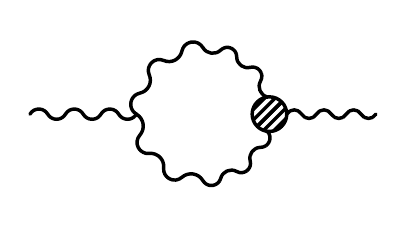} 
	\includegraphics[width=0.19\textwidth]{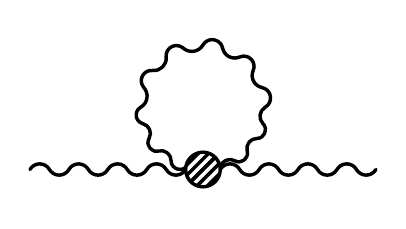}
	\includegraphics[width=0.19\textwidth]{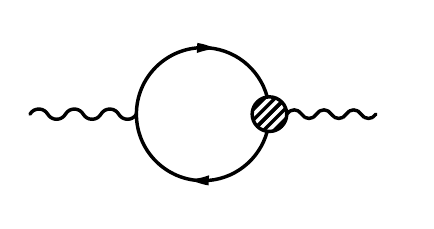} 
	\caption{Representative Feynman diagrams contributing to the $WW$, $ZZ$, $\gamma Z$, and $\gamma\gamma$ two-point functions in SMEFT. }
	\label{fig:twopoint_diags}
\end{figure}

 \begin{figure}[t]
	\centering
	\includegraphics[width=0.25\textwidth]{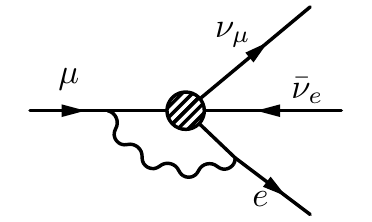} 
	\includegraphics[width=0.25\textwidth]{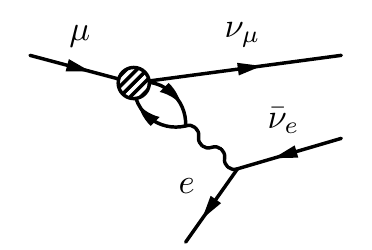} 
	\caption{Representative Feynman diagrams contributing to the decay of the muon at one loop and involving four-fermion operators.}
	\label{fig:fermi_4ferm_diags_duplicate}
\end{figure}

The conclusion of this counting exercise is that there is a large overlap between the set of operators appearing
in the NLO counterterms  in the different schemes.  The main difference is that a handful of four-fermion operators
related to muon decay appear in the LEP and $\alpha_\mu$ schemes but not in the $\alpha$ scheme.  

The number of Wilson  coefficients contributing to observables in the different schemes is process dependent and is determined by the structure of  the LO amplitude.  For instance, consider a process involving a $\gamma \ell \ell$ vertex, where $\ell$ is a charged
lepton and $\gamma$ is a photon. In the $\alpha$ scheme, the square of the bare vector-coupling
vertex plus SMEFT counterterms (other than from field strength renormalisation) reads
\begin{align}
\frac{4 M_W^2 s_w^2}{v_\alpha^2}\left(1 + \frac{ 2\Delta e^{(4,1,\alpha)}}{v_\alpha^2} + 2\Delta e^{(6,1,\alpha)} \right) \,.
\end{align}
In the $\alpha_\mu$ scheme, on the other hand, the bare vertex plus associated
counterterms read
\begin{equation}
\label{eq:dE61_vmu}
\begin{split}
& \frac{4 M_W^2 s_w^2}{v_\mu^2}\bigg\{ \frac{2 \Delta e^{(4,1,\alpha)}}{v_\mu^2} + 2 \Delta e^{(6,1,\alpha)} - 4 \Delta e^{(4,1,\alpha)}\Delta r^{(6,0)} \bigg\}  \\
&+ \frac{4 M_W^2 s_w^2}{v_\mu^2}\bigg\{1 - \vmu^2\Delta r^{(6,0)} -
 \frac{1}{\vmu^2}\Delta r^{(4,1)} - \Delta r^{(6,1)}+2\Delta r^{(6,0)}\Delta r^{(4,1)} \bigg\}\, .
 \end{split}
\end{equation}
The two results are equal to each other if Eq.~(\ref{eq:DeltaR}) is used to relate $v_\mu$ to $v_\alpha$,
but when the numerical value of $v_\mu$ is used as an input the terms on the second line of Eq.~(\ref{eq:dE61_vmu}) contribute a large number of coefficients compared to what one has in the $\alpha$ scheme. The same set of coefficients contributes to muon decay calculated in the $\alpha$ scheme, or in the LEP scheme when $M_W$ appears in a  
tree-level vertex.

\subsection{Perturbative convergence}
\label{subsec:PertConv}

Generally speaking, one uses renormalisation schemes that avoid sensitivity to large logarithms of light fermion masses in fixed-order corrections, and also tadpole contributions to finite parts of observables in cases where some parameters are renormalised in the \msbar~scheme and some in the on-shell scheme~\cite{Cullen:2019nnr}. As long as those two issues are dealt with, top-quark loops are the main source of enhanced NLO corrections in the finite parts of counterterms. These can be especially important when associated with the counterterm $\Delta s_w$, since they involve inverse powers of $s_w^2\sim 0.25$ through the relation
\begin{align}
2 \Delta s_w = - 2 \frac{c_w^2}{s_w^2}(\Delta M_W-\Delta M_Z) \approx -7(\Delta M_W-\Delta M_Z) \, ,
\end{align}
where the factor of 2 is chosen to match that in Eq.~(\ref{eq:dV2_vhat}). 

In the SM, enhanced corrections from top-loop contributions to $\Delta s_w$ related to the renormalisation scheme are easy to trace. First, by analysing the one-loop Feynman diagrams in the large-$m_t$ limit, one can show that in the $\alpha_\mu$ scheme
\begin{align}
\label{eq:Vmu_LMT_SM}
\Delta v_{\mu}^{(4,1,\mu)}\bigg|_{m_t\to \infty} \equiv  \Delta v_{\mu,t}^{(4,1,\mu)} = 2\Delta M_{W,t}^{(4,1,\mu)} \,.
\end{align}
The subscript ``$t$" here and below refers to the large-$m_t$ limit of the given quantity, i.e.\ the terms containing positive powers of $m_t$ in the limit $m_t\to\infty$. Second, using Eqs.~(\ref{eq:dV2_vhat}, \ref{eq:dr_exp_coeffs}), along with the fact that the SM contributions to $\Delta e$ are subleading in the large-$m_t$ limit, the $\alpha$-scheme result is
\begin{align}
\label{eq:Valpha_LMT_SM}
\Delta v_{\alpha,t}^{(4,1,\alpha)} = - \Delta r_t^{(4,1)} +   2\Delta M_{W,t}^{(4,1,\alpha)}  \, , 
\end{align}
where 
\begin{align}
\label{eq:delta_r41}
\frac{\Delta r_t^{(4,1)}}{\valpha^2} = - \frac{2\Delta s_{w,t}^{(4,1,\alpha)}}{\valpha^2}  \equiv - \frac{c_w^2}{s_w^2} \frac{ \Delta \rho_t^{(4,1)}}{v_\alpha^2} \approx - 3.4 \% \, ,
\end{align}
and we have defined
\begin{align}
\label{eq:drho41}
\frac{\Delta \rho_t^{(4,1)}}{\valpha^2} \equiv \frac{3}{16\pi^2} \frac{m_t^2}{\valpha^2}\approx 1\%\,.
\end{align}
The numerical values above use $\mu=M_W$ to evaluate the running parameter $v_\alpha$, along with the inputs in Table~\ref{tab:Inputs}. Finally, using Eqs.~(\ref{eq:Vmu_LMT_SM}, \ref{eq:Valpha_LMT_SM}), the counterterms for $v_T$ in the large-$m_t$ limit in the two schemes can be written as
\begin{align}
\label{eq:GenWeak}
\frac{1}{v_{T,0}^2}  \bigg |_{m_t\to\infty} = \frac{1}{v_\sigma^2}\left[1+\frac{1}{v_\sigma^2}\left(\Delta r_t^{(4,1)}\delta_{\alpha\sigma} - 2\Delta M_{W,t}^{(4,1)} \right) \right] \, ,
\end{align}
where $\delta_{\alpha\sigma}$ is the Kronecker delta, and we have used that $\Delta M_W^{(4,1,\alpha)}=\Delta M_W^{(4,1,\mu)} = \Delta M_W^{(4,1)}$, see Eq.~(\ref{eq:61diffs}).

For the heavy boson decays considered in this work, the tree-level decay rates all scale as $1/v_T^2$.  Therefore, Eq.~(\ref{eq:GenWeak}) produces a simple pattern for the NLO corrections in the $\alpha$ and $\alpha_\mu$ schemes. In the $\alpha_\mu$ scheme, the tadpole and divergent contributions in $\Delta M_{W,t}^{(4,1)}$ cancel against other such contributions in physical observables, producing one-loop corrections proportional to $\Delta \rho_t^{(4,1)}\sim 1\%$ in the large-$m_t$ limit. In the $\alpha$ scheme, the $\Delta M^{(4,1)}_{W,t}$ term is accompanied by a factor of $\Delta r_t^{(4,1)}$, which produces a correction of roughly $-3.4\%$ compared to the $\alpha_\mu$ scheme. One indeed sees this pattern in the NLO SM corrections to $W$~decays, $Z$~decays, and Higgs~decays into fermions, shown in Tables~\ref{tab:wlnu_nlo}, \ref{tab:hbb},  and \ref{tab:zll_nlo}. Input-scheme dependent NLO corrections to weak vertices are thus better behaved in the $\alpha_\mu$ scheme, and the numerical differences between the two schemes are nearly process independent.\footnote{On the other hand, if the bare vertex contains a photon, then examining Eq.~(\ref{eq:dE61_vmu}) shows that the situation is reversed and $+3.4\%$ correction is associated with the $\alpha_\mu$ scheme.} 

We now ask whether a simple relation between the dominant NLO corrections in the $\alpha$ and $\alpha_\mu$ schemes also exists in SMEFT.  To this end, we first define
\begin{align}
\label{eq:W_alpha}
\frac{M_{W,0}^2}{v_{T,0}^2} z_W \bigg|_{m_t\to \infty} &  \equiv  
\frac{M_{W}^2}{v_{\sigma}^2}\left[1 + v_\sigma^2 K_W^{(6,0,\sigma)} + \frac{1}{v_\sigma^2}K_{W}^{(4,1,\sigma)} + K_{W}^{(6,1,\sigma)} \right]  \, ,
\end{align}
where $z_W$ is the squared wavefunction renormalisation factor of the $W$-boson field.  After replacing the bare quantities 
on the left-hand side by their renormalised counterparts, it is straightforward to determine the $K_W^{(i,j,\sigma)}$ in terms of 
$\Delta M_{W,t}^{(i,j,\sigma)}$, $\Delta z_{W,t}^{(i,j,\sigma)}$, and $\Delta v_{\sigma,t}^{(i,j,\sigma)}$.  This yields
$K_{W}^{(6,0,\sigma)} =-\Delta v_{\sigma,t}^{(6,0,\sigma)}$ at tree level, and substituting in the 
explicit results for the counterterms  leads to the following one-loop expressions in the $\alpha$ scheme:
\begin{align}
\label{eq:KWalpha}
K_{W}^{(4,1,\alpha)} &= \Delta r^{(4,1)}_t  \, ,  \nonumber \\
K_{W}^{(6,1,\alpha)} &= -\frac{1}{2} \dot{K}_{W}^{(6,0,\alpha)}\ln\frac{\mu^2}{m_t^2} + \Delta r_t^{(4,1)}\bigg[\frac{1}{s_w^2}C_{HD}+ \frac{3}{c_w s_w}C_{HWB} \nonumber \\
& + 2 C_{\substack{Hq \\ 33}}^{(3)}+ \frac{2\sqrt{2}(1-2c_w^2)}{c_w^2}\frac{M_W}{m_t}C_{\substack{uW\\ 33}}  \bigg] \, ,
\end{align}
where 
\begin{align}
\label{eq:WaLogs}
&\dot{K}_W^{(6,0,\alpha)} =-4\Delta r_t^{(4,1)}\bigg[C_{HD} +\frac{2s_w}{c_w} C_{HWB} + 2 C_{\substack{Hq \\ 33}}^{(1)} -2 C_{\substack{Hu \\ 33}} \nonumber \\
& \hspace{2.5cm} - \frac{2\sqrt{2}s_w}{c_w^2}\frac{M_W}{m_t}\left(c_w C_{\substack{uB\\ 33}} +\frac{5}{3}s_w C_{\substack{uW\\ 33}} \right)  \bigg] \,.
\end{align}
In the $\alpha_\mu$ scheme one has instead
\begin{align}
\label{eq:KWmu}
K_{W}^{(4,1,\mu)} & = 0 \, , \nonumber \\
K_{W}^{(6,1,\mu)} & = -\frac{1}{2} \dot{K}_{W}^{(6,0,\mu)}\ln\frac{\mu^2}{m_t^2} + \Delta \rho_t^{(4,1)}\sum_{j=1,2} \bigg[C_{\substack{Hl \\ jj}}^{(3)}  - C^{(3)}_{\substack{lq \\ jj33}}\bigg] \, ,
\end{align}
where 
\begin{align}
\dot{K}_{W}^{(6,0,\mu)} &= -4\Delta\rho_t^{(4,1)} \sum_{j=1,2} \bigg[C_{\substack{Hl \\ jj}}^{(3)}  - C^{(3)}_{\substack{lq \\ jj33}}\bigg] \, .
\end{align} 

One sees that the SMEFT expansion of $K_W$ is tadpole free, finite, and independent of the renormalisation scale up to NLO. This is not an accident -- it gives the flavour-independent part of the large-$m_t$ limit of $W$ decay into fermions.   Furthermore, rearranging
the above expressions yields the following result for the $v_T$ counterterms:\footnote{We omit here Wilson coefficient counterterms $\delta C_i$, which contribute only divergent parts and thus do not play a role in the discussion of perturbative convergence.}
\begin{align}
\label{eq:weak_vev}
\Delta v_{\sigma,t}^{(4,1,\sigma)} & = -K_{W}^{(4,1,\sigma)} + 2\Delta M_{W,t}^{(4,1)} \, , \nonumber \\
\Delta v_{\sigma}^{(6,0,\sigma)} & = -K_{W}^{(6,0,\sigma)} \, , \nonumber \\
\Delta v_{\sigma,t}^{(6,1,\sigma)} & = -K_{W}^{(6,1,\sigma)}+ \left[2 \Delta M_{W,t}^{(6,1,\sigma)}
 +2\Delta M_{W,t}^{(4,1)}K_W^{(6,0,\sigma)}+\Delta z_{W,t}^{(6,1,\sigma)} \right] \,.
\end{align}  
 The SM part of Eq.~(\ref{eq:weak_vev}) is identical to Eq.~(\ref{eq:GenWeak}). The dimension-six parts are the generalisation to SMEFT.  In each case, the counterterm for $v_T$ is split into two distinct parts: a physical piece that is a finite, gauge and scale-independent quantity (the $K_W$), plus remaining terms which contain tadpoles and divergent parts that cancel against other such terms in physical observables.  While at one-loop in the SM it was simple to identify the physical factor $\Delta r^{(4,1)}$ in 
 the $\alpha$ scheme by studying the counterterm $v_\alpha^{(4,1,\alpha)}$ alone, in SMEFT it is helpful to choose an observable process in order to split the counterterm into the two distinct parts. While the choice of $W$ decay is not unique,
it leads directly to the SM results obtained from studying $v_T$ alone.

We can now use our expressions for the counterterms for $v_T$ in Eq.~(\ref{eq:weak_vev}) to check whether, as in the SM,  a simple pattern emerges for input-scheme dependent SMEFT corrections to weak vertices. As an example, consider the following expression, which gives a flavour-independent correction to $Z$-boson decays into fermions:
\begin{align}
\label{eq:kZ_def}
z_Z  \frac{M_{Z,0}^2}{v_{T,0}^2}\left(1- v_{T,0}^2 \frac{C_{HD}}{2}\right) \bigg|_{m_t\to \infty} = \frac{M_Z^2}{v_\sigma^2}\left[1 + v_\sigma^2 k_{Z}^{(6,0,\sigma)} + \frac{1}{v_\sigma^2}k_{Z}^{(4,1,\sigma)} + k_{Z}^{(6,1,\sigma)} \right]  \, ,
\end{align}
where $z_Z$ is the wavefunction renormalisation factor squared of the $Z$-boson field.\footnote{Compared to Eq.~(\ref{eq:W_alpha}) an additional factor of $C_{HD}$ arises for $Z$-boson decays. This arises from the relations between the $W/Z$-mass and the Lagrangian parameters in SMEFT and can be seen by considering the flavour independent part of Eq.~(5.25) in addition to Eq.~(5.27) in~\cite{Alonso:2013hga}.} The expression on the right-hand side is finite, tadpole free, and scale-independent up to NLO. Writing the counterterms  for $v_T$ using Eq.~(\ref{eq:weak_vev}), one has
\begin{align}
\label{eq:k_Z_alpha_SM}
k_{Z}^{(6,0,\sigma)} & = K_{W}^{(6,0,\sigma)}+ k_{Z}^{(6,0)}  \, , \nonumber \\
k_{Z}^{(4,1,\sigma)} & = K_{W}^{(4,1,\sigma)} +k_{Z}^{(4,1)}  \, , \\
k_{Z}^{(6,1,\sigma)} & = K_{W}^{(6,1,\sigma)} +2 k_{Z}^{(4,1)} K_{W}^{(6,0,\sigma)} +k_{Z}^{(6,1)} \nonumber \, .
\end{align}
Here we have split each term in the perturbative expansion further into scheme dependent and independent parts (the latter being denoted without the $\sigma$ superscript). Both the scheme dependent and independent parts are separately scale independent and tadpole free. The results for the scheme-independent pieces are
\begin{align}
k_{Z}^{(6,0)} &= -\frac{C_{HD}}{2} \, , \nonumber \\
k_{Z}^{(4,1)} &= 2\left(\Delta M_{Z}^{(4,1)}- \Delta M_{W}^{(4,1)}\right) = \Delta \rho_t^{(4,1)} \, , \\
k_{Z}^{(6,1)} &= 2\Delta \rho_t^{(4,1)} C_{\substack{Hq \\ 33}}^{(3)} -\frac{ \dot{k}_{Z}^{(6,0)}}{2}\ln\frac{\mu^2}{m_t^2} \, , \nonumber
\end{align}
where 
\begin{align}
\dot{k}_{Z}^{(6,0)} & = -4\Delta \rho_t^{(4,1)}\left[C_{HD}+2C_{\substack{Hq \\ 33}}^{(1)}-2C_{\substack{Hu \\ 33}}  \right]  \, .
\end{align}
Inverse powers of $s_w$ appear only in the $\alpha$ scheme and are absorbed into the factors $K_W^{(i,j,\alpha)}$, so the scheme-independent coefficients $k_Z^{(i,j)}$ have an expansion in $\Delta \rho_t^{(4,1)}$. In the SM, it is evident that the scheme-dependent corrections $k_Z^{(4,1,\sigma)}$ follow the pattern discussed after Eq.~(\ref{eq:GenWeak}). In SMEFT, scheme-dependent corrections appear in the combination $K_W^{(6,1,\sigma)}+ 2k_Z^{(4,1)} K_W^{(6,0,\sigma)}$ in the last line of Eq.~(\ref{eq:k_Z_alpha_SM}). Moreover, the $K_W^{(6,1,\sigma)}$ pieces are explicitly $\mu$-dependent, and one normally chooses the scale in a process-dependent way.  For these reasons, the numerical pattern of scheme-dependent NLO corrections to weak vertices in SMEFT in the $\alpha$ and $\alpha_\mu$ schemes is not nearly as regular as in the SM; this is best seen by comparing results for a range of processes, which we leave to Section~\ref{sec:HeavyBosonDecays}.

We have focussed the above discussion on the $\alpha$ and $\alpha_\mu$ schemes.  Corrections in the LEP scheme are derived from those in the $\alpha_\mu$ scheme by using Eq.~(\ref{eq:MW_prediction}) to eliminate $M_W$ in favour of $\hat{M}_W$. The result simplifies considerably in the large-$m_t$ limit. To derive it, we first note that the large-$m_t$ limit of the expansion coefficients of $\Delta r$ defined in Eq.~(\ref{eq:dr_exp_coeffs}) can be written in terms of the $K_W$ from Eq.~(\ref{eq:W_alpha}) according to
\begin{align}
\label{eq:DR_From_KW}
\Delta r_t^{(i,j)} = K_W^{(i,j,\alpha)} - K_W^{(i,j ,\mu)} \, .
\end{align}
We can convert these into expansion coefficients of $\hat{\Delta}r_t$ using Eq.~(\ref{eq:rhat_coeffs_ders}). The only non-trivial SMEFT piece is the NLO coefficient, for which we find
\begin{align}
\hat{\Delta} r_t^{(6,1)} & =  \Delta r_t^{(6,1)} + \frac{1}{c_{2w}}\left[\frac{c_w}{s_w}C_{HWB} -\Delta r^{(6,0)}- K_W^{(6,0,\alpha)} \right] K_W^{(4,1,\alpha)} \, \Bigg|_{M_W = \mwhat}  \, .
\end{align}
Inserting these results into Eq.~(\ref{eq:MW_prediction}) gives the following large-$m_t$ corrections to the $W$-boson mass in the LEP scheme within the SM
\begin{align}
\label{eq:DelMWSM_LMT_LEP}
\hat\Delta_{W,t}^{(4,1,\mu)} =\frac{1}{2} \frac{\hat{c}_w^2}{\hat{c}_{2w}}\Delta \rho_t^{(4,1)}  \, ,
\end{align}
while the SMEFT result is
\begin{align}
\label{eq:DelMW_LMT_LEP}
\hat{\Delta}_{W,t}^{(6,1,\mu)} &= \frac{s_w^2}{2 c_{2w}} \left(K_W^{(6,1,\mu)} - K_W^{(6,1,\alpha)}\right)   \\
& +  \frac{s_w^2}{2c_{2w}^2}\bigg[K_W^{(6,0,\alpha)} -\frac{c_w}{s_w}C_{HWB}+ \left(1-\frac{s_w^2}{2} -\frac{2 c_w^2 s_w^2}{c_{2w}}\right)\Delta r^{(6,0)}   \bigg] K_{W}^{(4,1,\alpha)} \, \Bigg|_{M_W = \mwhat} \,. \nonumber
\end{align}
As an example, let us use this to write the factor of $M_W^2$ in Eq.~(\ref{eq:W_alpha}) in terms of $\hat{M}_W^2$. Denoting the resulting LEP-scheme expansion coefficients as $\hat{K}_W^{(i,j,\mu)}$, one has  the NLO SM result
\begin{align}
\label{eq:DelWt}
\hat{K}_W^{(4,1, \mu)}= 2\frac{1}{\vmu^2}\hat{\Delta}_{W,t}^{(4,1,\mu)} \approx  1.5 \%  \, .
\end{align}
The tree-level  SMEFT result  is
\begin{align}
\label{eq:KW_LEP_LO}
\vmu^2 \hat{K}_W^{(6,0,\mu)} = \frac{1}{c_{2w}}\left(c_w^2 K_W^{(6,0,\mu)} - s_w^2 K_W^{(6,0,\alpha)} \right)
\approx 1.4  K_W^{(6,0,\mu)} - 0.4K_W^{(6,0,\alpha)}  \, , 
\end{align}
while the NLO contribution is 
\begin{align}
\label{eq:ugly_LEP}
\hat{K}_W^{(6,1,\mu)} & = \frac{1}{c_{2w}}\left(c_w^2 K_W^{(6,1,\mu)} - s_w^2 K_W^{(6,1,\alpha)} \right) 
+\frac{c_w^2}{c_{2w}^2} K_W^{(4,1,\alpha)} \bigg\{\left(1- \frac{c_w^2 s_w^2}{c_{2w}} \right)C_{HD} \nonumber \\
&+ 3\frac{s_w}{c_{w}}\left(1-\frac{4}{3}\frac{c_w^2s_w^2}{c_{2w}}\right) C_{HWB}
-2s_w^2 \left(1-  \frac{s_w^2}{c_{2w}}  \right)K_W^{(6,0,\mu)} \bigg\}   \Bigg|_{M_W = \mwhat}\, .
\end{align}
For other processes, the numerical factors multiplying the  $\hat{\Delta}_W$ terms are dictated by the dependence of the bare vertex on $M_W$, and are therefore rather process dependent.

\section{Derived parameters}
\label{sec:Derived}

The simplest observables are  ``derived parameters'', where an input parameter in one scheme is calculated
as a SMEFT expansion in another.  For the schemes considered here there are three such quantities: 
$\alpha$ in the $\alpha_\mu$ scheme, $G_F$ in the $\alpha$ scheme, or $M_W$ in the LEP scheme. All of these
 are functions of the expansion coefficients  (and their derivatives, in the case of the LEP scheme) 
 of  $\Delta r$ defined in Eq.~(\ref{eq:DeltaR}).  In this section we briefly examine the latter two cases, and also define
 the procedure for estimating higher-order corrections in the SMEFT expansion through scale variations used
 throughout the remainder of the paper.

The SMEFT expansion for $G_F$ in the $\alpha$ scheme is obtained from Eq.~(\ref{eq:DeltaR}) and yields
\begin{align}
\label{eq:GF_alpha}
	G_{F,\alpha} & = \frac{1}{\sqrt{2} v_\alpha^2}
	\left[1+ \valpha^2\Delta r^{(6,0)} +\frac{1}{\valpha^2}\Delta r^{(4,1)} +\Delta r^{(6,1)} \right] \,.
\end{align}
The tree-level result (LO) evaluates to 
\begin{align}
\label{eq:GFpredict_LO_alphalite}
\begin{split}
	\frac{G_{F,\alpha}^\text{LO}}{G_F} =&   1.034	+ v_\alpha^2
	\Bigg[3.859 C_{HWB}
	+1.801 C_{HD}  
	+1.034 \sum_{j=1,2}	C_{\substack{Hl \\ jj}}^{(3)}
	-1.034 C_{\substack{ll \\ 1221}}
	 \Bigg]  \, ,
\end{split}
\end{align}
and the sum of tree-level and one-loop corrections (NLO) is 
\begin{align}
\label{eq:GFpredict_NLO_alphalite}
\begin{split}
	\frac{G_{F, \alpha}^\text{NLO}}{G_F} =& 0.992 + v_\alpha^2
	\Bigg[
	3.733 C_{HWB}
	+1.756 C_{HD} 
	+1.064 \sum_{j=1,2}	C_{\substack{Hl \\ jj}}^{(3)}
	-1.039 C_{\substack{ll \\ 1221}}
	\\&
	-0.167 C_{\substack{Hu \\ 33}} 
	+0.142 C_{\substack{Hq \\ 33}}^{(1)}
	-0.083 C_{\substack{Hq \\ 33}}^{(3)}
	+0.062 C_{\substack{uB \\ 33}}
	+0.020 C_{\substack{uW \\ 33}} \\ 
	&
	+0.018 C_{\substack{ll \\ 1122}}
	-0.016 \sum_{j=1,2} C_{\substack{lq \\ jj22}}^{(3)} 
	+0.010 C_{W}
	-0.006 \sum_{j=1,2} \bigg(  C_{\substack{Hu \\ jj}}  +  C_{\substack{Hq \\ jj}}^{(3)} \bigg)
	\\&
	+0.004 \sum_{j=1,2}  C^{(1)}_{\substack{Hl \\ jj}}
	+0.003 \bigg( C^{(1)}_{\substack{Hl \\ 33}} + \sum_{i=1,2,3} C_{\substack{He \\ ii}} + \sum_{i=1,2,3} C_{\substack{Hd \\ ii}} -  \sum_{j=1,2} C^{(1)}_{\substack{Hq \\ jj}} \bigg)
	\\ &
	+0.002 \bigg( C_{HB} + C_{HW} +C_{H\Box} - C^{(3)}_{\substack{Hl \\ 33}}  \bigg)
 \Bigg]  \, , 
\end{split}
\end{align}
where in both cases we have used $\mu=M_Z$, so that $v_\alpha = v_\alpha(M_Z)$ and $C_i = C_i(M_Z)$ in the above equations.
In the SM, the LO prediction for $G_F$ differs by $3.4\%$ from the measured value while at NLO the difference is
$-0.8\%$. Evidently, the large-$m_t$ limit contribution in Eq.~(\ref{eq:delta_r41}) accounts for the bulk of the NLO correction. 
The LO SMEFT result contains 5 Wilson coefficients which alter the result, while the NLO one contains 
the full set of 33 Wilson coefficients identified in Table~\ref{tab:NumWC}. 

SMEFT expansions of physical quantities such as $G_{F,\alpha}$ contain a residual dependence on the renormalisation scale $\mu$ due to the truncation of the full series at a fixed order in perturbation theory.  In the SM this is due to the running of $\alpha$, while in SMEFT the Wilson coefficients $C_i$ also run.  It is often useful to use the stability of the results under variations of the scale $\mu$ about a default value as an estimate of uncalculated,  higher-order corrections in the perturbative expansion. The Wilson coefficients
are unknown numerical quantities that we wish to extract from data, so in order to implement their running we must 
calculate their value at arbitrary scales $\mu$ given their value at a default scale choice~$\mu^{\rm def}$. For our purposes,
it is sufficient to use the fixed-order solution to the  RG equation in this calculation, which reads
\begin{align}
\label{eq:C_Running}
C_i(\mu) & = C_i(\mu^{\rm def}) + \ln\left(\frac{\mu}{\mu^{\rm def}}\right) 
 \dot{C}_i(\mu^{\rm def}) \, ,
\end{align}
where $\dot{C}_i$ was defined in Eq.~\eqref{eq:RenCi}.  For the running of $\alpha$ we can also used the fixed-order 
solution to the RG equation given in Eq.~(\ref{eq:run_alpha}).
Throughout the paper, we estimate uncertainties from scale variations by using the afore mentioned equations to evaluate observables for the three scale choices $\mu \in \{\mu^{\rm def},2\mu^{\rm def},\mu^{\rm def}/2\}$.  Central values are given for $\mu=\mu^{\rm def}$, 
 and upper and lower uncertainties are determined by values of the observables at the other two 
 choices.\footnote{At NLO a large number of $\dot{C}_i$ must be evaluated; we have employed \texttt{DsixTools}~\cite{Celis:2017hod, Fuentes-Martin:2020zaz} for this purpose.}  
 
 Let us apply this method to the calculation of $M_W$ in the LEP scheme, which is obtained by evaluating Eq.~(\ref{eq:MW_prediction}).
Compared to $G_{F,\alpha}$, the $W$-mass is sensitive to a different combination of $\Delta r$ as well as its derivatives with 
respect to $M_W$.   The LO result with $\mu=M_Z$ as the default value and scale uncertainties estimated as described above yields
\begin{align}
	\label{MWpredict_LO_alphalite}
	M_W^\text{LO} = &
	79.82^{+0.13}_{-0.13}\text{ GeV} +\mwhat \vmu^2 \Bigg[
	-0.795^{+0.038}_{-0.038}C_{HWB}
	-0.360^{+0.026}_{-0.026}C_{HD}   \nonumber \\
	&-0.220^{+0.008}_{-0.008} \sum_{j=1,2} 	C_{\substack{Hl \\ jj}}^{(3)} +0.22^{+0.003}_{-0.003}C_{\substack{ll \\ 1221}} 
	+0.000^{+0.038}_{-0.038}C_{\substack{Hq \\ 33}}^{(1)}
	+0.000^{+0.036}_{-0.036}C_{\substack{Hu \\ 33}}  \nonumber \\
	&
	+0.000^{+0.013}_{-0.013}C_{\substack{uB \\ 33}}
	+0.000^{+0.012}_{-0.012}C_{\substack{uW \\ 33}}
	+0.000^{+0.006}_{-0.006} \sum_{j=1,2} C_{\substack{lq \\ jj33}}^{(3)} +\ldots
	\Bigg] \, , 
\end{align}
where the $\dots$ indicate contributions where the difference between the upper and lower values obtained from scale variation is less that 1\% of $\hat{M}_W$ when the numerical choice $C_i=\vmu^{-2}$ is made.  At NLO we find
\begin{align}
	\label{MWpredict_NLO_alphalite}
	M_W^\text{NLO} = &
	80.47^{+0.01}_{-0.00}\text{ GeV} +\mwhat \vmu^2 \Bigg[
	-0.807^{+0.002}_{-0.000}C_{HWB}
	-0.381^{+0.004}_{-0.000}C_{HD} \\ \nonumber
	&-0.228^{+0.000}_{-0.000} \sum_{j=1,2} C_{\substack{Hl \\ jj}}^{(3)}
	+0.223^{+0.000}_{-0.000}C_{\substack{ll \\ 1221}}  
	+0.032^{+0.000}_{-0.010}C_{\substack{Hu \\ 33}}  \\ \nonumber
	&-0.028^{+0.009}_{-0.000}C_{\substack{Hq \\ 33}}^{(1)}
	+0.016^{+0.000}_{-0.003}C_{\substack{Hq \\ 33}}^{(3)}
	+0.012^{+0.000}_{-0.002}C_{\substack{uB \\ 33}} +\ldots
	\Bigg] \, ,\nonumber
\end{align}
where in this case the $\dots$ refer to contributions where both the central values and the difference in upper and lower scale uncertainties are both less than 1\% in magnitude.  For both the SM and SMEFT, the scale uncertainties are significantly larger at LO than at NLO. While the NLO corrections in SMEFT all lie within the scale uncertainties of the LO calculation, the same is not true of the SM, where scale variations in the SM at LO do not capture the behaviour of the higher-order corrections. 

We can understand the qualitative features of these results by studying them in the large-$m_t$ limit.  Using 
Eqs.~(\ref{eq:DelMWSM_LMT_LEP}, \ref{eq:DelMW_LMT_LEP}) for the NLO corrections in this limit, the numerical result 
at the scale $\mu=M_Z$ is
\begin{align}
	\label{MWpredict_NLO_alphalite_LMT}
	M_{W,t} ^\text{NLO} = 
	&
	80.36^{+0.00}_{-0.00}\text{ GeV} +\mwhat \vmu^2 \bigg[
	-0.799^{+0.001}_{-0.000}C_{HWB}
	-0.373^{+0.002}_{-0.000}C_{HD} \\ \nonumber
	&-0.226^{+0.000}_{-0.000} \sum_{j=1,2}	C_{\substack{Hl \\ jj}}^{(3)}
	+0.222^{+0.000}_{-0.000}C_{\substack{ll \\ 1221}}  
	+0.035^{+0.000}_{-0.008}C_{\substack{Hu \\ 33}}  \\ \nonumber
	&-0.035^{+0.007}_{-0.000}C_{\substack{Hq \\ 33}}^{(1)}
	+0.014^{+0.000}_{-0.003}C_{\substack{Hq \\ 33}}^{(3)}
	+0.012^{+0.000}_{-0.000}C_{\substack{uB \\ 33}} +\ldots
	\bigg] \,. 
\end{align}
This is clearly a good approximation to Eq.~(\ref{MWpredict_NLO_alphalite}), where as in that equation we have not included contributions of less than 1\%.  
The SM result is scale invariant in this limit, because the top quark is decoupled from the QED coupling $\alpha(\mu)$.

In the above results and throughout the paper we used the \msbar~definition of $\alpha$ in a five-flavour version of QED$\times$QCD. In the literature, one often uses the effective on-shell coupling $\alpha^{\rm O.S.}(M_Z)$ which is related to $\alpha(M_Z)$ using Eq.~\eqref{eq:ToOnShell}. When instead this choice is made, we find the following SM results for $G_F$ in the $\alpha$ scheme
\begin{align}
\label{eq:GFpredict_alphaMZ}
\frac{G_{F, \alpha^{\rm O.S.}}^\text{LO}}{G_F} = 1.025 \, ,  \qquad \frac{G_{F, \alpha^{\rm O.S.}}^\text{NLO}}{G_F} = 0.993 \, ,
\end{align}
while for $M_W$ in the LEP scheme we have
\begin{align}
\label{MWpredict_alphaMZ}
M_{W, \alpha^{\rm O.S.}}^\text{LO} = 79.97\text{ GeV} \, ,  \qquad M_{W, \alpha^{\rm O.S.}}^\text{NLO} = 80.46\text{ GeV} \, .
\end{align}
At LO these two quantities differ by $2-3$\% compared to Eqs.~(\ref{eq:GFpredict_LO_alphalite}, \ref{MWpredict_LO_alphalite}), while at NLO the differences in the two schemes for $\alpha$ are well below the percent level;  we have checked that this also
holds true in SMEFT. 

The NLO result for $M_W$ in SMEFT generalises the previous result  \cite{Dawson:2019clf} to include the full flavour
structure, and resums logarithms of light fermion masses related to the running of $\alpha$; a more detailed comparison is given Appendix~\ref{sec:DG_compare}. The current state-of-the-art in the SM \cite{Awramik:2003rn}  includes complete two-loop
corrections as well as a partial set of even higher-order corrections. Adjusted to our numerical inputs, the result derived from
the parametrisation in Eq.~(6) of that paper, which we refer to as ``NNLO", reads 
\begin{align}
M_W^{\rm NNLO} &= 80.36~{\rm GeV} \, ,
\end{align}
which is outside the uncertainties in the NLO result Eq.~(\ref{MWpredict_NLO_alphalite}). In order to gain insight into 
the structure of higher-order corrections, we have studied the split of the NNLO result into pure EW, and mixed 
EW-QCD components, which was  given in  \cite{Awramik:2003rn} for the unphysical value $M_h=100$~GeV.   
When adjusting our own inputs to that unphysical value, we find that the pure NNLO EW contributions are within our 
NLO uncertainty estimate, so that the discrepancy is due to mixed EW-QCD effects first appearing at NNLO and unrelated to
the running of $\alpha$.  The large-$m_t$ limit of these EW-QCD corrections can be obtained by making the following 
replacement in Eq.~(\ref{eq:DelMWSM_LMT_LEP}) \cite{Halzen:1990je}:
\begin{align}
\Delta \rho_t^{(4,1)} \to \Delta \rho_t^{(4,1)}\left[1 -\frac{\alpha_s}{\pi}\frac{2}{3}\left(2\zeta_2 +1 \right)\right] \, . 
\end{align}
Including this correction changes the central value in Eq.~(\ref{MWpredict_NLO_alphalite}) to 80.41~GeV, which agrees
with the NNLO result to better than the per-mille level.  Further improvements can be made through resummations of the type 
discussed in Section~\ref{sec:universal_corrections}.

\section{Heavy boson decays at NLO}
\label{sec:HeavyBosonDecays}

While the previous sections elucidated some general features of the different input schemes, the aim of this section is to study in detail three benchmark observables to complete NLO in the SMEFT expansion in each scheme: $W$ decay into leptons, $Z$~decay into charged leptons, and Higgs decay into bottom quarks. For the numerical analysis we focus on $W\to\tau\nu$
and $Z\to \tau\tau$, while in the analytic results submitted in the electronic version we keep the lepton species arbitrary.

We write the expansion coefficients of the decay rates to NLO in SMEFT for boson $X\in\{W,Z,h\}$ to fermion pair $f_1 f_2$ as
\begin{align}
\Gamma^{\rm s}_{Xf_1 f_2} = \Gamma_{Xf_1 f_2}^{{\rm s}(4,0)} + \Gamma_{Xf_1 f_2}^{{\rm s}(4,1)} + \Gamma_{Xf_1 f_2}^{{\rm s}(6,0)} +\Gamma_{Xf_1 f_2}^{{\rm s}(6,1)} \, ,
\label{eq:Gamma_def_pieces}
\end{align}
where the superscript $s(i,j)$ refers to dimension-$i$, $j$-loop contributions in input scheme $s \in \{\alpha, \alpha_\mu, {\rm LEP}\}$. To study convergence, 
it is convenient to work instead with expansion coefficients of the decay rate normalised to the LO SM result, namely
\begin{equation}
\Delta^{{\rm s}(i,j)}_{X f_1 f_2} = \frac{\Gamma^{{\rm s}(i,j)}_{Xf_1 f_2}}{\Gamma^{{\rm s}(4,0)}_{Xf_1f_2}} \, .
\label{eq:nlo_rat}
\end{equation}
Throughout the section numerical values for the decay rates are evaluated using the default value $\mu^{\rm def.}= m_\text{decay}$, where $m_\text{decay}$ is the mass of the decaying particle, and scale uncertainties are obtained by
varying the scale up and down by a factor of 2 about the default value, as in Section~\ref{sec:Derived}.

Obviously, results for three decays in three renormalisation schemes and involving a large number of 
SMEFT Wilson coefficients contain a plethora of information.  We have organised it as follows:
\begin{itemize}
\item Figures~\ref{fig:Wdecay_numRes}, \ref{fig:Hdecay_numRes} and \ref{fig:Zdecay_numRes} show Eq.~(\ref{eq:nlo_rat}) for the NLO SM corrections as well as corrections appearing at LO and NLO in SMEFT when the choice $C_i=1 \,{\rm TeV}^{-2}$ is made. They also show the large-$m_t$ limits of the NLO corrections in cases where top-loops contribute, and group the coefficients such that those appearing solely due to the choice of renormalisation scheme appear on the far right.  
\item In Tables~\ref{tab:wlnu_nlo}, \ref{tab:hbb}  and \ref{tab:zll_nlo} we show the size of the NLO corrections to the SM and SMEFT coefficients which appear at tree-level in the different schemes,  for the default scale choices.  
\item In Appendix~\ref{sec:NumRes}, we give results for the numerically most important contributions to the decay rates at LO and NLO in the SMEFT expansion, including uncertainties as estimated from scale variations.
\end{itemize}
The following subsections serve to explain and highlight the most noteworthy patterns emerging from these results.

\subsection{$W \to \ell \nu$ decays}
\label{subsec:Wdecay}
%
\begin{figure}[!t]
	\centering
	\includegraphics[scale=0.7]{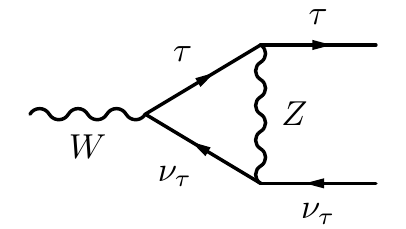} \quad
	\includegraphics[scale=0.7]{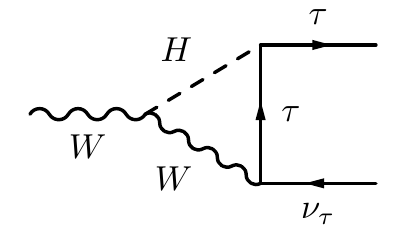} \quad
	\includegraphics[scale=0.7]{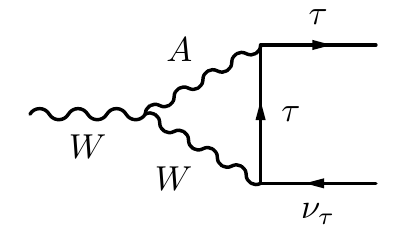} \\
	\includegraphics[scale=0.7]{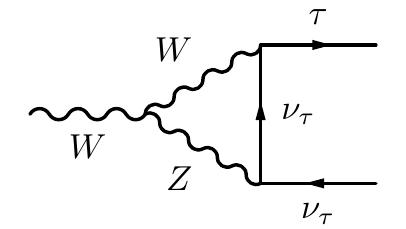} \quad
	\includegraphics[scale=0.7]{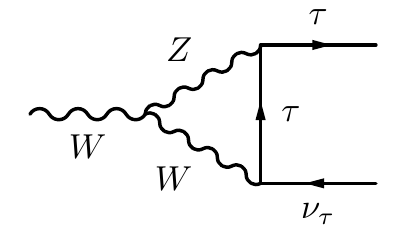} \quad
	\includegraphics[scale=0.7]{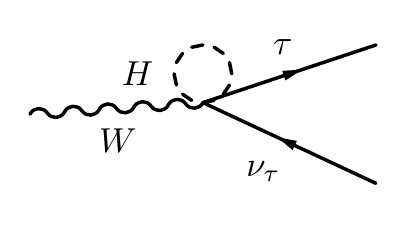} \\
	\includegraphics[scale=0.7]{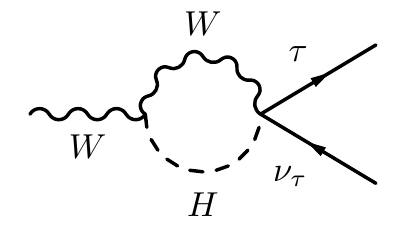} \quad
	\includegraphics[scale=0.7]{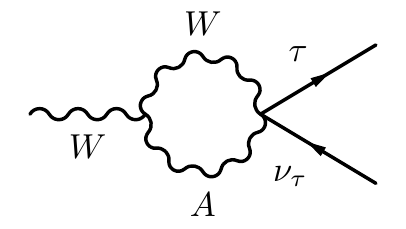} \quad
	\includegraphics[scale=0.7]{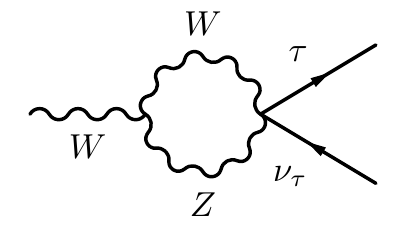} \\
	\caption{Representative virtual corrections for $W$ decay into leptons at NLO.}
	\label{fig:Wdecay}
\end{figure}
%
%
 \begin{figure}[!t]
          \includegraphics[scale=0.45]{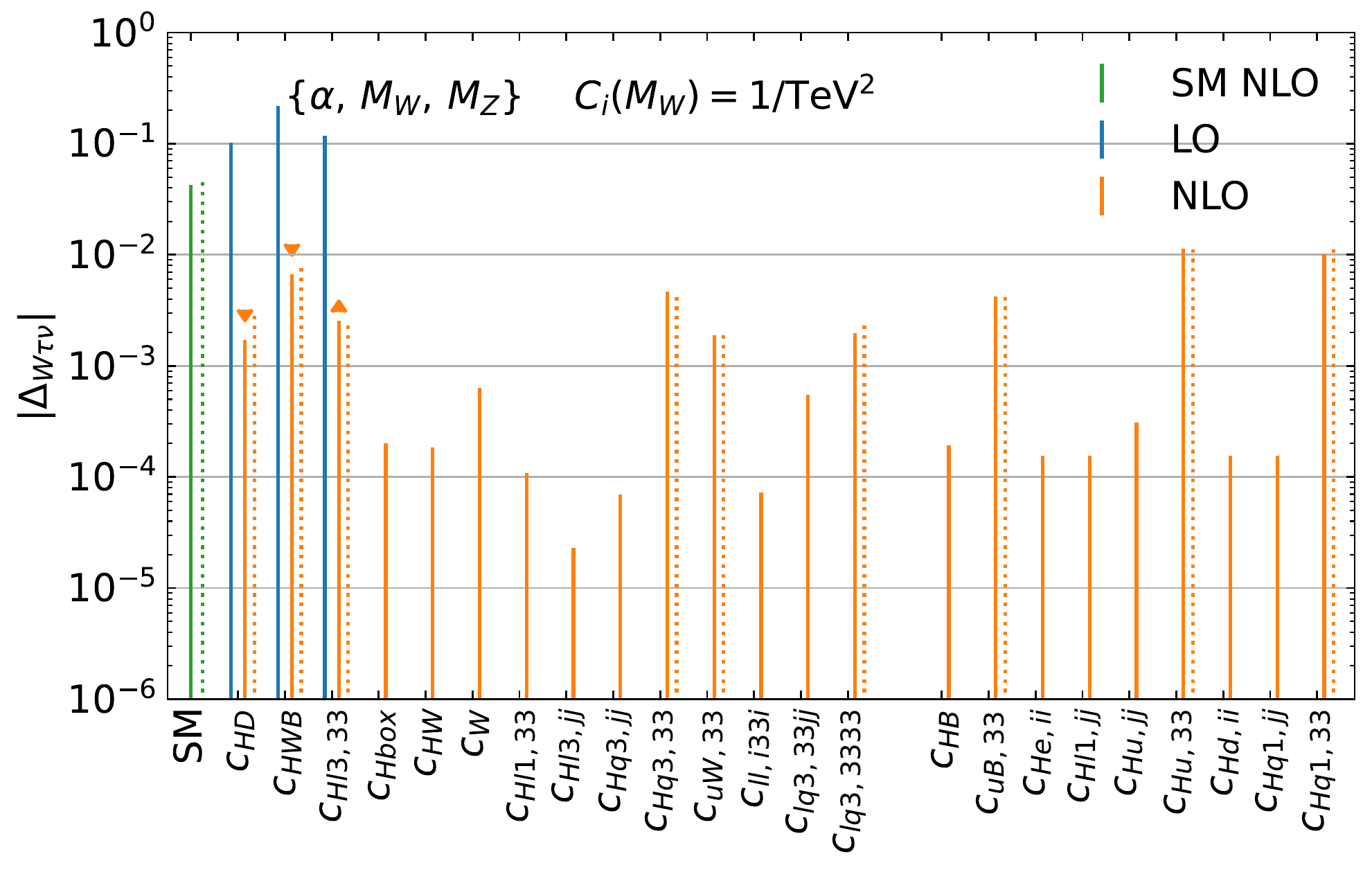} \\
          \includegraphics[scale=0.45]{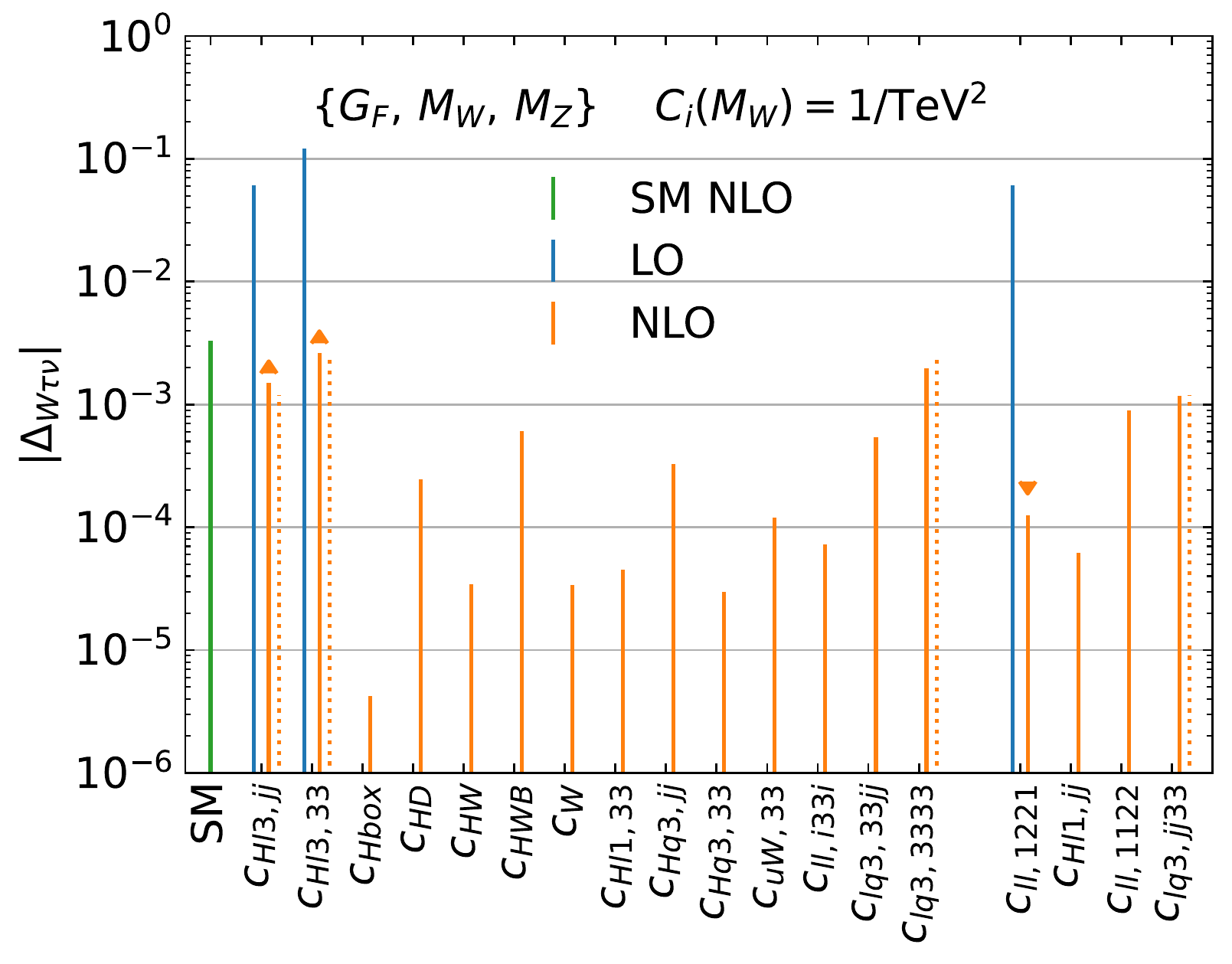} \\
          \includegraphics[scale=0.45]{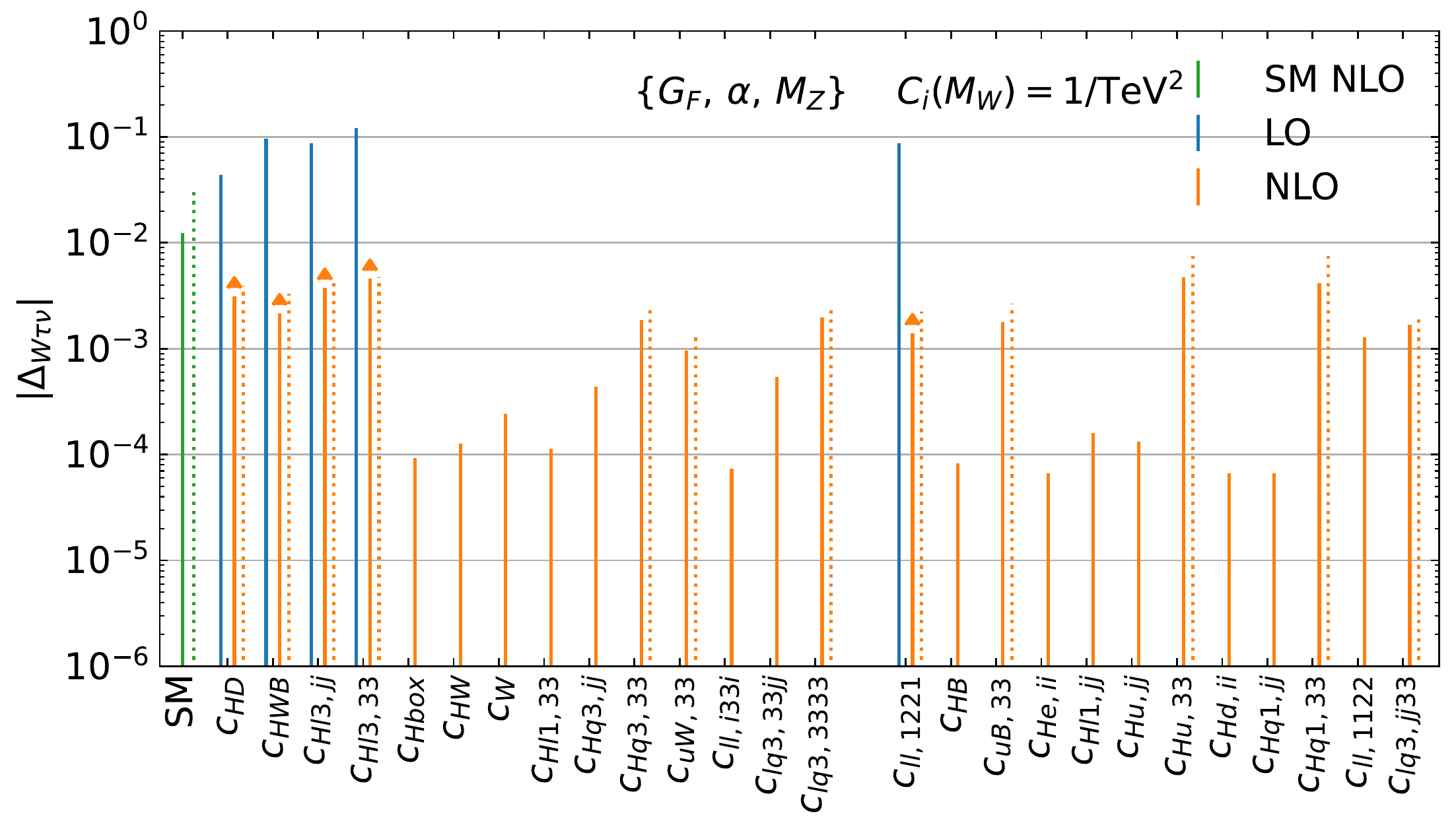} \\
        \caption{LO and NLO corrections $\Delta^{{\rm s}(i,j)}_{W\tau\nu}$, as defined in Eq.~(\ref{eq:nlo_rat}), for the decay $W \to \tau \nu$ in the three schemes. 
Note that ``NLO" in the legends only refers to the NLO corrections and that we write superscripts in the Wilson coefficient names as $C_{Hq3}\equiv C_{Hq}^{(3)}$.
The flavour indices $i$ and $j$ run over values $j\in 1,2$, and $i\in 1,2,3$.
Operators which appear only through counterterms in a particular scheme are shown on the right. 
The dashed lines indicate the large-$m_t$ limit of the NLO corrections.
For operators appearing at LO the orange triangles indicate if the sign of the NLO correction is the same as (triangle pointing up) or different from (triangle pointing down) the sign of the LO contribution.}
        \label{fig:Wdecay_numRes}
 \end{figure}

The tree-level decay rate for $W\to \tau \nu$ decays, written in terms of $v_T$, takes the form
\begin{align}
 \Gamma_{W\tau \nu}^{(4,0)} +  \Gamma_{W\tau \nu}^{(6,0)}  = \frac{M_W}{12\pi}
 \frac{\MWsq}{v_T^2}\left(1 + 2 v_T^2  C_{\substack{Hl \\ 33}}^{(3)} \right) \, .
 \label{eq:W_decay_LO}
\end{align}
Renormalisation-scheme dependence thus enters the result through the counterterms for $M_W$ and $v_T$.

The NLO decay rate is calculated by evaluating virtual corrections such as those shown in Figure~\ref{fig:Wdecay},
and then adding together with UV counterterms and real emission diagrams with an extra photon in the final
state to get a finite result.
The size of NLO SM corrections in the different schemes is easily understood using the large-$m_t$ analysis in Section~\ref{subsec:PertConv}.  In that limit, the NLO corrections in the $\alpha_\mu$ scheme vanish, while those in 
$\alpha$ scheme are roughly $-3.4\%$, a pattern which agrees well with the full results in Table~\ref{tab:wlnu_nlo}. 
The SM LEP scheme corrections in the large-$m_t$ limit are
\begin{align}
\label{eq:LOW_LEP}
\frac{\hat{M}_W}{12\pi}
 \frac{\hat{M}_W^2}{v_\mu^2}\left(1 +\frac{3}{2}\frac{\hat{c}_w^2}{\hat{c}_{2w}} \frac{\Delta \rho_t^{(4,1)}}{\vmu^2}\right)\approx 
 \frac{\hat{M}_W}{12\pi}
 \frac{\hat{M}_W^2}{v_\mu^2}\left(1 + 0.02 \right) \,,
\end{align}
so that the NLO correction is again very close to the result in the table. Note that in Eq.~(\ref{eq:LOW_LEP}) we have consistently expressed  all powers of the $W$ mass in terms of  $\hat{M}_W$, whether they come from the 2-body phase space or directly from the amplitude, which
accounts the factor of $3/2$ compared to Eq.~(\ref{eq:DelWt}).  
Absolute values of the decay rates at LO and NLO are given in Appendix~\ref{app:Wdecay}.  In that notation, one finds the following 
ratios in the SM at NLO
\begin{align}
\frac{\Gamma^{\alpha}_{W,\text{NLO}}}{\Gamma^{\alpha_\mu}_{W,\text{NLO}}} = 0.992 \, , 
\qquad \frac{\Gamma^{{\rm LEP}}_{W,\text{NLO}}}{\Gamma^{\alpha_\mu}_{W,\text{NLO}}} =1.003 \,. 
\end{align}
The first ratio agrees quite well with the estimate $G^{\rm NLO}_{F,\alpha}/G_F$ using Eq.~(\ref{eq:GFpredict_NLO_alphalite}), while the second is consistent with the estimate $(M_W^{\rm NLO})^3/M_W^3$ using Eq.~(\ref{MWpredict_NLO_alphalite}).  Once the NLO corrections are included the results between the schemes show (better than) percent-level agreement. 

In Figure~\ref{fig:Wdecay_numRes} the corrections in SMEFT are shown. The absolute size of the SMEFT corrections is determined by the choice $C_i ={\rm TeV}^{-2}$.  For that choice,  SMEFT contributions are suppressed by $v_\sigma^2\times {\rm TeV}^{-2} \approx 6\%$, and are anywhere between 10\% to below per-mille level of the SM tree-level result depending on the coefficient.
The NLO SMEFT results contain a large number of Wilson coefficients.  We have organised the coefficients in Figure~\ref{fig:Wdecay_numRes} such that those 
appearing only due to the renormalisation of $v_T$ or $M_W$ up to NLO are separated out onto the right part of the figure, while those appearing also in the bare matrix elements or wavefunction renormalisation factors and thus common to all
schemes are on the left.  In the $\alpha_\mu$ scheme the coefficients $\Delta v_\mu^{(6,1,\mu)}$ appearing
in Eq.~(\ref{eq:dVmu_coeffs}) have a large overlap with those appearing in $W$-boson couplings, and as a result
only four-fermion coefficients as well as those that modify $Z$ couplings to leptons, $C_{\substack{Hl \\ jj}}^{(1)}$, with $j=1,2$,
are particular to that scheme.  In the $\alpha$ scheme, on the other hand, the renormalisation of $v_T$ brings in
sensitivity to coefficients related to the renormalisation of $M_Z$ and $e$,  which are listed in Eqs.~(\ref{eq:MZ_coeffs}) 
and~(\ref{eq:E_coeffs}).  The LEP scheme is sensitive to the full set of coefficients contained in $\Delta r$, through the
 renormalisation of $M_W$, and therefore contains the overlap of the coefficients in the other two schemes.
 Taken as a whole, the number of Wilson coefficients contributing at NLO for the central scale choice is $39$ in the LEP scheme, $35$ in the $\alpha$ scheme and $25$ in the $\alpha_\mu$ scheme.

As in the SM, the numerically dominant NLO SMEFT corrections are related to top-quark loops. In the $\alpha$ and 
$\alpha_\mu$ schemes, the scheme-dependent corrections in the large-$m_t$ limit are nearly all contained in the factors $K_W$ given in Eqs.~(\ref{eq:KWalpha}, \ref{eq:KWmu}).  For the
default input choices, the SMEFT contributions evaluate to
\begin{align}
\label{eq:KW_mu_MW}
\vmu^2 K_{W}^{(6,0,\mu)} & + K_{W}^{(6,1,\mu)} = 
\vmu^2 \bigg[\sum_{j=1,2} \bigg(-C_{\substack{Hl \\ jj}}^{(3)}(1+0.0193)  + 0.0193 C^{(3)}_{\substack{lq \\ jj33}}\bigg) +C_{\substack{ll \\ 1221}}(1+0.0) \bigg]
      \, , \nonumber \\ 
       \valpha^2 K_{W}^{(6,0,\alpha)} & + K_{W}^{(6,1,\alpha)} = 
\valpha^2 \bigg[ 1.74C_{HD}\left(1-0.0275\right) + 3.73 C_{HWB}\left(1-0.0354\right)  \nonumber \\
&   +0.206 \left( C_{\substack{Hq \\ 33}}^{(1)} - C_{\substack{Hu \\ 33}}\right)    -0.0674 C_{\substack{Hq \\ 33}}^{(3)}       
-0.0727 C_{\substack{uB\\ 33}} -0.0334 C_{\substack{uW\\ 33}} 
       \bigg] \, .
\end{align}
For coefficients appearing at LO, the NLO corrections are the second term in the parentheses, facilitating a comparison with 
Table~\ref{tab:wlnu_nlo}. Results also for coefficients first appearing at NLO  can be found in Eq.~(\ref{eq:W_NLO_alpha_exact}) and Eq.~(\ref{eq:W_NLO_mu_exact}).  
We see the large-$m_t$ limit corrections are a good approximation to the full ones.  
Interestingly, for the coefficients appearing at LO, there is no large hierarchy between the size of NLO corrections in the 
$\alpha$ scheme  compared to the $\alpha_\mu$ scheme,  even though the analytic result for $K_W^{(6,1,\alpha)}$ contains 4 (3) inverse powers of $s_w$ in the case of $C_{HD}$ ($C_{HWB}$).  In fact, the largest corrections are from  $C_{\substack{Hq \\ 33}}^{(1)}$ and  $C_{\substack{Hu \\ 33}}$,
 which appear only due to the scale-dependent logarithmic terms from Eq.~(\ref{eq:WaLogs}).  This illustrates the important point that, unlike  the SM,  the NLO corrections are strongly scale dependent in SMEFT.
 
The SMEFT corrections in the LEP scheme can be derived from results in the $\alpha_\mu$ scheme using Eq.~\eqref{eq:MW_prediction} to write $M_W$ in terms of $\hat{M}_W$.  The expansion coefficients arising after converting the factor of $M_W^2$ in the large-$m_t$ limit,
$\hat{K}_W^{(6,j,\mu)}$, were given in Eqs.~(\ref{eq:KW_LEP_LO}, \ref{eq:ugly_LEP}).  
In order to calculate the decay rate one must also write the factor of  $M_W$ arising from 2-body phase space in terms of 	
$\hat{M}_W$.  We have checked that after doing so the large-$m_t$ limit corrections to the coefficients appearing
in $\hat{K}_W$ are a good numerical approximation to the full ones.  

In addition to the corrections related to the flavour-independent corrections, there are also contributions from the coefficient $C_{\substack{Hl \\ 33}}^{(3)}$, which specifically modifies the $\tau\nu W$ coupling. The large-$m_t$ limit correction to $\Delta^{{\rm LEP}(6,1)}_{W,t}$ due to this coefficient is given by
\begin{align}
 - 2 \Delta\rho_t^{(4,1)}C_{\substack{Hl \\ 33}}^{(3)}
\left(1+ 2\ln\frac{\mu^2}{m_t^2}\right) \left(1 +3 \hat{\Delta}_{W,t}^{(4,1,\mu)} \right) \, .
\end{align}
The corresponding results in the $\alpha$ and $\alpha_\mu$ schemes are obtained from the above by setting 
$\hat{\Delta}_{W,t}^{(4,1,\mu)}$
to zero.  Numerically, one finds that the NLO corrections to $C_{\substack{Hl \\ 33}}^{(3)}$ are about 4\% in the LEP scheme,
and 2\% in the $\alpha$ and $\alpha_\mu$ schemes, in rough agreement with Table~\ref{tab:wlnu_nlo}.  
Compared to the other schemes, the NLO corrections to the coefficients appearing at tree-level in the LEP scheme show a rather irregular pattern due to the complicated dependence on the Weinberg angle. 

While the size of the NLO corrections studied above is rather scale dependent,
 the sum of the LO  and NLO contributions is independent of the scale (up to uncalculated NNLO terms in the SMEFT expansion)  and is thus much less sensitive.  To study this effect in detail,  in Appendix~\ref{app:Wdecay} we give numerical results in the three schemes including scale variations at LO and NLO. It is seen that in SMEFT, the dominant NLO corrections are typically within the uncertainties of  the LO calculation as estimated through scale variations, and that the scale uncertainties in the NLO results are substantially smaller than in the LO ones.

\begin{table}[t]
\centering
\begin{tabular}{c|ccccccc}
$W \rightarrow  \tau \nu$ & SM       & $C_{HD}$ & $C_{HWB}$ & $C^{(3)}_{\substack{Hl\\jj}}$ & $C_{\substack{ll\\1221}}$  & $C^{(3)}_{\substack{Hl\\33}}$  \\[3mm]
 \hline
$\alpha$                                & $-4.2\%$ & $-1.7\%$ & $-3.0\%$  & ---        & ---                   & $2.2\%$                         \\
$\alpha_\mu$              & $-0.3\%$ & ---      & ---       & $2.5\%$    & $-0.2\%$     & $2.2\%$       \\
LEP                                     &      $2.0\%$      &       $8.1\%$      &    $3.2\%$         &  $5.1\%$                                 &       $2.5\%$  &   $4.6\%$                                                 
\end{tabular}
\caption{NLO corrections to prefactors of LO Wilson coefficients in the three schemes. Negative corrections indicate a reduction in the magnitude of the numerical coefficient of a given Wilson coefficient. The flavour index $j$ refers to $j\in 1,2$.}
\label{tab:wlnu_nlo}
\end{table}

\subsection{$h \to b\bar{b}$ decays}
\label{sec:higgsbb}

 \begin{table}[t]
\begin{center}
\begin{tabular}{cl|rrrrrrr}
\multicolumn{2}{l|}{$h \to b\bar{b}$} & SM  
& $C_{H\Box}$ & $C_{HD}$  & $ C_{\substack{dH \\ 33}}$  & $C_{HWB}$ & $C_{\substack{Hl \\ jj}}^{(3)} $
 & $C_{\substack{ll\\ 1221}}$  \\[3mm]
\hline 
&NLO QCD   & 20.3\% & 20.3\% & 20.3\%  &20.3\% &  20.3\% & - & -  \\
$\alpha$&NLO EW & -5.2 \% &  2.1\%  & -11.0\% & 4.2\% &  -6.7\%  & - & -  \\
\cline{2-9}
&NLO correction & 15.1\%  & 22.4\% & 9.3\%   &24.5\%   & 13.6\% & - & -  \\[1mm] 
\hline\hline
 &NLO QCD   & 20.3\% & 20.3\% & 20.3\%  &20.3\% & - & 20.3\% & 20.3\%  \\
$\alpha_\mu$ &NLO EW & -0.8 \% & 2.1\%  &  2.0\%   & 1.9\%  & -& 0.9\%  & -0.8\% \\
\cline{2-9}
&NLO correction & 19.5\% &  22.4\% &  22.3\%  &22.2\% & -& 21.2\%     &19.5\% \\ 
\hline\hline
 &NLO QCD   & 20.3\% & 20.3\% & 20.3\%  &20.3\% & - & 20.3\% & 20.3\%  \\
LEP &NLO EW & -0.7 \% & 2.1\%  &  1.6\%   & 1.9\%  & -& 0.7\%  & -0.9\% \\
\cline{2-9}
&NLO correction & 19.5\% &  22.3\% &  21.9\%  &22.2\% & -& 21.0\%     &19.3\% \\ 
\end{tabular}
\caption{\label{tab:CorSplit} 
NLO corrections to prefactors of LO Wilson coefficients in the three schemes, split into QCD and EW corrections. The flavour index $j$ refers to $j\in 1,2$. }
\label{tab:hbb}
\end{center}
\end{table}

The tree-level decay rate for $h\to b\bar{b}$ decay is given by
\begin{equation}
 \Gamma_{hb\bar{b}}^{(4,0)} +  \Gamma_{hb\bar{b}}^{(6,0)}   = \frac{3 m_b^2 M_h}{8 \pi v_T^2} \left[ 1 + v_T^2 \left(2  C_{H\Box} - \frac{1}{2}C_{HD}-\sqrt{2} \frac{v_T}{m_b} C_{\substack{dH\\33}}  \right) \right]   \, .
 \label{eq:Hbb_tree_level}
\end{equation}
The decay $h\to b\bar{b}$ has two important differences with respect to the decays $W\to \ell \nu$ and $Z\to \ell \ell$ (to be discussed in Section~\ref{sec:ZtoLL}). First, we retain the $b$-quark mass and, second, the strong coupling $\alpha_s(\mu)$ plays a role in the results already at NLO. 
The Higgs mass $M_h$ is evaluated on-shell, but the NLO corrections do not involve its counterterm since it appears through phase space rather than through the amplitude.  Therefore, the input-scheme dependence to NLO arises mainly through the counterterm for $v_T$.\footnote{Results in the $\alpha_\mu$ and LEP scheme differ because one must eliminate $M_W$ in favour of $\hat{M}_W$ in the  NLO SM correction, but this is a small effect numerically.}

The decay $h\to b\bar{b}$ receives both QCD and EW corrections at NLO. The two effects are additive 
and to study the EW input scheme dependence of the results it is useful to quote the QCD and EW corrections separately, as in Table~\ref{tab:CorSplit}. To this order, the QCD corrections are scheme independent.  In the $\alpha$ scheme the EW corrections are rather large and depend heavily on the Wilson coefficient considered, ranging from -11\% to 4\% and thus inducing significant shifts to QCD alone, while in the $\alpha_\mu$ and LEP schemes
the corrections are smaller are more uniform. 

We can understand the qualitative features of the NLO EW corrections using the large-$m_t$ limit.  To this end, we use Eq.~(\ref{eq:weak_vev}) to write the NLO decay rate in this limit as	
\begin{align}
 \Gamma^s_{hb\bar{b}} \bigg |_{m_t\to\infty}= \frac{3 m_b^2 M_h}{8 \pi v_\sigma^2} \bigg[& 1 + v_\sigma^2\left(K_h^{(6,0)} + K_W^{(6,0,\sigma)}\right) + \frac{1}{v_\sigma^2}\left(K_h^{(4,1)} + K_{W}^{(4,1,\sigma)}  \right) \nonumber \\
 &+ K_h^{(6,1)}+\Delta K_h^{(6,1,\sigma)}  \bigg] \, ,
 \label{eq:K_h_def}
\end{align}
where $K_h^{(6,0)}$ is the SMEFT contribution in  Eq.~(\ref{eq:Hbb_tree_level}), and the 
scheme-dependent part of the NLO SMEFT correction is
\begin{align}
\label{eq:KH_61_alpha}
\Delta K_h^{(6,1,\sigma)} = K_{W}^{(6,1,\sigma)} + 2 K_h^{(4,1)}K_W^{(6,0,\sigma)}
 +\frac{1}{\sqrt{2}}\frac{v_\sigma}{m_b} K_{W}^{(4,1,\sigma)}C_{\substack{dH\\ 33}} \, .
\end{align}
 Large-$m_t$ limit results in the $\alpha$ scheme have been given previously in \cite{Cullen:2019nnr}, while those in the $\alpha_\mu$ scheme can be extracted from \cite{Gauld:2015lmb}.  We make use of those results in what follows, thus employing the ``vanishing gauge coupling limit", which in this case amounts
to taking the limit $M_W\ll M_h$ in addition to $m_t\to \infty$.  The LEP and $\alpha_\mu$ scheme results are identical in this limit.

In the SM, the scheme-independent NLO correction in the large-$m_t$ limit is given by 
\begin{align}
\frac{1}{v_\sigma^2}K_h^{(4,1)} = \frac{1}{3 v_\sigma^2 }\Delta \rho_t^{(4,1)} \left(1+ \frac{7(N_c-3)}{3} \right) \approx 0.003 \, .
\end{align}
It follows from the discussion in Section~\ref{subsec:PertConv} that the large-$m_t$ limit corrections in the $\alpha_\mu$ scheme 
are tiny, while those in the $\alpha$ scheme are well approximated by $K_{W}^{(4,1,\alpha)} \approx -3.4\%$.  
Clearly, this mimics the features of the exact NLO EW corrections given in Table~\ref{tab:hbb}.

In SMEFT, the scheme-independent\footnote{In fact there is mild dependence
on the scheme through the numerical value for $v_\sigma$.}  NLO correction in the  large-$m_t$ limit is given by
\begin{align}
\label{eq:KH61}
\frac{K_h^{(6,1)}}{K_h^{(4,1)}}& = C_{HD}\left(-1+ 6 L_t \right) +   2\sqrt{2} \frac{M_W}{m_t}\left(-7+6 L_t\right)C_{\substack{uW\\ 33}}  
+4\left(1+ 6 L_t \right) C^{(3)}_{\substack{Hq\\ 33}}  \nonumber \\
& + \frac{3}{2\sqrt{2}}\frac{v_\sigma}{m_b}\left(-1+ 10 L_t \right)C_{\substack{dH\\ 33}} + \dots  \, , 
\end{align}
where $L_t=\ln(\mu^2/m_t^2)$ and we have set $N_c=3$. The $\dots$ refer to Wilson coefficients which 
contain no overlap with those appearing in the scheme-dependent pieces in Eq.~(\ref{eq:KH_61_alpha}). 
In the $\alpha$ scheme the numerical value of the NLO corrections at $\mu=M_h$ is 
\begin{align}
\label{eq:Higgs_corrections_uni_nonuni}
 & \frac{1}{\valpha^2}\left( K_h^{(6,1)} +  \Delta K_h^{(6,1,\alpha)} \right) =
\bigg\{ -C_{HD}(1.6 + 9.7) + (0.0 - 17) C_{HWB}    \nonumber \\
& -(3.7+6.8) C_{\substack{Hq \\ 33}}^{(3)} 
+ (0.0 - 8.8)(C_{\substack{Hu \\ 33}} -C_{\substack{Hq \\ 33}}^{(1)} )
+(0.0 - 3.1)C_{\substack{uB \\ 33}} + (-4.6 + 0.42)C_{\substack{uW \\ 33}} \nonumber\\
& -\frac{\sqrt{2}\valpha}{{m_b}}\left(1.8 + 1.7\right) C_{\substack{dH \\ 33}} \bigg\}\times 10^{-2}   + \dots \, ,
\end{align}
where the $\dots$ refer to coefficients not appearing in  $\Delta K_{W}^{(6,1,\alpha)}$, and 	
the order of the numbers inside the parentheses multiplying the Wilson coefficients on the right-hand side	
of the above equation matches the order of the two terms on the left-hand side.  In most cases the scheme-dependent 	
parts contained in $\Delta K_{W}^{(6,1,\alpha)}$ dominate over the scheme-independent ones.  For  coefficients	
not appearing already at NLO, one can verify that the results above are close to the exact NLO results  in 	
Eq.~(\ref{eq:hbb_vhat}).  Combined with the LO result in Eq.~(\ref{eq:hbb_vhat_LO}), one infers NLO EW 	
corrections of $-9\%$ for $C_{HD}$, $-5\%$ for $C_{HWB}$ in the $\alpha$ scheme.  In the $\alpha_\mu$ 	
scheme, one has 	
\begin{align}	
\frac{1}{\vmu^2} \Delta K_h^{(6,1,\mu)}& =	
\bigg\{0.6C_{\substack{ll \\ 1221}} + \sum_{j=1,2} \bigg[-0.9 C_{\substack{Hl \\ jj}}^{(3)}  +0.3 C^{(3)}_{\substack{lq \\ jj33}}\bigg] \bigg\}\times 10^{-2} \,.	
\end{align}	
Contributions from $C_{HWB}$ are completely absent in the $\alpha_\mu$ scheme, while the NLO EW correction to $C_{HD}$	
from the above result and Eq.~(\ref{eq:hbb_vmu_LO}) is $3$\% in the large-$m_t$ limit.  This explains the pattern	
of results seen for these coefficients in Table~\ref{tab:hbb}. It makes clear that in this case factors of $K_W^{(i,j,\alpha)}$ 	
work much the same in SMEFT as in the SM, producing  sizeable NLO EW corrections compared to the $\alpha_\mu$ scheme. 
 
The full set of NLO corrections in the different schemes is shown in Figure~\ref{fig:Hdecay_numRes}. In the numerical
results in Appendix~\ref{sec:hbb_num} we follow  \cite{Cullen:2019nnr} and leave in symbolic form enhancement factors of  $m_b/v_\sigma$ which disappear when Minimal Flavour Violation is assumed.  We have not done this in the figure, which
explains, for instance, the very large contribution from $C_{\substack{dH \\ 33}}$.  In contrast to the case of $W$ decay,
in some cases there are large differences between the large-$m_t$ limit and full corrections; this occurs when 
a Wilson coefficient receives both EW and QCD corrections, the latter invariably being the larger effect.  From the 
perspective of EW input-scheme dependent corrections, the most important feature of the figure is the number of Wilson 
coefficients appearing.  In particular, there are far more in the $\alpha$ scheme, 42 in total, than in the 
$\alpha_\mu$ or LEP schemes, both of which receive contributions from the same 29 Wilson coefficients.  
The main reason is that the renormalisation of $v_T$ in the $\alpha$ scheme involves the large set of 
flavour-specific couplings to fermions identified given in Eq.~(\ref{eq:MZ_coeffs}), while in the $\alpha_\mu$ and 
LEP schemes $M_Z$ does not enter the tree-level amplitude and many of these coefficients are therefore absent.
 
 \begin{figure}[thb]
          \includegraphics[scale=0.45]{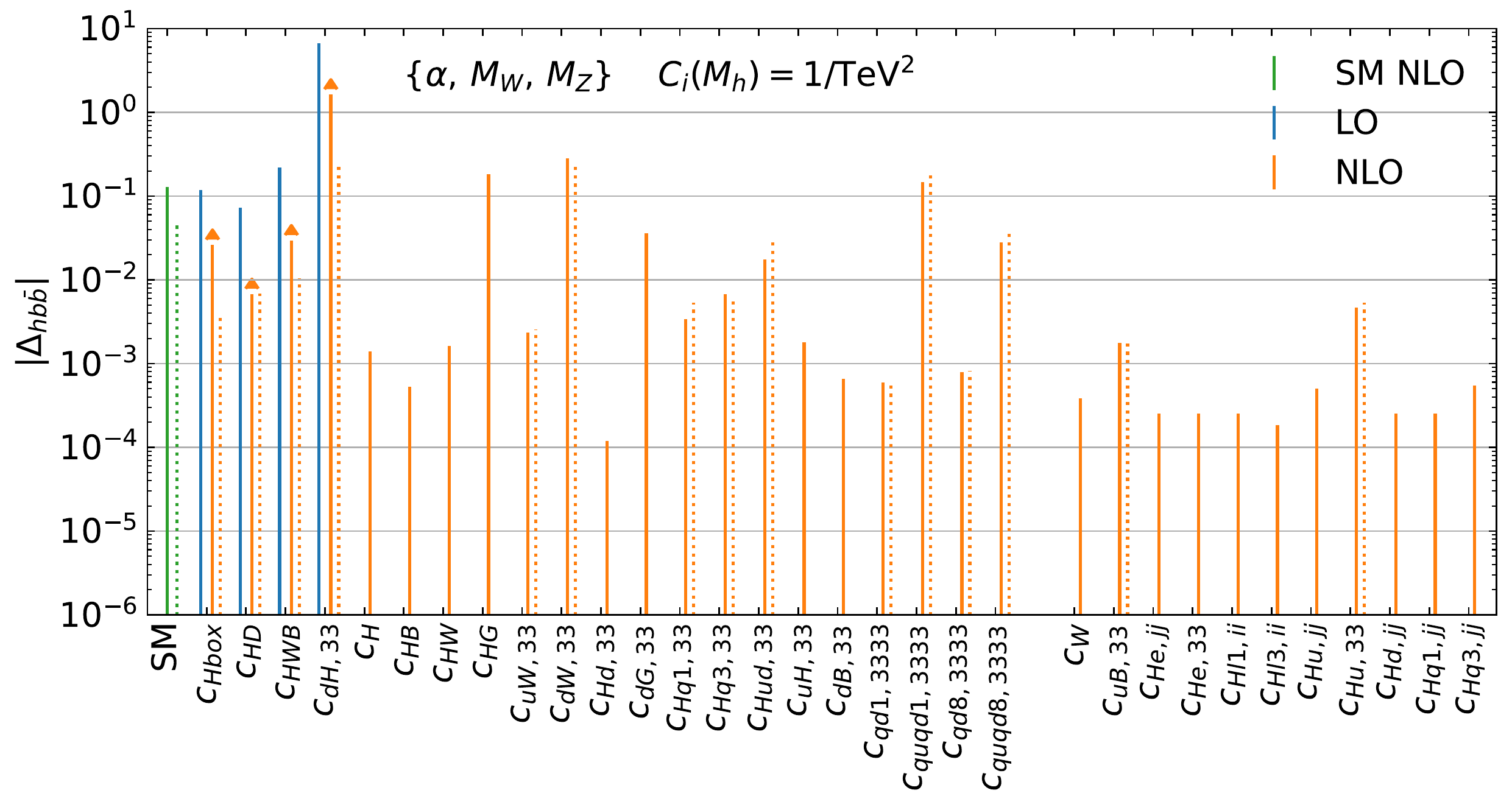} \\
          \includegraphics[scale=0.45]{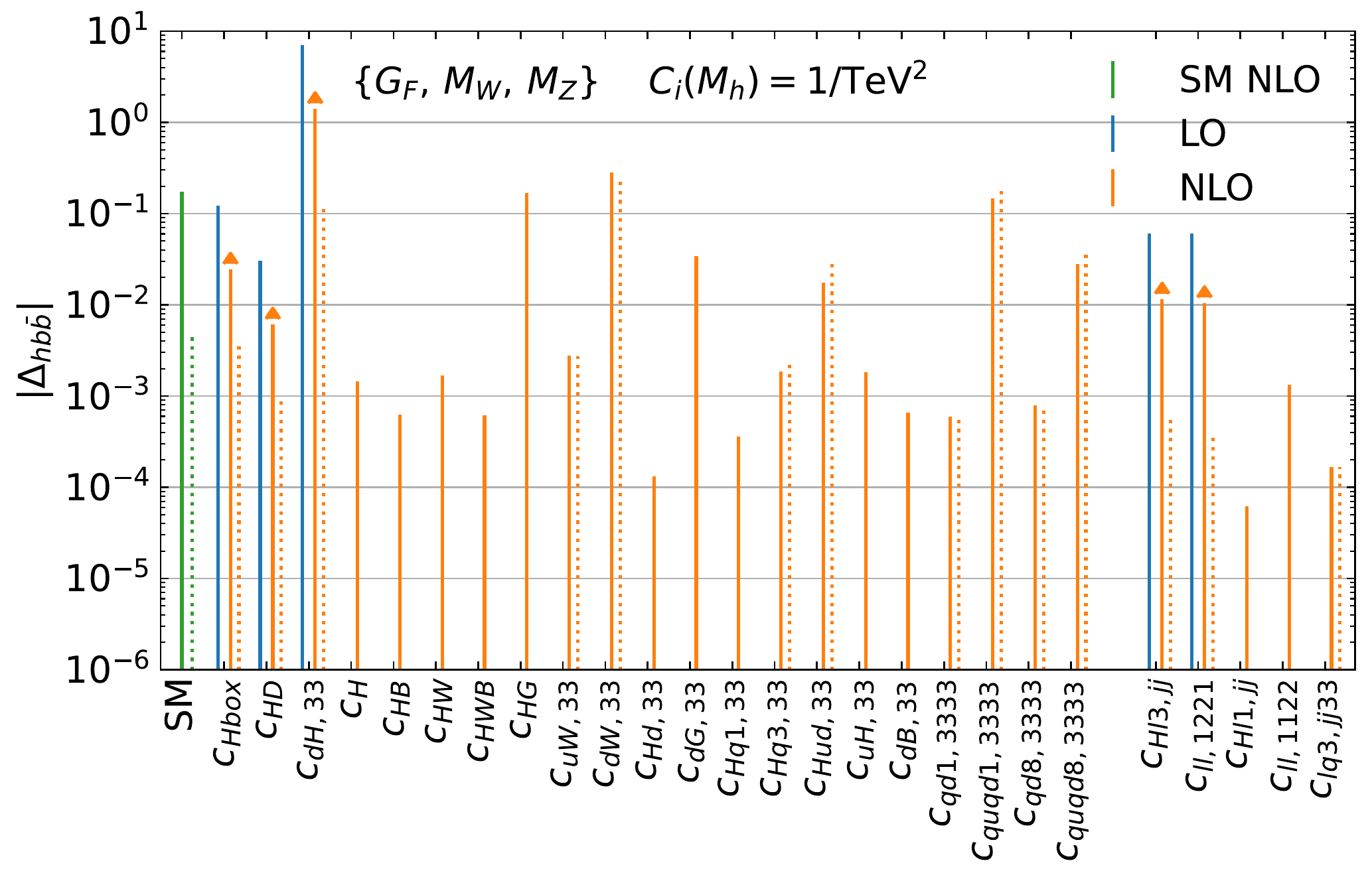} \\
          \includegraphics[scale=0.45]{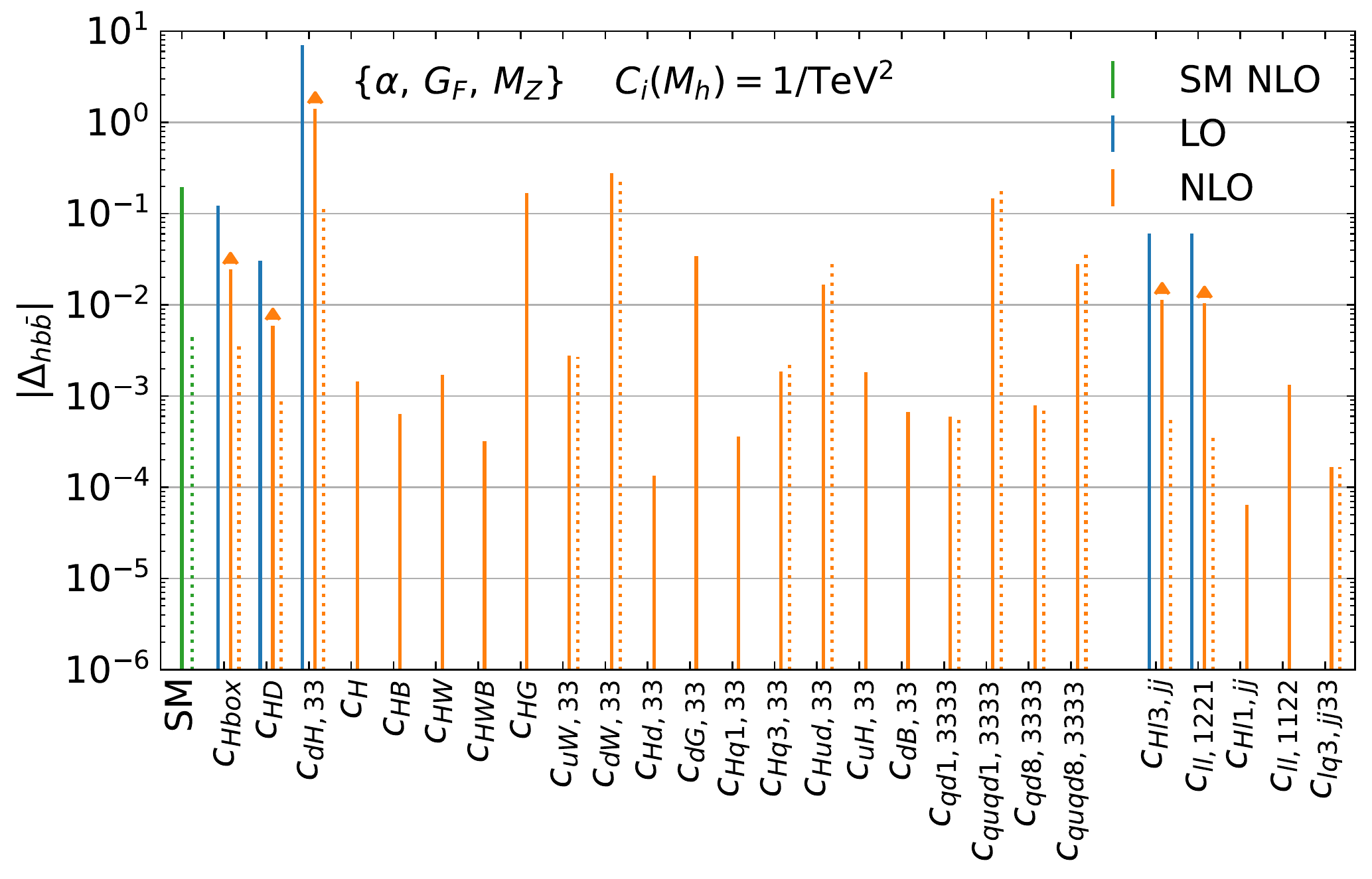} \\
        \caption{As in Figure~\ref{fig:Wdecay_numRes}, but for the decay $h\to b \bar{b}$.}
        \label{fig:Hdecay_numRes}
 \end{figure}

\subsection{$Z \to \ell\ell$ decays}
\label{sec:ZtoLL}

The tree-level decay rate for $Z\to \tau\tau$  decay, written in terms of $v_T$, takes the form
\begin{align}
\label{eq:Gam_Z_tree}
\Gamma_{Z\tau\tau}^{(4,0)} + \Gamma_{Z\tau\tau}^{(6,0)} & = \frac{M_Z}{24 \pi }\bigg\{\left[\frac{M_Z^2}{v_T^2}\left(1- \frac{v_T^2}{2}C_{HD} \right)\right] \left(g_\tau^{(4,0)}  + v_T^2 g_\tau^{(6,0)}  \right) \nonumber \\
& + 2M_Z^2\left[c_{2w}\left(C^{(1)}_{\substack{Hl\\33}} + C^{(3)}_{\substack{Hl\\33}}\right)-2 s_w^2C_{\substack{He\\33}}\  \right] \bigg\} \, ,
\end{align}
where 
\begin{align}
g_\tau^{(4,0)} &= 1 -4s_w^2 + 8s_w^4 \nonumber \, , \\
g_\tau^{(6,0)} & = 2\left(1-4 s_w^2\right)\left(c_w^2 C_{HD} + 2c_w s_w C_{HWB} \right) \,.
\end{align}
The term inside the square brackets in the first line of Eq.~(\ref{eq:Gam_Z_tree}) is independent of the fermion species
into which the $Z$ decays and was considered in Eq.~(\ref{eq:kZ_def}). 
The function $g_\tau$ depends on the charge and weak isospin of the $\tau$ lepton,
and the terms on second line are specific to $Z\tau\tau$ couplings in SMEFT.  The LO decay rate depends on
the full set of parameters $M_W,M_Z,v_T$, and so scheme-dependent corrections involve the full set of coefficients
identified in Section~\ref{sec:Counting}.

The NLO decay rates in the three schemes are shown in Figure~\ref{fig:Zdecay_numRes}. In the $\alpha$ scheme the
set of coefficients appearing in the renormalisation of $v_T$ is the same as that for renormalising $M_Z$ and $M_W$,
so it does not introduce any unique coefficients at NLO.  In the LEP and $\alpha_\mu$ schemes, on the other hand,
the renormalisation of $v_T$ introduces a set of 4-fermion coefficients shown on the right-hand side of the figure 
that would not otherwise appear in the decay rate.  In this case the number of coefficients appearing at NLO 
is quite large: 63 in the $\alpha$ scheme, and 67 in the $\alpha_\mu$ and LEP schemes.

In order to understand the dominant corrections we study the large-$m_t$ limit. Let us first consider the corrections to the 
SMEFT coefficients specific to $Z\tau\tau$ couplings, given in the second line of Eq.~(\ref{eq:Gam_Z_tree}).  
In order to evaluate them in the three schemes, we can use 
\begin{align}
\Delta M_{Z,t}^{(4,1)}= \hat{ \Delta} M_{Z,t}^{(4,1,\mu)} =-\Delta \rho_t^{(4,1)}\ln\frac{\mu^2}{m_t^2}+\dots \, , 
\end{align}
 where the $\dots$ signify tadpole contributions which cancel against those in the bare matrix elements.  Along with the LEP 
 scheme result 
 \begin{align}
\hat{\Delta} s_{w,t}^{(4,1,\mu)}=  -\frac{\hat{c}_w^2}{2\hat{c}_{2w}}\Delta \rho_t^{(4,1)} \approx - 0.7\Delta \rho_t^{(4,1)} \approx -0.4 \Delta s_{w,t}^{(4,1,\mu)}  \,,
\end{align}
it is then easy to show that in the large-$m_t$ limit we can replace the tree-level expressions involving 
$C_{\substack{He\\33}}$ by 
\begin{align}
M_Z^2 s_w^2C_{\substack{He\\33}} \to M_Z^2 s_w^2C_{\substack{He\\33}} \left(1+\frac{1}{\vmu^2}
\left[\frac{c_w^2}{s_w^2} -2\ln\frac{\mu^2}{m_t^2}\right]\Delta \rho_t^{(4,1)} \right)\approx  M_Z^2 s_w^2C_{\substack{He\\33}} \left(1 +0.06\right)\, , \nonumber \\
 M_Z^2 s_w^2C_{\substack{He\\33}} \to M_Z^2 \hat{s}_w^2C_{\substack{He\\33}} \left(1+\frac{1}{\vmu^2}
\left[-\frac{\hat{c}_w^2}{\hat{c}_{2w}} -2\ln\frac{\mu^2}{m_t^2}\right]\Delta \rho_t^{(4,1)} \right)\approx  
M_Z^2 \hat{s}_w^2C_{\substack{He\\33}} \left(1 +0.01\right)\, , \nonumber \\
\end{align}
where the first result is for the $\alpha_\mu$ (or $\alpha$ scheme after $\mu\to\alpha$) and the second line is for the LEP scheme.
The results are a good approximation to the exact ones shown in  Table~\ref{tab:zll_nlo}. The fairly large difference between the LEP and $\alpha_\mu$ scheme makes clear that the corrections can be quite sensitive to the exact dependence on e.g.\ $s_w$ in the tree-level results.  
We have checked that the corrections to the remaining coefficients appearing in the second line of Eq.~(\ref{eq:Gam_Z_tree}) are
also well-approximated by the large-$m_t$ limit. 

The NLO corrections related to the first line of Eq.~(\ref{eq:Gam_Z_tree}) are more complicated.  To study them, we first note that the large-$m_t$ limit corrections to the function $g_\tau$ can be written in the $\alpha$ and $\alpha_\mu$ 
schemes as 
\begin{align}
\label{eq:gtau_NLO}
 g_\tau = g_\tau^{(4,0)}  + v_\sigma^2 g_\tau^{(6,0)}  + \frac{1}{v_\sigma^2}g_\tau^{(4,1)} + g_\tau^{(6,1)} 
 +\left(K_W^{(6,0,\sigma)}g_\tau^{(4,1)}  -K_W^{(4,1,\sigma)} g_\tau^{(6,0)}\right) \, .
\end{align}
The scheme-independent function $g_\tau^{(4,1)}$ is obtained by replacing $s_w\to s_w(1+\Delta s_{w})$ 
and isolating the SM corrections; it thus reads
\begin{align}
g_\tau^{(4,1)}= -4 c_w^2(1-4s_w^2)\Delta \rho_t^{(4,1)} \, .
\end{align}
The function $g_\tau^{(6,1)}$ is obtained in the same way, except for in that case one must also include corrections from $Z-\gamma$ mixing to get a finite and tadpole-free result.  The explicit result is 
\begin{align}
g_{\tau}^{(6,1)} &=  -\frac{1}{2} \dot{g}_{\tau}^{(6,0)} \ln \frac{\mu^2}{m_t^2} + 
g_{\tau}^{(4,1)}\left( -\frac{C_{HWB}}{2c_ws_w}+2 C_{\substack{Hq \\ 33}}^{(3)} -2 \sqrt{2}\frac{M_W}{M_T}  C_{\substack{uW\\ 33}}\right)  \nonumber\\
& -12  c_{2w} \Delta \rho_t^{(4,1)} \left(c_w^2 C_{HD} +2 c_w s_w C_{HWB}\right)\,,
\end{align}
where
\begin{align}
\dot{g}_{\tau}^{(6,0)} = -4 g_{\tau}^{(4,1)}\bigg[C_{HD} +\frac{s_w}{c_w} C_{HWB}  
+ 2 C_{\substack{Hq \\ 33}}^{(1)} -2 C_{\substack{Hu \\ 33}}
- \frac{\sqrt{2}s_w}{c_w^2}\frac{M_W}{m_t}\left(c_w C_{\substack{uB\\ 33}} +\frac{5}{3}s_w C_{\substack{uW\\ 33}} \right)  \bigg] \,.
\end{align}
We can now obtain the NLO corrections to the first line of  Eq.~(\ref{eq:Gam_Z_tree}) in the  large-$m_t$ limit in 
the $\alpha$ scheme  through the replacement
\begin{align}
 \frac{M_Z^2}{v_T^2}\left(1- \frac{v_T^2}{2}C_{HD} \right) \left(g_\tau^{(4,0)}  + v_T^2 g_\tau^{(6,0)}  \right) \to \frac{M_Z^2}{v_\alpha^2}\left(g_\tau^{(4,0)} + \valpha^2 K_{Z}^{(6,0,\alpha)} + \frac{1}{\valpha^2}K_{Z}^{(4,1,\alpha)} + K_{Z}^{(6,1,\alpha)}\right)  \,,
\end{align}
where the coefficients $K_Z$ are obtained by expanding out Eqs.~(\ref{eq:kZ_def}) and~(\ref{eq:gtau_NLO}).
The SM result in the $\alpha$  scheme is then given by
\begin{align}
g_\tau^{(4,0)}+\frac{1}{\valpha^2} K_Z^{(4,1,\alpha)}&  =g_\tau^{(4,0)}+ \frac{1}{\valpha^2}\left( g_\tau^{(4,0)} K_{W}^{(4,1,\alpha)} +g_\tau^{(4,0)}k_{Z}^{(4,1)} + g_\tau^{(4,1)}\right) \nonumber \\
& \approx g_\tau^{(4,0)}(1 -0.034 + 0.009 - 0.006)  \, ,
\end{align}
where the order of numerical terms on the second line matches the first, and $g_\tau^{(4,0)}\approx 0.51$.
In the $\alpha_\mu$ scheme $K_{W}^{(4,1,\mu)} =0$, and in the LEP scheme one replaces
$g_\tau^{(4,1)} \to   -\frac{s_w^2}{c_{2w}}g_\tau^{(4,1)} \approx  -0.40 g_\tau^{(4,1)}$.  This accounts for the 
SM corrections in the $\alpha$ and $\alpha_\mu$ schemes given in Table~\ref{tab:zll_nlo}, which as in Higgs and 
$W$ decay follows the pattern identified in Section~\ref{subsec:PertConv}.

Turning to SMEFT, the LO corrections in the $\alpha$ scheme are contained in 
\begin{align}
\label{eq:KZ_60_num}
K_Z^{(6,0,\alpha)} & = g_\tau^{(4,0)}K_W^{(6,0,\alpha)} - g_\tau^{(4,0)}\frac{C_{HD}}{2} + g_\tau^{(6,0)}
\nonumber \\
&\approx g_\tau^{(4,0)}K_W^{(6,0,\alpha)} - 0.25 C_{HD} + \left(0.17C_{HD} + 0.18 C_{HWB} \right) \, \nonumber\\
&  \approx
0.80C_{HD}+2.0 C_{HWB} \, , 
\end{align}
where the order of the terms on the second line matches that in the first.  In the $\alpha_\mu$ scheme one replaces 
$\alpha\to\mu$ in the above equation; in that case it is clear that the tree-level contributions from 
$C_{HD}$ and $C_{HWB}$ are quite small, since $K_W^{(6,0,\mu)}$ contains neither of these coefficients.
 At NLO in SMEFT, we can write
\begin{align}
K_Z^{(6,1,\sigma)}= K_Z^{(6,1)} + \Delta K_Z^{(6,1,\sigma)} \, , 
\end{align}
where the first term is independent of the scheme.  In terms of component objects, one finds
\begin{align}
\Delta K_Z^{(6,1,\sigma)} & = g_\tau^{(4,0)} K_W^{(6,1,\sigma)} + 2 g_\tau^{(4,0)}K_W^{(6,0,\sigma)} k_Z^{(4,1)}+2 g_\tau^{(4,1)}K_W^{(6,0,\sigma)} \, , \nonumber \\
K_Z^{(6,1)} &= g_\tau^{(4,0)} k_Z^{(6,1)} + g_\tau^{(6,1)} + g_\tau^{(6,0)}k_Z^{(4,1)}+g_\tau^{(4,1)}k_Z^{(6,0)} \,.
\end{align}
One can use explicit expressions for the component functions given above to evaluate these numerically. As  an example, let us consider the contributions from $C_{HWB}$ and $C_{HD}$ in the $\alpha_\mu$ scheme.  These are contained 
solely in the scheme-independent factor, which at the scale $\mu=M_Z$ 
\begin{align}
\label{eq:KZ_univ}
\frac{1}{\vmu^2} K_Z^{(6,1)} = -0.049 C_{HD} -0.042 C_{HWB} +\dots
\end{align}
where the $\dots$ refer to contributions from other $C_i$, which are less than 1\% in the units above.  Comparing with the 
second line of Eq.~(\ref{eq:KZ_60_num}), this implies NLO corrections of 60\% for $C_{HD}$ and $-20\%$ for $C_{HWB}$,
which are indeed close to the huge corrections in the exact results in Table~\ref{tab:zll_nlo}.  In the $\alpha$ scheme these coefficients
also contribute through the scheme dependent piece.  The numerical result is
\begin{align}
\label{eq:DKZ_alpha}
\frac{1}{\valpha^2} \Delta K_Z^{(6,1,\alpha)}  =&  -0.027 C_{HD}-0.064 C_{HWB}  \\
&+ g_\tau^{(4,0)} \left[0.17C_{\substack{Hq \\ 33}}^{(1)}   -0.17  C_{\substack{Hu \\ 33}} 
- 0.067 C_{\substack{Hq \\ 33}}^{(3)}   - 0.061  C_{\substack{uB\\ 33}} 
-0.023 C_{\substack{uW\\ 33}} \right] \,.\nonumber 
\end{align}
Even though the contributions on the first line contain up to four (three) inverse powers of $s_w$ in the case of  $C_{HD}$ ($C_{HWB}$), there is  no clear hierarchy compared to the scheme-independent pieces in Eq.~(\ref{eq:KZ_univ}).  
Combining them with the LO numbers in
Eq.~(\ref{eq:KZ_60_num}), we account for the pattern seen in Table~\ref{tab:zll_nlo}.  
Clearly, this pattern is quite complicated and is not driven by the scheme-dependent factors $K_W$ as in the SM.
On the other hand, the coefficients on the second line only appear through $K_W^{(6,1,\alpha)}$, and as seen from the exact
results in Eq.~(\ref{eq:Ztautau_numres_alpha}) we see that this factor indeed absorbs the dominant corrections from them, much like $K_W^{(4,1,\alpha)}$ in the SM. 

The LEP scheme results can be obtained from those in the $\alpha_\mu$ scheme by employing Eq.~(\ref{eq:LEP_convert}).  In the large-$m_t$ limit the only non-trivial conversions are on the functions $g_\tau$, which contain $M_W$ dependence already at  tree level.  For instance, calling the LEP-scheme functions $\hat{g}_\tau$, we have the LO SMEFT result
\begin{align}
\hat{g}_\tau^{(6,0)} = 4(1- 4s_w^2)\frac{c_w^2 s_w^2}{c_{2w}}\left[\frac{1}{2}C_{HD} + \frac{1}{c_w s_w}C_{HWB} - K_W^{(6,0,\mu)} \right] \, ,
\end{align}
and the LEP-scheme version of Eq.~(\ref{eq:KZ_60_num}) becomes
\begin{align}
\hat{K}_Z^{(6,0,\mu)} =&   \hat{g}_\tau^{(4,0)}K_W^{(6,0,\mu)} -\hat{g}_\tau^{(4,0)}\frac{C_{HD}}{2}+\hat{g}_\tau^{(6,0)}\nonumber \\
&\approx -0.29 C_{HD} -0.21 C_{HWB}-0.59\left( \sum_{j=1,2} C ^{(3)}_{\substack{Hl \\jj}}  -   C_{\substack{ll\\1221}}\right) \,.
\end{align}
Compared to the $\alpha_\mu$ scheme, the LO result for the coefficient $C_{HD}$ is significantly larger, and those from 
the operators contained in $K_W^{(6,0,\mu)}$ are slightly smaller, which roughly explains the pattern for those coefficients seen in  LEP scheme results Table~\ref{tab:zll_nlo}.  The result for $C_{HWB}$ is slightly increased, but remains small and for that  reason still receives a substantial NLO correction. 

We have derived the complete large-$m_t$ limit results and verified that they provide a good approximation to the full one, but the explicit expression for the function $\hat{g}^{(6,1,\mu)}_\tau$ is somewhat lengthy and we do not reproduce it here.  
In Section~\ref{subsec:Z_NLO} we show detailed LO and NLO results including uncertainties estimated from scale variations.  It is clear that in cases where the NLO corrections are large, namely for certain operators in the $\alpha_\mu$ and the LEP schemes,   the uncertainties are underestimated, while in the $\alpha$-scheme the uncertainty estimates are more reliable.  This example highlights very clearly that the issue of NLO corrections in SMEFT is considerably more scheme and process-dependent than in  the SM. The general rule that NLO corrections to weak decays 
are smaller in the LEP and $\alpha_\mu$ schemes than in the $\alpha$ scheme familiar from the 
SM does not transfer over to SMEFT.

 \begin{figure}[thb]
          \includegraphics[scale=0.45]{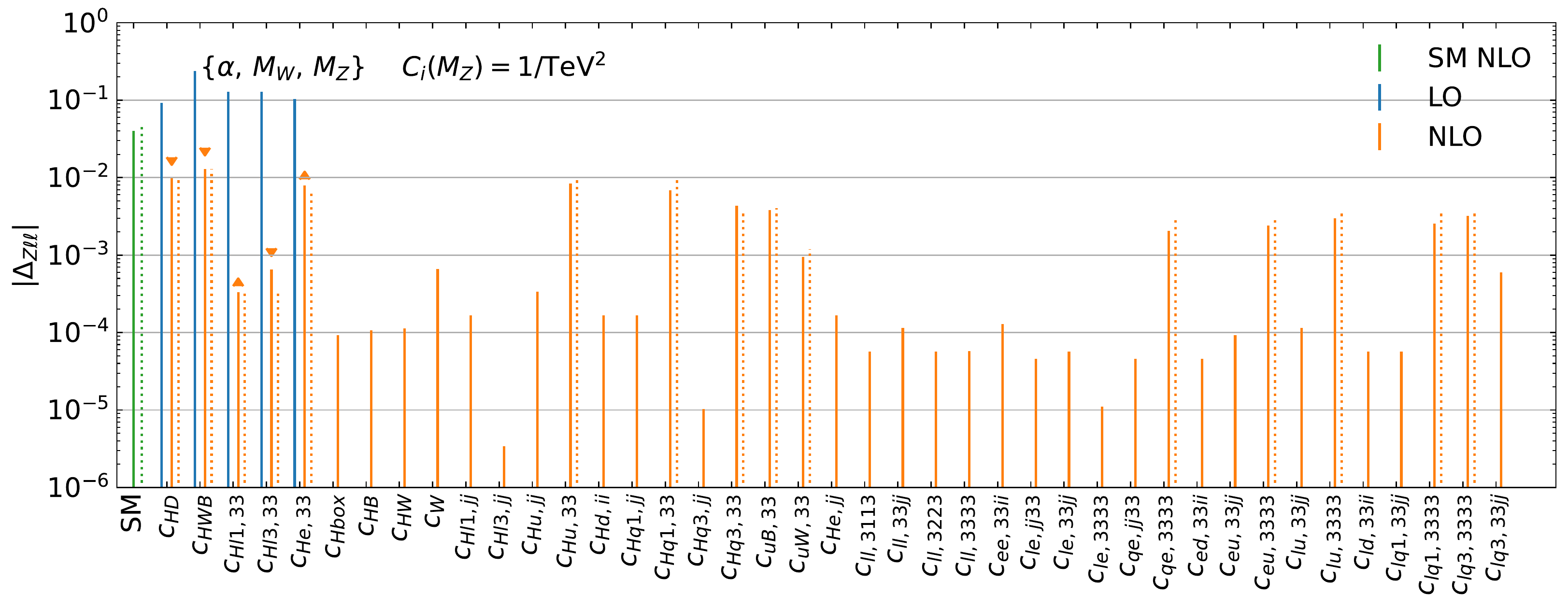} \\
          \includegraphics[scale=0.45]{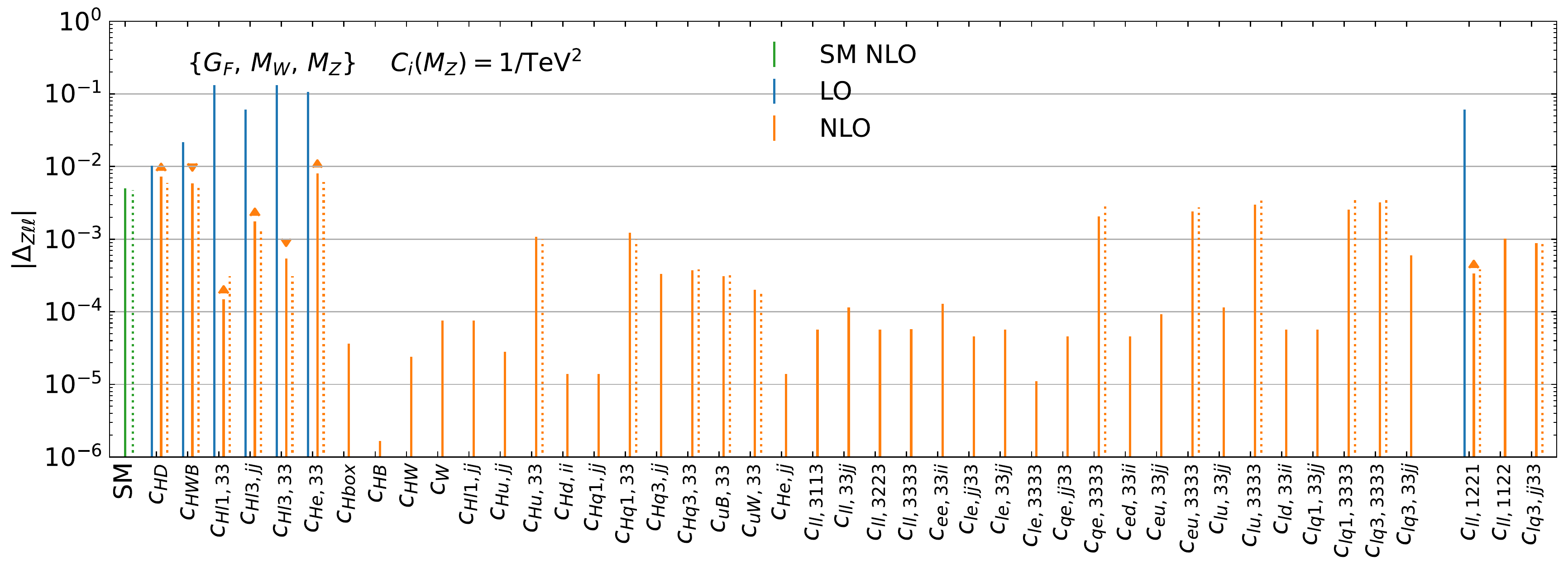} \\
          \includegraphics[scale=0.45]{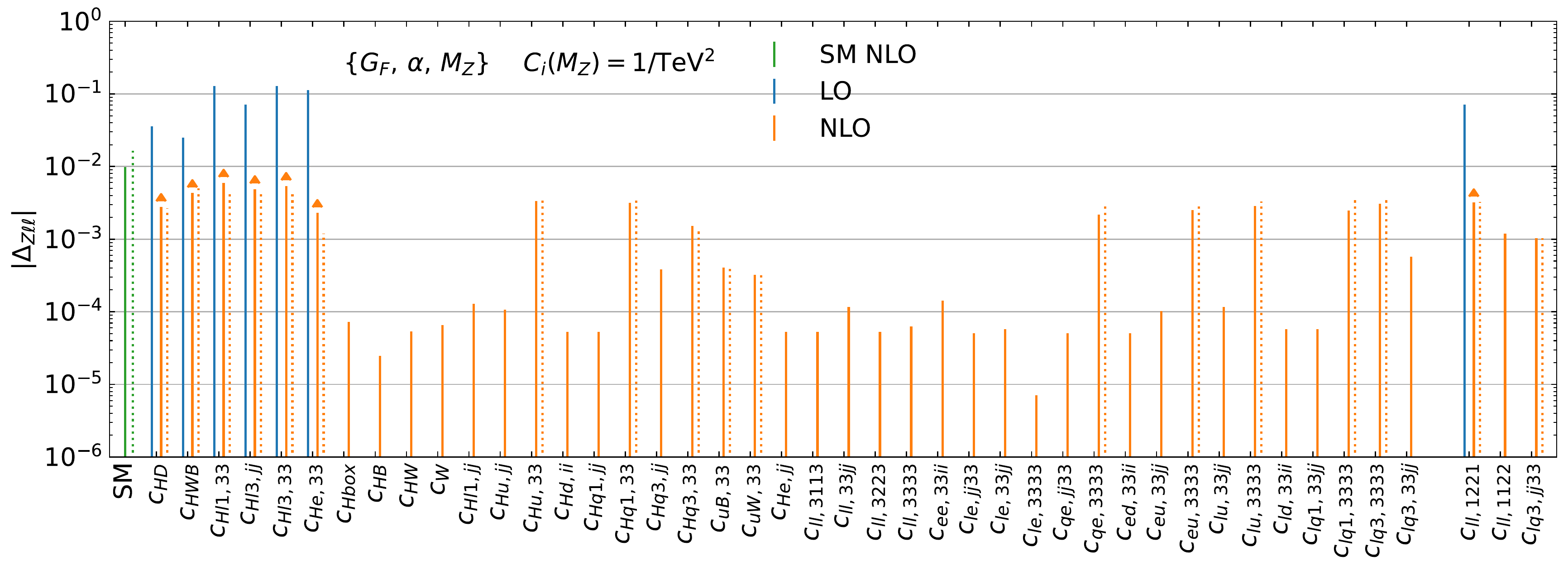} \\
        \caption{As in Figure~\ref{fig:Wdecay_numRes}, but for the decay $Z\to \tau \tau$.}
        \label{fig:Zdecay_numRes}
 \end{figure}
 
%
\begin{table}
\centering
\begin{tabular}{c|ccccccccc}
$Z \rightarrow \tau \tau$ & SM      & $C_{HD}$ & $C_{HWB}$ & $C_{\substack{He\\33}}$ & $C^{(1)}_{\substack{Hl\\33}}$   & $C^{(3)}_{\substack{Hl\\33}}$& $C^{(3)}_{\substack{Hl\\jj}}$ & $C_{\substack{ll\\1221}}$ \\[3mm] \hline
$\alpha$                      & $-4.0\%$ & $-10.6\%$ & $-5.4\%$   & $7.7\%$                  & $0.3\%$         & $-0.5\%$                  & ---   & ---                       \\
$\alpha_\mu$                  & $<0.1\%$ & $71.1\%$  & $-27.2\%$  & $7.6\%$                  & $0.1\%$      & $-0.4\%$   & $2.9\%$   & $0.6\%$                    \\
LEP    &    $1.0\%$      &        $7.8\%$   &       $17.4\%$     &         $2.0\%$        &        $4.7\%$  &  $4.2\%$    &    $6.9\%$  &     $4.5\%$                                          
\end{tabular}
\caption{NLO corrections to prefactors of LO Wilson coefficients in the three schemes. Negative corrections indicate a reduction in the magnitude of the numerical coefficient of a given Wilson coefficient, while $<0.1\%$ indicates changes below $0.1\%$, both positive and negative. The flavour index $j$ refers to $j\in 1,2$.}
\label{tab:zll_nlo}
\end{table}
\FloatBarrier
\section{Universal corrections in SMEFT}
\label{sec:universal_corrections}

A recurring theme of the previous sections was that EW corrections are dominated by top loops. 
While the numerical patterns in EW input-scheme dependent top-loop corrections in the SM are quite regular, those in SMEFT are more process and Wilson-coefficient dependent.  The purpose of this section is to show that the dominant scheme-dependent
EW corrections in SMEFT can nonetheless be taken into account by a certain set of simple substitutions in the 
LO results, similarly to the well-studied case of the SM.

Let us begin the discussion with the SM, where an important feature is that weak vertices in the $\alpha$ scheme receive
corrections proportional to $\Delta r_t^{(4,1)}$, related to the renormalisation of $v_T$. It is
simple to resum such corrections to all orders in perturbation theory.  Using the large-$m_t$ limit
result in Eq.~(\ref{eq:Valpha_LMT_SM}), and keeping for the moment only the $\Delta r_t^{(4,1)}$ terms (i.e.\ terms enhanced in the limit $c_w^2/s_w^2\gg 1$, in which case the $\Delta M_{W,t}$ piece is subleading), 
we have
\begin{align}
	\label{eq:def_vtilde}
	\frac{1}{v_{T,0}^2}  &\approx 
	\frac{1}{\valpha^2}\left[1 +\frac{1}{v_{T,0}^2}\Delta r_t^{(4,1)} \right]
	\approx \frac{1}{\valpha^2}\left[1+ \frac{1}{\valpha^2}\Delta r_t^{(4,1)} + \frac{1}{\valpha^2 v_{T,0}^2} \left(\Delta r_t^{(4,1)}\right)^2 +\dots \right] \nonumber \\
	& = \frac{1}{\valpha^2} \left[1-\frac{1}{\valpha^2}\Delta r_t^{(4,1)}\right]^{-1} \equiv \frac{1}{\tilde{v}_\alpha^2} \,.
\end{align}
This resums the $\Delta r_t^{(4,1)}$ terms to all orders.   Adding back the subleading terms away from the double limit $m_t, \,  c_w^2/s_w^2\gg 1$ by matching with the one-loop result yields
\begin{align}
	\label{eq:SM_resummed}
	\frac{1}{v_{T,0}^2}  = \frac{1}{\tilde{v}_\alpha^2}\left[1 - \frac{1}{\tilde{v}_\alpha^2} \left(\Delta v_\alpha^{(4,1,\alpha)}+ \Delta r_t^{(4,1)}\right)\right] \,.
\end{align}
Expressing the counterterm for $v_T$ as an expansion in $\tilde{v}_\alpha$ rather than $\valpha$ will obviously lead to 
a quicker convergence between orders.  For example, the SM prediction to NLO for the derived quantity $G_F$ in such 
a ``$\tilde{\alpha}$ scheme" is 
\begin{align}
	\label{eq:GF_alphatilde}
	G_{F,\tilde{\alpha}}^{\rm NLO} & = \frac{1}{\sqrt{2} \tilde{v}_\alpha^2}
	\left[1+ \frac{1}{\tilde{v}_\alpha^2} \left(\Delta r^{(4,1)} - \Delta r_t^{(4,1)} \right) \right] \, .
\end{align}
Numerically, including uncertainties from scale variation using the procedure described in Section~\ref{sec:Derived},
\begin{align}
\label{eq:GF_SM_resummed}
	\frac{G_{F,\tilde{\alpha}}^{\text{LO} }}{G_F} = 1.000^{+0.007}_{-0.007} \, , \qquad  \frac{G_{F,\tilde{\alpha}}^{\text{NLO} }}{G_F} = 0.994^{+0.000}_{-0.000}  \, , 
\end{align}
where $G_{F,\tilde{\alpha}}^{\text{LO} }$ refers to the first term in Eq.~(\ref{eq:GF_alphatilde}). 
This shows considerably improved convergence compared to the fixed-order $\alpha$ scheme expression in  
Eq.~(\ref{eq:GF_alpha}), and scale variations in the LO result give a good estimate of the NLO corrections. 

To the best of our knowledge, a resummation of the type described above was first derived in \cite{Consoli:1989fg}, at 
the level of the $W$-boson mass in the LEP scheme (and also including subleading two-loop terms in the limit $s_w\to 0$). 
In that case, similar reasoning using Eq.~(\ref{eq:DeltaR}) as a starting point leads to the resummed LO prediction
\begin{align}
\left(M_W^{\widetilde{ \rm LO}}\right)^2	= \tilde{M}_W^2 \equiv \frac{M_Z^2}{2}\left(1+\sqrt{1-\frac{4\pi \alpha \vmu^2}{M_Z^2 \left(1-\frac{1}{\vmu^2}\Delta r_t^{(4,1)} \right)}} \right) \, .
\end{align}
The NLO result within the resummation formalism,  modified to avoid double counting, is
 \begin{align}
	M_W^{\widetilde{ \rm NLO}}=  \tilde{M}_W \left[1-\frac{1}{2}\frac{\hat{s}_w^2}{\hat{c}_{2w}}\frac{1}{\vmu^2}\Delta \tilde{r}^{(4,1)}  \right] , \, \qquad \Delta \tilde{r}^{(4,1)} = \Delta r^{(4,1)} - \Delta r_t^{(4,1)} \, .
\end{align}
Evaluating numerically and including uncertainties from scale variation leads to 
\begin{align}
\label{eq:MW_SM_resummed}
	M_W^{\widetilde{ \rm LO}}= 80.33^{+0.13}_{-0.13} \text{GeV} \, , \qquad M_W^{\widetilde{ \rm NLO}}= 80.44^{+0.01}_{-0.00} \text{GeV} \, ,
\end{align}
which again shows improved perturbative convergence compared to the fixed-order results in 
Eqs.~(\ref{MWpredict_LO_alphalite}, \ref{MWpredict_NLO_alphalite}).  

Resummations are especially useful for derived parameters, which are known to a high level of 
experimental and perturbative accuracy.  However, when viewed as a subset of corrections to EW vertices contributing
to scattering amplitudes or decay rates in a specific input scheme, the corrections beyond NLO contained in the resummed formulas 
are typically negligible compared to process-dependent experimental and perturbative uncertainties.  
For instance, the central values of the LO resummed results in Eqs.~(\ref{eq:GF_SM_resummed}, \ref{eq:MW_SM_resummed})
can be split up as
\begin{align}
\frac{G_{F,\tilde{\alpha}}^{\text{LO} }}{G_F} & = 1.034 - 0.035 + 0.001  =1.000\, ,\nonumber \\
M_W^{\widetilde{ \rm LO}}&= (79.82 + 0.54 - 0.03)~\text{GeV} = 80.33\,\text{GeV}   \, , 
\end{align}
where in both cases the  sequence of three numbers after the first equality are the fixed-order LO, the fixed-order NLO correction, and the beyond NLO corrections, respectively.  Clearly, the NLO expansions of the resummed formulas approximate the full results
at sub percent-level precision, so a fixed-order implementation suffices for practical applications.

Universal NLO corrections to weak vertices implied by resummation can be obtained through a procedure of substitutions
on LO results.
The remaining, non-universal NLO corrections need to be calculated on a case-by-case basis,
but these are typically small compared to the ones already included at LO through the aforementioned substitutions.
While such procedures for universal corrections are well known in the SM (see for instance \cite{Denner:1991kt}), we give here a first implementation within 
SMEFT.  Step-by-step, it works as follows
\begin{itemize}
\item[(1)] Write the LO amplitude in terms of $v_T$, $M_W$, and $M_Z = M_W/c_w$.  
\item[(2)] Make EW-input scheme dependent replacements on the LO amplitudes.  In the $\alpha$ or $\alpha_\mu$ scheme, these read
\begin{align}
\frac{1}{v_T^2} & \rightarrow   \frac{1}{v_\sigma^2} \left[ 1 +v_\sigma^2 K_W^{(6,0,\sigma)}+ \frac{K_W^{(4,1,\sigma)}}{v_\sigma^2}    +  K_W^{(6,1,\sigma)} \right] \, , \nonumber \\
s_w^2 &\to  s_w^2\left(1 - \frac{1}{v_\sigma^2}\Delta r_t^{(4,1)} + \Delta v_\sigma^{(6,0,\sigma)}  \Delta r_t^{(4,1)} - 2 C_{\substack{Hq \\ 33}}^{(3)} \Delta r_t^{(4,1)}  \right) \, ,
\nonumber \\
c_w^2 & \to  c_w^2\left(1-  \frac{1}{v_\sigma^2} \Delta \rho_t^{(4,1)}  + 
\Delta v_\sigma^{(6,0,\sigma)}   \Delta \rho_t^{(4,1)} - 2 C_{\substack{Hq \\ 33}}^{(3)} \Delta \rho_t^{(4,1)}   \right) \,,
\label{eq:universal_shifts}
\end{align}
where as usual $\sigma\in\{\alpha,\mu\}$ and the $K_W$ are given in Eqs.~(\ref{eq:KWalpha}, \ref{eq:KWmu}).

In the LEP scheme, make the above replacements with $\sigma=\mu$ in the LO
amplitude. Subsequently, eliminate $M_W$ in favour of $\hat{M}_W$ using Eq.~(\ref{eq:MW_prediction}), 
in both the replacements and everywhere else in the LO
observable (so that factors of $M_W$ related to phase space are also taken into account).
\item[(3)] Expand the resulting expressions to NLO in a fixed-order SMEFT expansion before evaluating numerically.
\end{itemize}
We shall refer to results obtained from the above procedure as ``LO$_K$'' accurate.

In the SM, the substitutions in Eq.~(\ref{eq:universal_shifts}) are sufficient to capture NLO corrections proportional to 
$\Delta r_t^{(4,1)}$.  Beyond that, writing $M_Z=M_W/c_w$ before performing the shifts ensures that the 
large-$m_t$ limits of both $W$ and $Z$ decay are reproduced. 
In SMEFT, the substitution for $v_T$ is motivated by Eq.~(\ref{eq:weak_vev}), which splits the counterterm for $v_T$ into 
a ``physical",  $\mu$-independent order-by-order in perturbation theory and tadpole free part,  $K_W$, and an ``unphysical'' part, which
is tadpole dependent and divergent.  The physical part captures the most singular large-$m_t$ corrections as $s_w\to0$ in SMEFT, 
as well as $\mu$-dependent logarithms.  The substitutions for $s_w$ also capture such pieces of its counterterm, 
including a piece proportional to $C_{\substack{Hq \\ 33}}^{(3)}$ which is easily shown to be proportional to the NLO SM
result.  Finally, in both SMEFT and the SM, the shift for $c_w$ is chosen to maintain $s_w^2+c_w^2=1$.   While Eq.~(\ref{eq:universal_shifts}) is not unique, other reasonable choices would differ only by terms proportional to 
$\Delta \rho_t^{(4,1)}$  rather than $\Delta r_t^{(4,1)}$ and thus agree with the above to 
roughly the percent level.\footnote{Substitutions
for SMEFT vertices involving photons need to be considered on a case-by-case basis.  For instance,
a QED-type vertex in the $\alpha$ and LEP schemes is proportional to $e$ and spurious corrections
would be generated through the substitution procedure outlined above.}

\setlength{\tabcolsep}{3pt}
\renewcommand{\arraystretch}{1.5}
\begin{table}[t]
	\centering
	\begin{tabular}{c||cc|cc|cc}
		& \multicolumn{2}{c}{$W\rightarrow \tau \nu$}& \multicolumn{2}{c}{$Z\rightarrow \tau \tau$}& \multicolumn{2}{c}{$h\rightarrow b \bar{b}$}\\
		&$\alpha$ & LEP &$\alpha$ & LEP &$\alpha$ & LEP \\
		\hline \hline
		NLO&$0.992^{+0.001}_{-0.001}$&$1.003^{+0.000}_{-0.000}$&
		$0.992^{+0.001}_{-0.001}$ & $1.002^{+0.000}_{-0.000}$ & 
		$0.991^{+0.001}_{-0.001}$ & $1.000^{+0.000}_{-0.000}$ \\
		NLO$_{\mathrm{t}}$&$1.001^{+0.007}_{-0.007}$&$1.003^{+0.005}_{-0.005}$&
		$1.002^{+0.007}_{-0.007}$ & $1.003^{+0.002}_{-0.002}$ & 
		$1.013^{+0.007}_{-0.007}$ & $1.011^{+0.001}_{-0.001}$ \\ \hline
		LO&$1.036^{+0.008}_{-0.008}$&$0.983^{+0.005}_{-0.005}$&
		$1.034^{+0.008}_{-0.008}$ & $0.993^{+0.001}_{-0.001}$ & 
		$1.045^{+0.007}_{-0.007}$ & $1.008^{+0.001}_{-0.001}$  \\
		$\text{LO}_K$ &$1.001^{+0.007}_{-0.007}$&$1.003^{+0.005}_{-0.005}$&
		$1.002^{+0.007}_{-0.007}$	& $1.003^{+0.002}_{-0.002}$ & 
		$1.010^{+0.007}_{-0.007}$ & $1.008^{+0.001}_{-0.001}$ \\
	\end{tabular}
	\caption{SM results in the $\alpha$ and LEP~schemes. For each process, the results are normalised to the SM NLO results in the $\alpha_\mu$~scheme.  }
	\label{tab:resummingSM}
\end{table}
\renewcommand{\arraystretch}{1}

In Table~\ref{tab:resummingSM}, we compare various perturbative approximations to heavy-boson decay rates in 
the SM within the $\alpha$ and LEP schemes, in each case normalised to the NLO result in the $\alpha_\mu$ scheme at the default
scale choice. The LO and NLO results refer to fixed-order perturbation theory, NLO$_t$ refers to the large-$m_t$ limit of NLO, 
and LO$_K$ refers to the sum of LO and NLO corrections obtained through the above procedure.  
For the case of $W$ and $Z$ decay in the $\alpha$ scheme, the convergence between 
LO$_K$ and NLO is greatly improved compared to pure fixed order, and varying the scale
in the LO$_K$ results gives a good estimate for the residual corrections contained in the full NLO result.
Also in Higgs decay LO$_K$ is a marked improvement over LO, although
in that case the results in all schemes are subject to a roughly -1\% scheme-independent
correction which is unrelated to the large-$m_t$ limit and not captured through scale variations.

We next turn to SMEFT, focusing on cases where LO$_K$ results involve corrections proportional to $\Delta r_t^{(4,1)}$.
In Table~\ref{tab:resummingAlphaScheme} we show heavy-boson decay rates in 
SMEFT in the $\alpha$ scheme, listing the prefactors of Wilson coefficients appearing in $K_W^{(6,1,\alpha)}$.  
In this case, the NLO$_t$ (but not LO$_K$) results use the large-$m_t$ limit of
Eq.~(\ref{eq:C_Running}) for scale variations of the Wilson coefficients.  We see that 
also in SMEFT, the LO$_K$ description improves perturbative convergence compared to pure
fixed order, taking into account especially the dominant scheme-dependent corrections.  This works 
best for $W$ decay, where the central values of LO$_K$ reproduce the NLO$_t$ results by construction, and perturbative
uncertainties are reduced compared to LO while still showing a good overlap with the NLO results.   In Higgs decay, 
Wilson coefficients that receive significant scheme-independent corrections as shown Eq.~(\ref{eq:KH61}), such 
as  $C_{\substack{uW\\33}}$, display the
biggest deviations from the NLO$_t$ and NLO results at LO$_K$ accuracy, although scale variations generally
give a good indication of the size of the missing pieces.  The case of $Z$ decay is similar, although in contrast to Higgs and 
$W$ decay the form of the LO amplitude in Eq.~(\ref{eq:Gam_Z_tree}) implies that the shifts of $s_w$  
in Eq.~(\ref{eq:universal_shifts}) also play a role.  
This latter effect is even more important in 
$Z$ decay in the $\alpha_\mu$ scheme; as shown in Table~\ref{tab:resummingZdecay_vmu}, LO$_K$ accuracy
largely takes into account the very large corrections to $C_{HD}$ and $C_{HWB}$ (as well as the more moderate
but still significant corrections to $C_{\substack{He\\33}}$) seen in Table~\ref{tab:zll_nlo}.  The LO$_K$ 
results for Higgs and $W$  decay in the $\alpha_\mu$ scheme, and for all decays in the LEP scheme, show similar
levels of improvement as the cases discussed above -- detailed tables can be found in Appendix~\ref{app:universal_corrections}.


\setlength{\tabcolsep}{2pt}
\renewcommand{\arraystretch}{1.5}
\begin{table}[t]
\small
\centering
\begin{tabular}{c||c|c|c|c|c|c|c}
$W\to \tau \nu$	& $C_{HD}$ & $C_{HWB}$ & $C^{(3)}_{\substack{Hq\\33}}$ & $C_{\substack{Hu\\33}}$  & $C^{(1)}_{\substack{Hq\\33}}$ & $C_{\substack{uB\\33}}$& $C_{\substack{uW\\33}}$  \\[3mm]
	\hline \hline
	NLO
	&$1.713^{+0.000}_{-0.011}$&$3.621^{+0.000}_{-0.011}$&$-0.079^{+0.018}_{-0.012}$	&$-0.195^{+0.038}_{-0.000}$	&$0.172^{+0.000}_{-0.033}$	&$-0.072^{+0.008}_{-0.000}$	&$-0.032^{+0.005}_{-0.000}$	\\
	$\text{NLO}_{t}$
	&$1.694^{+0.000}_{-0.009}$&$3.601^{+0.001}_{-0.008}$&$-0.067^{+0.019}_{-0.004}$	&$-0.206^{+0.034}_{-0.000}$	&$0.206^{+0.000}_{-0.030}$	&$-0.073^{+0.005}_{-0.000}$	&$-0.033^{+0.004}_{-0.000}$	\\ \hline
	LO&$1.742^{+0.120}_{-0.120}$&$3.733^{+0.131}_{-0.131}$&$0.000^{+0.008}_{-0.008}$	&$0.000^{+0.182}_{-0.182}$	&$0.000^{+0.189}_{-0.189}$	&$0.000^{+0.066}_{-0.066}$	&$0.000^{+0.059}_{-0.059}$	\\
	
	$\text{LO}_K$
	&$1.694^{+0.016}_{-0.033}$&$3.601^{+0.021}_{-0.031}$&$-0.067^{+0.011}_{-0.000}$	&$-0.206^{+0.029}_{-0.000}$	&$0.206^{+0.000}_{-0.032}$	&$-0.073^{+0.007}_{-0.000}$	&$-0.033^{+0.005}_{-0.000}$		\\
	
\multicolumn{4}{c}{}	\\[1mm]                                
$h\to b \bar{b}$	& $C_{HD}$ & $C_{HWB}$ & $C^{(3)}_{\substack{Hq\\33}}$ & $C_{\substack{Hu\\33}}$  & $C^{(1)}_{\substack{Hq\\33}}$ & $C_{\substack{uB\\33}}$& $C_{\substack{uW\\33}}$  \\[3mm]
	\hline \hline
	NLO&$1.106^{+0.002}_{-0.018}$&$3.482^{+0.005}_{-0.016}$&$-0.116^{+0.025}_{-0.000}$	&$-0.079^{+0.033}_{-0.000}$	&$0.058^{+0.000}_{-0.034}$	&$-0.030^{+0.008}_{-0.000}$	&$-0.040^{+0.009}_{-0.000}$	\\
	$\text{NLO}_{t}$&$1.129^{+0.002}_{-0.012}$&$3.560^{+0.005}_{-0.011}$&$-0.105^{+0.030}_{-0.000}$	&$-0.088^{+0.028}_{-0.000}$	&$0.088^{+0.000}_{-0.027}$	&$-0.031^{+0.006}_{-0.000}$	&$-0.042^{+0.006}_{-0.000}$	
	\\
	\hline 
	LO&$1.242^{+0.089}_{-0.089}$&$3.733^{+0.128}_{-0.128}$&$0.000^{+0.112}_{-0.112}$	&$0.000^{+0.183}_{-0.183}$	&$0.000^{+0.188}_{-0.188}$	&$0.000^{+0.066}_{-0.066}$	&$0.000^{+0.094}_{-0.094}$	\\

	$\text{LO}_K$&$1.134^{+0.004}_{-0.021}$&$3.536^{+0.014}_{-0.024}$&$-0.068^{+0.125}_{-0.110}$	&$-0.088^{+0.034}_{-0.000}$	&$0.088^{+0.000}_{-0.027}$	&$-0.031^{+0.007}_{-0.000}$	&$0.004^{+0.036}_{-0.029}$	\\
	                           
\multicolumn{4}{c}{} \\[1mm]
$Z\to \tau \tau$	&$C_{HD}$ & $C_{HWB}$ & $C^{(3)}_{\substack{Hq\\33}}$ & $C_{\substack{Hu\\33}}$  & $C^{(1)}_{\substack{Hq\\33}}$ & $C_{\substack{uB\\33}}$& $C_{\substack{uW\\33}}$  \\[3mm]
	\hline \hline
	NLO
	&$1.406^{+0.002}_{-0.021}$&$3.867^{+0.003}_{-0.016}$&$-0.074^{+0.014}_{-0.001}$	&$-0.143^{+0.031}_{-0.000}$	&$0.117^{+0.000}_{-0.032}$	&$-0.065^{+0.007}_{-0.000}$	&$-0.016^{+0.008}_{-0.000}$	\\
	
	$\text{NLO}_{t}$
	&$1.419^{+0.002}_{-0.015}$&$3.876^{+0.004}_{-0.011}$&$-0.061^{+0.016}_{-0.002}$	&$-0.156^{+0.027}_{-0.000}$	&$0.156^{+0.000}_{-0.026}$	&$-0.067^{+0.006}_{-0.000}$	&$-0.019^{+0.007}_{-0.000}$	\\
	\hline
	LO&$1.573^{+0.109}_{-0.109}$&$4.088^{+0.144}_{-0.144}$&$0.000^{+0.008}_{-0.008}$	&$0.000^{+0.163}_{-0.163}$	&$0.000^{+0.172}_{-0.172}$	&$0.000^{+0.072}_{-0.072}$	&$0.000^{+0.064}_{-0.064}$	\\
	
	$\text{LO}_K$&$1.426^{+0.000}_{-0.013}$&$3.870^{+0.030}_{-0.040}$&$-0.061^{+0.008}_{-0.000}$	&$-0.173^{+0.050}_{-0.002}$	&$0.173^{+0.000}_{-0.042}$	&$-0.061^{+0.012}_{-0.000}$	&$-0.023^{+0.007}_{-0.000}$	\\

\end{tabular}

\caption{The numerical prefactors of the Wilson coefficients in the $\alpha$~scheme appearing in $K_W^{(6,1,\alpha)}$ for various perturbative approximations. The tree-level decay rate as well as $v_\alpha^2$ have been factored out and the results have been evaluated at the scale of the process. We show the results for $W$~decay (top), $h$~decay (center) and $Z$~decay (bottom).
}
\label{tab:resummingAlphaScheme}
\end{table}
\renewcommand{\arraystretch}{1}

\setlength{\tabcolsep}{2pt}
\renewcommand{\arraystretch}{1.5}
\begin{table}[t]
\centering
\small
\begin{tabular}{c||c|c|c|c|c|c|c}
$Z \to \tau \tau$	&$C^{(3)}_{\substack{Hl\\jj}}$& $C^{(3)}_{\substack{lq\\jj33}}$ & $C_{\substack{ll\\1221}}$ & $C^{(3)}_{\substack{Hq\\33}}$  & $C_{HD}$ & $C_{HWB}$  & $C_{\substack{He\\33}}$   \\[3mm]
	\hline \hline
	NLO
	&$-1.029^{+0.001}_{-0.000}$&$0.015^{+0.000}_{-0.001}$ &$1.006^{+0.000}_{-0.000}$  &$0.006^{+0.000}_{-0.002}$ &$-0.289^{+0.009}_{-0.007}$ &$0.258^{+0.003}_{-0.008}$ &$-1.897^{+0.006}_{-0.002}$ 	\\	
	$\text{NLO}_{t}$
	&$-1.021^{+0.001}_{-0.000}$&$0.015^{+0.004}_{-0.005}$ &$1.006^{+0.002}_{-0.002}$ &$0.006^{+0.000}_{-0.002}$  &$-0.266^{+0.006}_{-0.005}$ &$0.272^{+0.002}_{-0.002}$ &$-1.864^{+0.005}_{-0.001}$ 	\\
	\hline
	LO&$-1.000^{+0.015}_{-0.015}$&$0.000^{+0.026}_{-0.026}$  &$1.000^{+0.004}_{-0.004}$  &$0.000^{+0.001}_{-0.001}$  &$-0.169^{+0.011}_{-0.011}$&$0.355^{+0.012}_{-0.012}$	  &$-1.764^{+0.046}_{-0.046}$   \\
$\text{LO}_K$&$-1.021^{+0.012}_{-0.010}$&$0.015^{+0.000}_{-0.001}$  &$1.006^{+0.004}_{-0.004}$  &$0.006^{+0.001}_{-0.000}$ &$-0.260^{+0.017}_{-0.017}$ &$0.267^{+0.009}_{-0.009}$ &$-1.838^{+0.048}_{-0.048}$   \\
\end{tabular}
\caption{The numerical prefactors of the $Z$~decay SMEFT Wilson coefficients in the $\alpha_\mu$~scheme appearing leading to dominant corrections at various perturbative approximations. The tree-level decay rate as well as $v_\mu^2$ have been factored out and the results have been evaluated at the scale of the process.}
\label{tab:resummingZdecay_vmu}
\end{table}
\renewcommand{\arraystretch}{1}

\FloatBarrier

\section{Conclusions}
\label{sec:conclusions}

We have performed a systematic study of three commonly used EW input schemes to NLO in dimension-six SMEFT.  
After introducing a unified notation which makes transparent the connections between  the $\alpha$, $\alpha_\mu$, and 
LEP schemes, thus facilitating both NLO calculations in the schemes directly or conversions 
between them, we studied the structure of the SMEFT expansion in the different schemes.  This was done at the generic level 
in Section~\ref{sec:salient}, at the level of derived parameters such as the $W$-boson mass in the LEP scheme or $G_F$ in the
$\alpha$ scheme in  Section~\ref{sec:Derived}, and at the level of heavy boson decay rates in 
Section~\ref{sec:HeavyBosonDecays}.  In all cases these NLO calculations are either original or 
generalise previous results to include the full flavour structure of SMEFT.   They will be useful for benchmarking 
automated tools for NLO EW corrections in SMEFT, when they become available, and we have therefore included
the analytic results as computer files in the electronic submission of this work.

In the SM, the dominant differences between EW input schemes are mainly taken into account by 
NLO top-loop corrections to the sine of the Weinberg angle, $s_w$. As an example, for decay rates of heavy bosons, 
these appear in our formalism through the renormalisation of the Higgs vacuum expectation value $v_T$, 
and given that such decay rates scale as $1/v_T^2$ a regular pattern of roughly -3.5\% corrections in the $\alpha$ scheme
compared to the $\alpha_\mu$ and LEP schemes is observed.  In SMEFT, the dominant corrections related to
the renormalisation of $v_T$ still arise from top loops, but these involve $\mu$-dependent logarithmic corrections related
to the running of Wilson coefficients, in addition to more complicated dependence on the Weinberg angle than in the SM,
and as a result the numerical results across Wilson coefficients and processes are not nearly as regular.  Nonetheless, 
we identified the analytic structure of the dominant scheme-dependent NLO corrections in SMEFT, and gave in Section~\ref{sec:HeavyBosonDecays} a simple procedure for including these universal NLO 
corrections in the LO results.  Once these are taken into account, residual NLO corrections in different schemes are
of similar size; these corrections can be approximated by calculating process dependent top-loop corrections, or 
eliminated altogether through an exact NLO calculation.  

We end with a comment on theory uncertainties and the choice of an EW input scheme in fits of SMEFT Wilson coefficients 
from data.  Observables in SMEFT exhibit scheme-dependent sensitivity to the full set of SMEFT Wilson coefficients because
input parameters across schemes are related through SMEFT expansions. However, once a comprehensive set of observables is combined and dominant scheme-dependent corrections have been taken into account, 
there is no strong argument in favour of one scheme or another, and the consistency of
Wilson coefficients obtained from global fits to data in different input schemes provides a valuable check on the 
robustness of such analyses.

\section*{Acknowledgements}
We thank Pier Paolo Giardino and Sally Dawson for comparison with~\cite{Dawson:2019clf,Dawson:2022bxd}.
A.B.~gratefully acknowledges support from the Alexander-von-Humboldt foundation as a Feodor Lynen Fellow and 
the hospitality and support of the Mainz Institute for Theoretical Physics, where parts of this project were completed.
B.P. is grateful to the Weizmann Institute of Science for its kind hospitality and support through the SRITP and the Benoziyo Endowment Fund for the Advancement of Science.

 \appendix

\section{The $\alpha_\mu$ scheme at NLO}
\label{sec:NLO_vmu}

The $\alpha_\mu$ scheme is 
defined by the renormalisation condition that the relation in Eq.~(\ref{eq:v_mu_def}), $\vmu = \left(\sqrt{2}G_F\right)^{-\frac{1}{2}}$,
holds to all orders in perturbation theory.
The Fermi constant~$G_F$ is a Wilson coefficient 
appearing in the effective Lagrangian
\begin{align}
\label{eq:L_eff}
{\cal L}_{\rm eff} = {\cal L}_{\rm QED} + {\cal L}_{\rm QCD} + {\cal L}_\mu \, ,
\end{align}
where 
\begin{align}
 {\cal L}_\mu =  
-2\sqrt{2} G_F Q_\mu, \qquad Q_\mu =  \left[\bar{\nu}_\mu \gamma_\mu P_L  \mu \right]\times 
\left[ \bar{e}\gamma^\mu  P_L \nu_e \right] \,.
\end{align}
The four-fermion operator $Q_\mu$ mediates tree-level muon decay,
and radiative corrections are obtained through Lagrangian insertions of a 
five-flavour version of QED$\times$QCD, where the top-quark is integrated out.
We will work only to NLO in the couplings, so QCD couplings will not appear and we can drop the QCD Lagrangian in 
what follows.

The Fermi constant $G_F$ is calculated by matching SMEFT onto
the effective Lagrangian above, by integrating out the heavy electroweak bosons 
and the top quark.  In practice, this is done by ensuring that renormalised Green's functions match order by order in perturbation theory,  to leading order in the 
EFT expansion parameter $m_\mu/M_W\ll 1$.   The matching can be performed with any convenient choice of external states.
We work with massless fermions, and set all external momenta to zero.  
In that case the loop corrections to the bare tree-level amplitude in the EFT are scaleless and vanish, so the renormalised amplitude is just given by the tree-level one plus UV counterterms.  
The main task is thus to evaluate the renormalised NLO matrix element for the muon decay in SMEFT.  

To write the matrix element for the process $\mu \to \nu_\mu e \bar{\nu}_e$, 
we first  define the spinor product 
\begin{align}
\label{eq:Sdef}
S_\mu = \left[\bar{u}(p_{\nu_\mu})\gamma_\nu P_L u(p_\mu) \right]\times 
\left[  \bar{u}(p_{e})\gamma^\nu P_L v(p_{\bar{\nu}_e})\right]  \, ,
\end{align}
where $P_L = (1-\gamma_5)/2$ and it is understood that the arguments 
of the Dirac spinors $u$ and $v$ are evaluated at $p_i=0$.  Furthermore, we define expansion
coefficients of the bare one-loop amplitude in terms of the bare parameter $v_{T,0}$ as 
\begin{align}
\label{eq:A_mu_bare}
{\cal A}_{\rm bare} &= -\frac{2}{\vTbare^2}
\left({\cal A}_{\rm bare}^{(4,0)} +\vTbare^2 {\cal A}_{\rm bare}^{(6,0)}+ \frac{1}{\vTbare^2}{\cal A}_{\rm bare}^{(4,1)} + {\cal A}_{\rm bare}^{(6,1)} \right) S_\mu+\dots \, .
\end{align}
The $\dots$ in the above equations refer either to spinor structures with different chirality structure, which we do not interfere with the tree-level SM result and 
can thus be neglected, or matrix elements of evanescent operators. 
Evanescent operators, which vanish in four dimensions as a result of their $\gamma$-matrix structure, no longer vanish in dimensional regularization where we work in $d$~dimensions.  The definition of the evanescent operators depends on the definition of the $\gamma_5$ matrix in $d$~dimensions~\cite{Herrlich:1994kh}.  We choose to define $\gamma_5$ in naive dimensional regularization,  where it anti-commutes with the other $\gamma$ matrices, $\{ \gamma_5 , \, \gamma_\mu\}=0$.
For the muon decay only one evanescent operator appears in the one-loop diagrams with a four-fermion interaction and a boson connecting the two fermion bilinears.  
It is defined in the chiral basis as~\cite{Dekens:2019ept}
\begin{equation}
\begin{split}
P_R \gamma^\mu \gamma^\nu \gamma^\lambda P_L \otimes P_R \gamma_\mu \gamma_\nu \gamma_\lambda P_L 
&= 4 (4 - \epsilon ) P_R \gamma^\mu P_L \otimes P_R \gamma_\mu P_L + E_{LL} \, ,
\end{split}
\end{equation}
where $P_R = (1+\gamma_5)/2$ and the $\otimes$ indicates 
a direct product of $\gamma$ matrices (as in Eq.~(\ref{eq:Sdef}) after removing the external spinors).
The scheme choice for the evanescent operators impacts the finite pieces at one-loop when multiplied with $1/\epsilon$ terms. 
The evanescent operator $E_{LL}$ itself can be removed by an appropriate counterterm. 
The renormalised amplitude in the $\alpha_\mu$ scheme to one-loop order then takes the form 
\begin{align}
\label{eq:A_mu_exp}
{\cal A} &= -\frac{2}{\vmu^2}
\left({\cal A}^{(4,0,\mu)} +\vmu^2 {\cal A}^{(6,0,\mu)}+ \frac{1}{\vmu^2}{\cal A}^{(4,1,\mu)} + {\cal A}^{(6,1,\mu)} \right) S_\mu+\dots \nonumber \\
& \overset{!}{=} -\frac{2}{\vmu^2} S_\mu+ \dots \, .
\end{align}
In the second line of Eq.~(\ref{eq:A_mu_exp}) we have indicated that after imposing the renormalisation conditions in the $\alpha_\mu$ 
scheme $G_F$ does not receive any corrections at higher orders. 
Expanding $\vTbare^2$ in Eq.~\eqref{eq:A_mu_bare} using Eq.~(\ref{eq:vT_elim_vmu}) and enforcing the above equality determines the expansion coefficients $\Delta v_\mu^{(i,j,\mu)}$ in Eq.~(\ref{eq:vT_elim_vmu}).
The tree-level results are
\begin{align}
 {\cal A}^{(4,0,\mu)} & = 1 \, ,  \\
  {\cal A}^{(6,0,\mu)} & = C_{\substack{Hl \\ 11}}^{(3)} + C_{\substack{Hl \\ 22}}^{(3)}
  - C_{\substack{ll \\ 1221}} - \Delta v_\mu^{(6,0,\mu)} \, .
 \end{align}
This implies that 
\begin{align}
\label{eq:vmu60}
\Delta v_\mu^{(6,0,\mu)} = C_{\substack{Hl \\ 11}}^{(3)} + C_{\substack{Hl \\ 22}}^{(3)}
  - C_{\substack{ll \\ 1221}} \,.
\end{align}
At one loop, on the other hand, one finds that 
\begin{align}
\Delta v_\mu^{(4,1,\mu)}  =& \,{\cal A}^{(4,1)}_{\rm bare} + \frac{1}{2} \Delta Z_f^{(4,1,\mu)}
\, ,\\
\Delta v_\mu^{(6,1,\mu)} =& \, {\cal A}^{(6,1)}_{\rm bare}+ \frac{1}{2} \Delta Z_f^{(6,1,\mu)} 
+ \Delta v_\mu^{(6,0,\mu)}\left(\frac{1}{2} \Delta Z_f^{(4,1,\mu)} -2   \Delta v_\mu^{(4,1,\mu)}\  \right) \nonumber \\
&+\delta C_{\substack{Hl \\ 11}}^{(3)} + \delta C_{\substack{Hl \\ 22}}^{(3)}
  - \delta C_{\substack{ll \\ 1221}} \,.
\end{align}
In the above, the $\delta C$ are given in Eq.~\eqref{eq:RenCi} and we have defined the combination of on-shell wavefunction renormalisation factors for the external fermions
\begin{align}
\Delta Z_f = \Delta Z_\mu^{L} +\Delta Z_{\nu_\mu}^{L*} +  
\Delta Z_e^{L*} +\Delta Z_{\nu_e}^{L}  \,  ,
\end{align}
where the superscript $L$ has been used to indicate left-handed fermions and the $\Delta Z_f$ are expanded as usual 
\begin{align}
\Delta Z_f = \frac{1}{\vmu^2} \Delta Z_f^{(4,1,\mu)}+  \Delta Z_f^{(6,1,\mu)} \,.
\end{align}
At one loop, $\Delta Z_f$ receives contributions from photon graphs, which vanish, and
heavy-particle graphs ($Z$ and $W$ exchanges), which give   finite 
contributions that must be taken into account.
The explicit results for the one-loop coefficients in Eq.~(\ref{eq:vT_elim_vmu})
are relatively compact, and we list them here for convenience.  In the SM, 
one has
\begin{align}
\label{eq:dVmu_SM_result}
16\pi^2 \, \Delta v_\mu^{(4,1,\mu)}&  =  -\frac{\MHsq}{2} - \MWsq- \frac{\MZsq}{2}+ N_c \MTsq
+\frac{3 \MWsq }{\MHsq-\MWsq} A_0(\MHsq)
-2 N_c A_0(\MTsq) \nonumber \\
& +\left(9  - \frac{3 \MHsq }{\MHsq- \MWsq}   \right) A_0(\MWsq)+ 3 A_0(\MZsq)
+3\frac{c_w^2}{s_w^2} \left[ A_0(\MWsq) -A_0(\MZsq) \right] \nonumber \\
&+16\pi^2 \Delta v_{\mu,{\rm tad}}^{(4,1,\mu)} \, ,
\end{align}
where the tadpole contribution in unitary gauge is 
\begin{align}
16\pi^2 \MHsq \Delta v_{\mu,{\rm tad}}^{(4,1,\mu)} & = 8 M_W^4 + 4M_Z^4
-3 \MHsq A_0(\MHsq)+ 8 N_c  \MTsq A_0(\MTsq)\nonumber \\
& -12 \MWsq A_0(\MWsq)
-6\MZsq A_0(\MZsq) \,
\end{align}
and
\begin{align}
A_0(M^2) =  M^2 \left(\frac{1}{\epsilon} + 1 + \ln \frac{\mu^2}{M^2}\right) \,.
\end{align}
In SMEFT we find  
\begin{align}
\label{eq:dVmu_SMEFT}
16 \pi^2 \Delta v_\mu^{(6,1,\mu)} &= \frac{1}{\epsilon}
\bigg[ \MWsq \Bigg( \frac{2}{3}C_{H\Box}-\frac{28}{3} C_{\substack{Hl \\ 11}}^{(3)} 
-\frac{28}{3} C_{\substack{Hl \\ 22}}^{(3)} +
\frac{8}{3} C_{\substack{Hl \\ 33}}^{(3)} +
8 C_{\substack{Hq\\ 11}}^{(3)} +
8 C_{\substack{Hq\\ 22}}^{(3)} +
8 C_{\substack{Hq\\ 33}}^{(3)} \nonumber \\
&+ 12 \left( C_{\substack{ll \\ 1122}} - C_{\substack{ll \\ 1221}} \right) \Bigg)
 - 6 \MZsq C_{\substack{ll \\ 1221}}+
6 \MTsq \left(C_{\substack{Hl \\ 11}}^{(3)}
+ C_{\substack{Hl \\ 22}}^{(3)}   
-  C^{(3)}_{\substack{lq \\ 1133}}-  C^{(3)}_{\substack{lq \\ 2233}} \right)
\bigg]
 \nonumber \\ &
+16\pi^2 \Delta v_\mu^{(4,1, \mu)}\left( -2\Delta v_\mu^{(6,0,\mu)}+\frac{C_{HD} }{2}\right)   \nonumber \\ &
+ \MHsq\left(-C_{H\Box}+\frac{C_{HD}}{2}\right) 
 +5 \MZsq  C_{\substack{ll \\ 1221}}
\nonumber \\ &
+\MWsq\left(-C_{H\Box}-\frac{3 C_{HD}}{2}-12\frac{s_w }{c_w}C_{HWB}
+10 C_{\substack{Hl \\ 11}}^{(3)} + 10 C_{\substack{Hl \\ 22}}^{(3)} 
+ 10   \left( C_{\substack{ll \\ 1122}} - C_{\substack{ll \\ 1221}} \right)  
\right)
\nonumber \\ &
+3\MTsq\left(-\frac{ C_{HD}}{2}
+ C_{\substack{Hl \\ 11}}^{(3)} +  C_{\substack{Hl \\ 22}}^{(3)}  
+2  C_{\substack{Hq\\ 33}}^{(3)} 
- C^{(3)}_{\substack{lq \\ 1133}}-  C^{(3)}_{\substack{lq \\ 2233}} \right)
\nonumber \\
&+6 M_W^2\frac{A_0(\MHsq)-A_0(\MWsq)}{\MHsq-\MWsq}\left( C_{H\Box}-\frac{C_{HD}}{2}\right) 
\nonumber \\ &
+6 A_0(\MWsq)\left( C_{\substack{Hl \\ 11}}^{(1)}  +C_{\substack{Hl \\ 22}}^{(1)} 
+C_{\substack{Hl \\ 11}}^{(3)}  +C_{\substack{Hl \\ 22}}^{(3)} +  
2 C_{\substack{ll \\ 1122}}  \right) 
\nonumber \\ &
+6 c_w^2 A_0(\MZsq) \left( -C_{HD}-C_{\substack{Hl \\ 11}}^{(1)}  -C_{\substack{Hl \\ 22}}^{(1)} 
+C_{\substack{Hl \\ 11}}^{(3)}  +C_{\substack{Hl \\ 22}}^{(3)} +  
\left(-2+\frac{1}{c_w^2}\right) C_{\substack{ll \\ 1221}}  \right) 
\nonumber \\ &
+A_0(\MTsq)\left(3 C_{HD}- 6 C_{\substack{Hl \\ 11}}^{(3)}  -6C_{\substack{Hl \\ 22}}^{(3)}  -12C_{\substack{Hq \\ 33}}^{(3)} 
+6 C^{(3)}_{\substack{lq \\ 1133}}+ 6 C^{(3)}_{\substack{lq \\ 2233}}\right)
\nonumber \\
&+16\pi^2 \Delta v_{\mu,{\rm tad}}^{(6,1,\mu)} \, ,
\end{align}
where the tadpole contribution in unitary gauge is 
\begin{align}
16\pi^2 \MHsq  \Delta v_{\mu,{\rm tad}}^{(6,1,\mu)} & = 
+32\pi^2 \MHsq\Delta v_{\mu,{\rm tad}}^{(4,1, \mu)} C_{H\Box}
-8 M_W^4 \left(C_{HD} - 2 C_{HW}\right)
 \nonumber \\ & -8 \MWsq \MZsq
 \left(C_{HB}-C_{HW}\right)
 +2 M_Z^4\left(4 C_{HB}-C_{HD}+4 s_w c_w C_{HWB}\right)
 \nonumber \\ &
 - \MHsq A_0(\MHsq)\left( 4C_{H\Box}-4 C_{HD}-6 \frac{\vmu^2}{\MHsq}C_H \right)
 \nonumber \\ &
 +12 \MWsq A_0(\MWsq)\left(C_{HD}- 2 C_{HW} \right)-12 \MTsq A_0(\MTsq)\left(2 C_{HD}+
 \frac{\sqrt{2}\vmu}{m_t}  C_{\substack{uH \\ 33}}
\right)
  \nonumber \\ &
  -\MZsq A_0(\MZsq)\left(12 s_w^2 C_{HB} -3C_{HD}+12 c_w^2 C_{HW}+
  12 c_w s_w C_{HWB} \right)\,.
\end{align}
Note that the expansion coefficients are only gauge invariant when tadpoles
are included --  the split that we have given above is unique to unitary gauge.

\section{Numerical results for the decay rates}
\label{sec:NumRes}
Here we present numerical results for the decay rates considered in Section~\ref{sec:HeavyBosonDecays} in the three schemes. 
We use the notation 
\begin{align}
 \Gamma^{s}_{X,\text{LO}} & \equiv \Gamma^{s (4,0)}_{Xf_1 f_2} + \Gamma^{s (6,0)}_{Xf_1 f_2} \, , \nonumber \\
 \Gamma^{s}_{X,\text{NLO}}  & \equiv \Gamma^{s}_{X,\text{LO}} + \Gamma^{s (4,1)}_{Xf_1 f_2} + \Gamma^{s(6,1)}_{Xf_1 f_2} \, ,
\end{align}
where the quantities appearing on the right-hand side are defined in Eq.~(\ref{eq:Gamma_def_pieces}). Scale uncertainties
are obtained as explained in Section~\ref{sec:Derived}.
For brevity, we show only those coefficients which have an absolute numerical prefactor or absolute difference between the upper and lower scale uncertainties of greater than $1\%$ of the LO SM result after factoring out the appropriate~$v_\sigma^2$; results omitted for this reason are indicated by $\dots$ in the equations that follow.

 \subsection{$W \to \tau \nu$ decay}
 \label{app:Wdecay}
For $W$ decay in the $\alpha$-scheme we find
  \begin{align}
 		\Gamma^{\alpha}_{W,\text{LO}} =&  \,
 		234.6^{+1.8}_{-1.8} \text{ MeV} + v_\alpha^2 \Gamma^{\alpha(4, 0)}_{W \tau \nu_\tau} \bigg\{ 
 		3.733^{+0.132}_{-0.132}C_{HWB} 
 		+ 2.000^{+0.034}_{-0.034}C^{(3)}_{\substack{Hl\\33}} 
 		+ 1.742^{+0.120}_{-0.120}C_{HD}  
 		\nonumber \\	
 		& 
 		+0.000^{+0.189}_{-0.189}C^{(1)}_{\substack{Hq\\33}}
 		+0.000^{+0.182}_{-0.182}C_{\substack{Hu\\33}}
 		+0.000^{+0.066}_{-0.066}C_{\substack{uB\\33}}
 		+0.000^{+0.059}_{-0.059}C_{\substack{uW\\33}}  
 		\nonumber  \\
 		&
 		+0.000^{+0.046}_{-0.046}C^{(3)}_{\substack{lq\\3333}}
 		+0.000^{+0.008}_{-0.008}\bigg(C_{HB}+C_{HW}+C_{W}+
 		\sum_{i=1,2,3} C^{(3)}_{\substack{Hq\\ii}} 
 		+ \sum_{j=1,2} C^{(3)}_{\substack{lq\\33jj}}\bigg)
 		 \nonumber  \\ 
 		 &
 		+0.000^{+0.007}_{-0.007}C_{H\Box}
 		+0.000^{+0.005}_{-0.005}  \sum_{j=1,2} C_{\substack{Hu\\jj}} +\ldots \bigg\} \, ,	
\end{align}
 \begin{align}
 		\Gamma^{\alpha}_{W,\text{NLO}} = & \, 224.6^{+0.1}_{-0.2} \text{ MeV} + v_\alpha^2 \Gamma^{\alpha(4, 0)}_{W \tau \nu_\tau} \bigg\{ 
 		3.620^{+0.000}_{-0.011}C_{HWB} 
 		+2.043^{+0.000}_{-0.002}C^{(3)}_{\substack{Hl\\33}} 
 		\nonumber \\ &
 		+1.713^{+0.000}_{-0.011}C_{HD} 
 		-0.195^{+0.038}_{-0.000}C_{\substack{Hu\\33}} 
 		+0.172^{+0.000}_{-0.033}C^{(1)}_{\substack{Hq\\33}} 
 		-0.079^{+0.018}_{-0.002}C^{(3)}_{\substack{Hq\\33}} 
 		\nonumber\\ &
 		-0.072^{+0.008}_{-0.000}C_{\substack{uB\\33}}
 		-0.034^{+0.002}_{-0.000}C^{(3)}_{\substack{lq\\3333}}
 		-0.032^{+0.005}_{-0.000}C_{\substack{uW\\33}}
 		-0.011^{+0.000}_{-0.000}C_W 
 		\nonumber\\ &
 		+0.000^{+0.001}_{-0.026}C_{\substack{uu\\3333}}
 		+0.000^{+0.000}_{-0.023}C^{(1)}_{\substack{qq\\3333}}
 		+0.000^{+0.020}_{-0.000}C^{(1)}_{\substack{qu\\3333}}
 		\nonumber\\ &
 		+0.000^{+0.003}_{-0.012}C^{(3)}_{\substack{qq\\3333}}
 		+\ldots \bigg\} \, ,
 		\label{eq:W_NLO_alpha_exact}	
 \end{align} 
For the $\alpha_\mu$-scheme we obtain
 \begin{align}	
 		\Gamma^{\alpha_\mu}_{W,\text{LO}} &= 227.2^{+0.0}_{-0.0} \text{ MeV} + v_\mu^2 \Gamma^{\mu(4,0)}_{W \tau \nu_\tau} \bigg\{
 		2.000^{+0.031}_{-0.031}C^{(3)}_{\substack{Hl\\33}}
 		-1.000^{+0.015}_{-0.015} \sum_{j=1,2} C^{(3)}_{\substack{Hl\\jj}}
 		\nonumber \\ &
 		+1.000^{+0.004}_{-0.004}C_{\substack{ll\\1221}} 	
 		+0.000^{+0.044}_{-0.044}C^{(3)}_{\substack{lq\\3333}}
 		+0.000^{+0.026}_{-0.026} \sum_{j=1,2} C^{(3)}_{\substack{lq\\jj33}}
 		+0.000^{+0.011}_{-0.011}C_{\substack{ll\\1122}} 
 		\nonumber\\
 		&
 		+0.000^{+0.007}_{-0.007} \sum_{j=1,2} C^{(3)}_{\substack{lq\\33jj}}
 		+\ldots \bigg\} \, ,	
 \end{align}
 \begin{align}
 		\Gamma^{\alpha_\mu}_{W,\text{NLO}} =& 226.5^{+0.0}_{-0.0} \text{ MeV} + v_\mu^2 \Gamma^{\mu(4, 0)}_{W \tau \nu_\tau} \bigg\{ 2.043^{+0.000}_{-0.001}C^{(3)}_{\substack{Hl\\33}} 
 		-1.025^{+0.001}_{-0.000} \sum_{j=1,2} C^{(3)}_{\substack{Hl\\jj}}
 		\nonumber \\ &
 		+0.998^{+0.000}_{-0.000}C_{\substack{ll\\1221}} 
 		 -0.033^{+0.001}_{-0.000}C^{(3)}_{\substack{lq\\3333}}+0.019^{+0.000}_{-0.001} \sum_{j=1,2} C^{(3)}_{\substack{lq\\jj33}}
 		 \nonumber \\ &
 		 -0.015^{+0.000}_{-0.000}C_{\substack{ll\\1122}} 
 		+0.010^{+0.000}_{-0.000}C_{HWB} +\ldots \bigg\} \, .	
 	\label{eq:W_NLO_mu_exact}
 \end{align}
And finally for the LEP scheme, we find
\begin{align}
		\Gamma^\text{LEP}_{W,\text{LO}} =& 222.7^{+1.1}_{-1.1}\text{ MeV} +v_\mu^2 \Gamma^{\text{LEP}(4, 0)}_{W \tau \nu_\tau} \bigg\{
		-2.379^{+0.102}_{-0.102} C_{HWB}
		+2.000^{+0.019}_{-0.019} C^{(3)}_{\substack{Hl\\33}}
		\nonumber \\&
		-1.656^{+0.032}_{-0.032} \sum_{j=1,2} C^{(3)}_{\substack{Hl\\jj}}
		+1.656^{+0.001}_{-0.001} C_{\substack{ll\\1221}}
		-1.078^{+0.073}_{-0.073} C_{HD} 
		+0.000^{+0.114}_{-0.114} C^{(1)}_{\substack{Hq\\33}}
		\nonumber \\ &
		+0.000^{+0.109}_{-0.109} C_{\substack{Hu\\33}}
		+0.000^{+0.045}_{-0.045} C^{(3)}_{\substack{lq\\3333}} 
		+0.000^{+0.043}_{-0.043} \sum_{j=1,2} C^{(3)}_{\substack{lq\\jj33}} 
		+0.000^{+0.040}_{-0.040} C_{\substack{uB\\33}}
	\nonumber \\ &
		+0.000^{+0.037}_{-0.037} C_{\substack{uW\\33}} 
		+0.000^{+0.018}_{-0.018} C_{\substack{ll\\1122}}
		+0.000^{+0.007}_{-0.007} \sum_{j=1,2} C^{(3)}_{\substack{lq\\33jj}}
		+\ldots
		\bigg\} \, ,	
\end{align}
\begin{align}
		\Gamma^\text{LEP}_{W,\text{NLO}} =& 227.2^{+0.0}_{-0.0} \text{ MeV}+ v_\mu^2 \Gamma^{\text{LEP}(4, 0)}_{W \tau \nu_\tau} \bigg\{ 
		-2.455^{+0.008}_{-0.000} C_{HWB}
		+ 2.091^{+0.001}_{-0.001} C^{(3)}_{\substack{Hl\\33}} 
		\nonumber \\&
		-1.742^{+0.002}_{-0.000} \sum_{j=1,2} C^{(3)}_{\substack{Hl\\jj}} 
		+1.697^{+0.000}_{-0.001} C_{\substack{ll\\1221}} 
		-1.165^{+0.012}_{-0.001} C_{HD}
		+0.116^{+0.002}_{-0.031} C_{\substack{Hu\\33}}
		\nonumber \\ &
		-0.103^{+0.029}_{-0.002} C^{(1)}_{\substack{Hq\\33}}
		-0.033^{+0.002}_{-0.002} C^{(3)}_{\substack{lq\\3333}}
		+0.046^{+0.001}_{-0.010} C^{(3)}_{\substack{Hq\\33}}
		+0.044^{+0.000}_{-0.008} C_{\substack{uB\\33}} 
		\nonumber\\
		&
		-0.024^{+0.001}_{-0.000} C_{\substack{ll\\1122}}
		+0.019^{+0.000}_{-0.006} C_{\substack{uW\\33}}
		+0.032^{+0.001}_{-0.003} \sum_{j=1,2} C^{(3)}_{\substack{lq\\jj33}}
		\nonumber\\
		&
		+0.000^{+0.015}_{-0.001} C_{\substack{uu\\3333}}
		+0.000^{+0.014}_{-0.000} C^{(1)}_{\substack{qq\\3333}}
		+0.000^{+0.000}_{-0.011} C^{(1)}_{\substack{qu\\3333}}
		+\ldots \bigg\} \, .
\end{align}

 \subsection{$h \to b \bar{b}$ decay}
 \label{sec:hbb_num}
 
 To evaluate scale uncertainties for $h\to b\bar{b}$ we also require the running of $m_b(\mu)$ and $\alpha_s(\mu)$. 
 As with the running of $\alpha(\mu)$, we again use a one-loop fixed-order solution to the RG equations for $m_b(\mu)$ and 
 $\alpha_s(\mu)$ which are given by 
\begin{align}
m_b(\mu) &= m_b(M_h)\bigg[1+\gamma_b(M_h)\ln \left(\frac{\mu}{M_h}\right)\bigg] \, , \\
\alpha_s(\mu) &= \alpha_s(M_h)\bigg[1 - \frac{\alpha_s(M_h)}{2\pi}\beta_0\ln\bigg(\frac{\mu}{M_h}\bigg)\bigg] \, ,
\end{align}
where
\begin{equation}
\gamma_b(\mu) = -\frac{3}{2\pi}\Big[\alpha_s(\mu)C_F + \alpha(\mu) Q_b^2\Big] \, , 
\qquad
\beta_0 = \frac{11}{3}C_A - \frac{4}{3}T_F \, n_f \, ,
\end{equation}
with
\begin{align*}
C_F = \frac{4}{3} \, , \quad C_A = 3 \, , \quad T_F = \frac{1}{2} \, , \quad {\rm and } \, \quad  n_f=5 \, .
\end{align*}
\newline
In the $\alpha$ scheme we find 
\begin{align}
 		\Gamma_{h bb, \text{LO}}^{\alpha} = &\, 2.300^{+0.209}_{-0.209} \text{ MeV}  + 
 		\valpha^2 \Gamma_{h bb}^{\alpha(4,0)}\bigg\{
 		-1.414^{+0.099}_{-0.099} \frac{\valpha}{m_b}C_{\substack{dH \\ 33}}
 		+3.733^{+0.243}_{-0.243} C_{HWB}
 		\nonumber \\ &
 		+2.000^{+0.084}_{-0.084} C_{H\Box}
 		+1.242^{+0.034}_{-0.034} C_{HD}
 		+ 0.000^{+0.078}_{-0.078} \frac{\valpha}{m_b}  C^{(1)}_{\substack{quqd \\ 3333}}
 		+ 0.000^{+0.067}_{-0.067} \frac{\valpha}{m_b}  C_{\substack{dW\\ 33}}
 		\nonumber \\ &
 		+ 0.000^{+0.015}_{-0.015} \frac{\valpha}{m_b}  C^{(8)}_{\substack{quqd \\ 3333}}
 		+ 0.000^{+0.008}_{-0.008} \frac{\valpha}{m_b}  C_{\substack{Hud \\ 33}}
 		+ 0.000^{+0.397}_{-0.397} C_{HG}
 		+ 0.000^{+0.213}_{-0.213} C_{\substack{dB\\ 33}}
 		\nonumber \\ &
 		+0.000^{+0.189}_{-0.189} C_{\substack{Hq \\ 33}}^{(1)}
 		+0.000^{+0.183}_{-0.183} C_{\substack{Hu \\ 33}} 
 		+0.000^{+0.112}_{-0.112} C_{\substack{Hq \\ 33}}^{(3)}
 		+0.000^{+0.094}_{-0.094} C_{\substack{uW \\ 33}}
 		\nonumber \\ &
 		+0.000^{+0.066}_{-0.066} C_{\substack{uB \\ 33}} 
 		+0.000^{+0.027}_{-0.027} C_{HW} 
 		+0.000^{+0.027}_{-0.027}  C_{\substack{uH \\ 33}}
 		+0.000^{+0.013}_{-0.013}  C^{(8)}_{\substack{qd \\ 3333}}
 		\nonumber \\ &
 		+0.000^{+0.011}_{-0.011}  C_{\substack{Hd \\ 33}}
 		+0.000^{+0.009}_{-0.009}  C^{(1)}_{\substack{qd \\ 3333}}
 		+0.000^{+0.008}_{-0.008} \sum_{j=1,2} C_{\substack{Hq \\ jj}}^{(3)}
 		+0.000^{+0.008}_{-0.008} C_W
 		\nonumber \\ &
 		+0.000^{+0.006}_{-0.006} C_{HB}
 		+0.000^{+0.005}_{-0.005} \sum_{j=1,2} C_{\substack{Hu \\ jj}} 
 		+\ldots
 		\bigg\} \, ,
\label{eq:hbb_vhat_LO}
\end{align}
 \begin{align}
 		\Gamma_{h bb, \text{NLO}}^{\alpha} = & 2.647^{+0.036}_{-0.119} \text{ MeV}  + 
 		\valpha^2 \Gamma_{h bb}^{\alpha(4,0)}\bigg\{
 		-1.761^{+0.072}_{-0.030} \frac{\valpha}{m_b}C_{\substack{dH \\ 33}}
 		-0.060^{+0.012}_{-0.020} \frac{\valpha}{m_b}  C_{\substack{dW\\ 33}}
 		\nonumber \\&
 		+ 4.239^{+0.055}_{-0.159} C_{HWB}
 		+3.094^{+0.704}_{-0.953} C_{HG}
 		+0.031^{+0.031}_{-0.000} \frac{\valpha}{m_b}  C^{(1)}_{\substack{quqd \\ 3333}}
 		\nonumber\\ &
 		+2.448^{+0.031}_{-0.083} C_{H\Box}
 		+1.358^{+0.013}_{-0.042} C_{HD}
 		+0.009^{+0.003}_{-0.000} \frac{\valpha}{g_s m_b} C_{\substack{dG \\ 33}} 
 		\nonumber\\ &
 		+0.006^{+0.005}_{-0.001} \frac{\valpha}{m_b}  C^{(8)}_{\substack{quqd \\ 3333}}
 		-0.004^{+0.003}_{-0.001} \frac{\valpha}{m_b}  C_{\substack{Hud \\ 33}}
 		-0.116^{+0.014}_{-0.024} C_{\substack{Hq \\ 33}}^{(3)}
 		\nonumber\\ &
 		-0.079^{+0.012}_{-0.032} C_{\substack{Hu \\ 33}} 
 		+0.058^{+0.035}_{-0.013} C_{\substack{Hq \\ 33}}^{(1)}
 		-0.040^{+0.004}_{-0.025} C_{\substack{uW \\ 33}}
 		-0.031^{+0.005}_{-0.011} C_{\substack{uH \\ 33}} 
 		\nonumber\\ &
 		-0.030^{+0.001}_{-0.015} C_{\substack{uB \\ 33}} 
 		+0.028^{+0.007}_{-0.013} C_{HW} 
 		+0.024^{+0.000}_{-0.000} C_{H} 
 		-0.014^{+0.010}_{-0.000} C^{(8)}_{\substack{qd \\ 3333}}
 		\nonumber\\ &
 		-0.011^{+0.022}_{-0.079}  C_{\substack{dB\\ 33}} 
 		-0.010^{+0.007}_{-0.001}  C^{(1)}_{\substack{qd \\ 3333}}
 		+0.000^{+0.072}_{-0.049}  \frac{1}{g_s} C_{\substack{uG \\ 33}}
 		\nonumber\\ &
 		+  0.000^{+0.000}_{-0.020} C^{(3)}_{\substack{qq \\ 3333}}
 		+  0.000^{+0.000}_{-0.018} C_{\substack{uu \\ 3333}}
 		+  0.000^{+0.016}_{-0.000} C^{(1)}_{\substack{qu \\ 3333}}
 		+  0.000^{+0.000}_{-0.016} C^{(1)}_{\substack{qq \\ 3333}}
 		\nonumber \\
 		&+\ldots
 		\bigg\} \, .
 		\label{eq:hbb_vhat}
 \end{align}
Here and in other numerical results for $h \to b \bar{b}$, we have left enhancement factors such as $v_\alpha/m_b$ symbolic, 
with the exception of $C_{\substack{dB \\ 33}} $.
Scale variations of the LO SMEFT results fail to include the NLO results of the operators first appearing at LO in all schemes, where only one operator is within $2\sigma$ region. However, for operators first appearing at NLO the NLO result is typically included in the LO scale-variation band.  More reliable uncertainty estimates can be made by varying the renormalisation scales of the $b$-quark
mass and Wilson coefficients independently as in \cite{Cullen:2019nnr}.

In the $\alpha_\mu$ scheme one finds
\begin{align}
 		\Gamma_{h bb, \text{LO}}^{\alpha_\mu} = & \, 2.217^{+0.221}_{-0.221}\text{ MeV}  +
 		\vmu^2 \Gamma_{h bb}^{\alpha_\mu(4,0)}\bigg\{
 		-1.414^{+0.095}_{-0.095} \frac{\vmu}{m_b}C_{\substack{dH \\ 33}}
 		+2.000^{+0.095}_{-0.095} C_{H\Box}
 		\nonumber\\ &
 		+1.000^{+0.104}_{-0.104} C_{\substack{ll\\ 1221}}
 		-1.000^{+0.086}_{-0.086} \sum_{j=1,2} C_{\substack{Hl \\ jj}}^{(3)} 
 		-0.500^{+0.021}_{-0.021} C_{HD}
 		\nonumber\\&
 		+0.000^{+0.074}_{-0.074} \frac{\vmu}{m_b} C^{(1)}_{\substack{quqd\\ 3333}}
 		+0.000^{+0.063}_{-0.063} \frac{\vmu}{m_b} C_{\substack{dW\\ 33}}
 		+0.000^{+0.014}_{-0.014} \frac{\vmu}{m_b} C^{(8)}_{\substack{quqd\\ 3333}}
 		\nonumber\\ &
 		+0.000^{+0.007}_{-0.007} \frac{\vmu}{m_b} C_{\substack{Hud\\ 33}}
 		+0.000^{+0.397}_{-0.397} C_{HG}
 		+0.000^{+0.206}_{-0.206} C_{\substack{dB\\ 33}}
 		+0.000^{+0.115}_{-0.115} C^{(3)}_{\substack{Hq\\ 33}}
 		\nonumber\\&
 		+0.000^{+0.034}_{-0.034} C_{\substack{uW\\ 33}}	
 		+0.000^{+0.034}_{-0.034} C_{HW}
 		+0.000^{+0.026}_{-0.026} C_{\substack{uH\\ 33}}
 		+0.000^{+0.026}_{-0.026} \sum_{j=1,2} C_{\substack{lq \\ jj33}}^{(3)} 
 		\nonumber\\ &
 		+0.000^{+0.012}_{-0.012} C_{\substack{qd \\ 3333}}^{(8)}
 		+0.000^{+0.011}_{-0.011} C_{\substack{ll \\ 1122}}
 		+0.000^{+0.009}_{-0.009} C_{\substack{qd \\ 3333}}^{(1)}
 		+0.000^{+0.008}_{-0.008} C_{\substack{Hd \\ 33}}
 		+ ...
 		\bigg\} \, ,
\label{eq:hbb_vmu_LO}
\end{align}
 \begin{align}
 		\Gamma_{h bb, \text{NLO}}^{\alpha_\mu} =
 		& \, 2.650^{+0.043}_{-0.129} \text{ MeV} +
 		\vmu^2\Gamma_{h bb}^{\alpha_\mu(4,0)}\bigg\{
 		-1.728^{+0.068}_{-0.029} \frac{\vmu}{m_b} C_{\substack{dH \\ 33}}
 		-0.057^{+0.009}_{-0.018} \frac{\vmu}{m_b} C_{\substack{dW \\ 33}} 
 		\nonumber\\&
 		+3.094^{+0.698}_{-0.946} C_{HG}
 		+2.447^{+0.035}_{-0.084} C_{H\Box}	
 		+0.030^{+0.028}_{-0.000} \frac{\vmu}{m_b} C^{(1)}_{\substack{quqd\\ 3333}} 
 		\nonumber\\&
 		-1.212^{+0.054}_{-0.020}\sum_{j=1,2}C_{\substack{Hl \\ jj}}^{(3)}
 		+1.195^{+0.019}_{-0.060}C_{\substack{ll \\ 1221}} 
 		-0.612^{+0.020}_{-0.008}C_{HD}
 		\nonumber\\ &
 		+0.009^{+0.003}_{-0.000} \frac{\vmu}{g_s m_b} C_{\substack{dG\\ 33}}
 		+0.006^{+0.005}_{-0.001} \frac{\vmu}{m_b} C^{(8)}_{\substack{quqd\\ 3333}}
 		-0.004^{+0.002}_{-0.001} \frac{\vmu}{m_b} C_{\substack{Hud\\ 33}} 
 		\nonumber\\ &
 		-0.046^{+0.001}_{-0.008} C_{\substack{uW\\ 33}} 
 		-0.031^{+0.008}_{-0.038} C_{\substack{Hq\\ 33}}^{(3)} 
 		-0.030^{+0.004}_{-0.010} C_{\substack{uH\\ 33}}
 		\nonumber\\ &
 		+0.028^{+0.006}_{-0.015} C_{HW}
 		+0.024^{+0.000}_{-0.000} C_{H}
 		-0.022^{+0.007}_{-0.002} C_{\substack{ll\\ 1122}}
 		-0.013^{+0.009}_{-0.000} C_{\substack{qd\\ 3333}}^{(8)}
 		\nonumber\\ &
 		-0.011^{+0.016}_{-0.073} C_{\substack{dB\\ 33}}
 		+0.010^{+0.001}_{-0.001} C_{HB}
 		-0.010^{+0.001}_{-0.001} C_{HWB} 
 		\nonumber\\ &
 		+0.003^{+0.010}_{-0.000}\sum_{j=1,2} C_{\substack{lq\\jj33}}^{(3)} 
 		+0.000^{+0.059}_{-0.083} \frac{1}{g_s}C_{\substack{uG\\33}}
 		+0.000^{+0.000}_{-0.010} C_{\substack{qq\\3333}}^{(3)}  
 		+ ...	\bigg\} \, .
\label{eq:hbb_vmu_NLO}
\end{align}
In the LEP scheme one finds
\begin{align}
 		\Gamma_{h bb, \text{LO}}^\text{LEP} = & 2.217^{+0.221}_{-0.221}\text{ MeV} + \vmu^2	\Gamma_{h bb}^{\text{LEP}(4,0)}\bigg\{
 		-1.414^{+0.095}_{-0.095} \frac{\vmu}{m_b}C_{\substack{dH \\ 33}}
 		+2.000^{+0.095}_{-0.095} C_{H\Box}
 		\nonumber\\ &
 		+1.000^{+0.104}_{-0.104} C_{\substack{ll\\ 1221}} 
 		-1.000^{+0.085}_{-0.085} \sum_{j=1,2}C_{\substack{Hl \\ jj}}^{(3)} 
 		-0.500^{+0.020}_{-0.020} C_{HD}
 		\nonumber\\ &
 		+0.000^{+0.074}_{-0.074} \frac{\vmu}{m_b} C^{(1)}_{\substack{quqd\\ 3333}} 
 		+0.000^{+0.062}_{-0.062} \frac{\vmu}{m_b} C_{\substack{dW\\ 33}}
 		+0.000^{+0.014}_{-0.014} \frac{\vmu}{m_b} C^{(8)}_{\substack{quqd\\ 3333}}
 		\nonumber\\ &
 		+0.000^{+0.008}_{-0.008} \frac{\vmu}{m_b} C_{\substack{Hud\\ 33}} 
 		+0.000^{+0.397}_{-0.397} C_{HG}
 		+0.000^{+0.207}_{-0.207} C_{\substack{dB\\ 33}}
 		\nonumber\\ &
 		+0.000^{+0.115}_{-0.115} C^{(3)}_{\substack{Hq\\ 33}}
 		+0.000^{+0.034}_{-0.034} C_{\substack{uW\\ 33}} 	
 		+0.000^{+0.033}_{-0.033} C_{HW}
 		\nonumber\\ &
 		+0.000^{+0.026}_{-0.026} C_{\substack{uH\\ 33}}
 		+0.000^{+0.026}_{-0.026} \sum_{j=1,2} C_{\substack{lq \\ jj33}}^{(3)}  		
 		+0.000^{+0.012}_{-0.012} C_{\substack{qd \\ 3333}}^{(8)}
 		\nonumber\\ &
 		+0.000^{+0.011}_{-0.011} C_{\substack{ll \\ 1122}} 
 		+0.000^{+0.009}_{-0.009} C_{\substack{qd \\ 3333}}^{(1)} 
 		+0.000^{+0.008}_{-0.008} C_{\substack{Hd \\ 33}}
 		+ ...
 		\bigg\} \, ,
\label{eq:hbb_LEP_LO}
\end{align}
\begin{align}
\Gamma_{h bb, \text{NLO}}^\text{LEP} =& 2.650^{+0.049}_{-0.124}\text{ MeV}  +\vmu^2	\Gamma_{h bb}^{\text{LEP}(4,0)}\bigg\{
-1.728^{+0.068}_{-0.029} \frac{\vmu}{m_b} C_{\substack{dH \\ 33}}
-0.056^{+0.009}_{-0.018} \frac{\vmu}{m_b} C_{\substack{dW \\ 33}} 
\nonumber\\ &
+3.094^{+0.698}_{-0.946} C_{HG}
+2.447^{+0.035}_{-0.084} C_{H\Box}	
+0.030^{+0.028}_{-0.000} \frac{\vmu}{m_b} C^{(1)}_{\substack{quqd\\ 3333}} 
\nonumber\\ &
-1.210^{+0.054}_{-0.020} \sum_{j=1,2}C_{\substack{Hl \\ jj}}^{(3)}   
+1.193^{+0.019}_{-0.060} C_{\substack{ll \\ 1221}} 
-0.609^{+0.020}_{-0.008} C_{HD}
\nonumber\\ &
+0.009^{+0.003}_{-0.000} \frac{\vmu}{m_b} C_{\substack{dG\\ 33}}
+0.006^{+0.005}_{-0.001} \frac{\vmu}{m_b} C^{(8)}_{\substack{quqd\\ 3333}} 
-0.003^{+0.002}_{-0.001} \frac{\vmu}{m_b} C_{\substack{Hud\\ 33}}
\nonumber\\ &
-0.045^{+0.001}_{-0.008} C_{\substack{uW\\ 33}} 
-0.031^{+0.008}_{-0.038} C_{\substack{Hq\\ 33}}^{(3)}
-0.030^{+0.004}_{-0.010} C_{\substack{uH\\ 33}} 
\nonumber\\
&
+0.028^{+0.006}_{-0.015} C_{HW}
+0.024^{+0.000}_{-0.000} C_{H}
-0.022^{+0.007}_{-0.002} C_{\substack{ll\\ 1122}}
-0.013^{+0.009}_{-0.000} C_{\substack{qd\\ 3333}}^{(8)} 
\nonumber\\
&
-0.011^{+0.016}_{-0.074} C_{\substack{dB\\ 33}} 
+0.003^{+0.010}_{-0.000} \sum_{j=1,2} C_{\substack{lq\\jj33}}^{(3)}
+0.011^{+0.001}_{-0.001} C_{HB} 
\nonumber\\ &
+0.000^{+0.059}_{-0.083} C_{\substack{uG\\33}} 
+0.000^{+0.000}_{-0.010} C_{\substack{qq\\3333}}^{(3)}  + ...	\bigg\} \, .
\label{eq:hbb_LEP_NLO}
\end{align}

\subsection{$Z \to \tau \tau$ decay}
\label{subsec:Z_NLO}
 
We present results for $Z$-boson decay in the three different schemes, using $\mu=M_Z$ as the central scale.
In the $\alpha$-scheme we find 
\begin{align}		
 		\Gamma^{\alpha}_{Z, \text{LO}} =& \, 86.75^{+0.067}_{-0.067} \text{ MeV} + v_\alpha^2 \Gamma^{\alpha(4,0)}_{Z\tau\tau}\bigg\{
 		4.088^{+0.144}_{-0.144}C_{HWB}
 		+2.190^{+0.056}_{-0.056}C^{(1)}_{\substack{Hl\\33}}
 		\nonumber\\&
 		+2.190^{+0.038}_{-0.038}C^{(3)}_{\substack{Hl\\33}} 
 		-1.764^{+0.051}_{-0.051}C_{\substack{He \\ 33}} 
 		+ 1.573^{+0.109}_{-0.109}C_{HD}  
 		+ 0.000^{+0.172}_{-0.172}C^{(1)}_{\substack{Hq\\33}} 
 		\nonumber\\&
 		+0.000^{+0.163}_{-0.163}C_{\substack{Hu\\33}} 
 		+0.000^{+0.072}_{-0.072}C_{\substack{uB\\33}} 
 		+0.000^{+0.064}_{-0.064}C_{\substack{uW\\33}}
 		+0.000^{+0.060}_{-0.060}C^{(1)}_{\substack{lq\\3333}} 
 		\nonumber\\&
 		+0.000^{+0.057}_{-0.057}C_{\substack{lu\\3333}} 
 		+0.000^{+0.050}_{-0.050}C^{(3)}_{\substack{lq\\3333}} 
 		+0.000^{+0.048}_{-0.048}C_{\substack{qe\\3333}} 
 		+0.000^{+0.046}_{-0.046} C_{\substack{eu\\3333}}
 		\nonumber\\&
 		+0.000^{+0.008}_{-0.008} \bigg( \sum_{j=1,2} C^{(3)}_{\substack{lq\\33jj}} 
 		+ \sum_{i=1,2,3} C^{(3)}_{\substack{Hq\\ii}} + C_{HW} + C_{HB}  \bigg) 
 		+0.000^{+0.008}_{-0.008} C_W
 		\nonumber\\&
 		+0.000^{+0.007}_{-0.007} C_{H\Box}
 		+0.000^{+0.006}_{-0.006} \sum_{j=1,2} C_{\substack{Hu\\jj}} 
 		+ \ldots
 		\bigg\} \, ,	
 	\label{eq:Ztautau_numres_alpha_LO}
 \end{align}
 \begin{align}	
 		\Gamma^{\alpha}_{Z, \text{NLO}} =& \, 83.25^{+0.04}_{-0.06} \text{ MeV}
 		+ v_\alpha^2 \Gamma^{\alpha(4,0)}_{Z\tau\tau} \bigg\{
 		3.867^{+0.003}_{-0.016}C_{HWB}
 		+2.196^{+0.000}_{-0.004}C^{(1)}_{\substack{Hl\\33}}
 		\nonumber\\ &
 		+2.179^{+0.000}_{-0.001}C^{(3)}_{\substack{Hl\\33}} 
 		-1.899^{+0.008}_{-0.002}C_{\substack{He \\ 33}} 
 		+1.406^{+0.002}_{-0.021}C_{HD}  
 		-0.143^{+0.031}_{-0.000}C_{\substack{Hu\\33}}
 		\nonumber\\ &
 		+0.117^{+0.000}_{-0.032}C^{(1)}_{\substack{Hq\\33}}  
 		-0.074^{+0.014}_{-0.001}C^{(3)}_{\substack{Hq\\33}} 
 		-0.065^{+0.007}_{-0.000}C_{\substack{uB\\33}} 
 		-0.054^{+0.001}_{-0.001}C^{(3)}_{\substack{lq\\3333}}
 		\nonumber\\ &
 		-0.051^{+0.004}_{-0.000}C_{\substack{lu\\3333}}  
 		+0.043^{+0.000}_{-0.003}C^{(1)}_{\substack{lq\\3333}} 
 		+0.041^{+0.002}_{-0.006}C_{\substack{eu\\3333}}
 		-0.035^{+0.007}_{-0.002}C_{\substack{qe\\3333}} 
 		\nonumber\\ &
 		-0.016^{+0.008}_{-0.000}C_{\substack{uW\\33}} 
 		-0.011^{+0.001}_{-0.000}C_W
 		-0.010^{+0.000}_{-0.000}\sum_{j=1,2} C^{(3)}_{\substack{lq\\33jj}}  
 		+ 0.000^{+0.000}_{-0.021} C_{\substack{uu\\3333}}
 		\nonumber\\&
 		+ 0.000^{+0.000}_{-0.019} C^{(1)}_{\substack{qq\\3333}}
 		+ 0.000^{+0.016}_{-0.000} C^{(1)}_{\substack{qu\\3333}}
 		+ 0.000^{+0.002}_{-0.010} C^{(3)}_{\substack{qq\\3333}}
 		+ \ldots
 		\bigg\} \, .
 	\label{eq:Ztautau_numres_alpha}
 \end{align}
 %
 In the $\alpha_\mu$-scheme we obtain
 \begin{align}	
 		\Gamma^{\alpha_\mu}_{Z, \text{LO}} =& \, 83.91^{+0.00}_{-0.00} \text{ MeV}
 		+ v_\mu^2 \Gamma^{\mu(4, 0)}_{Z\tau\tau} \bigg\{	
 		2.190^{+0.057}_{-0.057}C^{(1)}_{\substack{Hl\\33}} 
 		+2.190^{+0.034}_{-0.034}C^{(3)}_{\substack{Hl\\33}} 
 		-1.764^{+0.046}_{-0.046}C_{\substack{He\\33}} 
 		\nonumber\\&
 		-1.000^{+0.015}_{-0.015}\sum_{j=1,2} C^{(3)}_{\substack{Hl\\jj}} 
 		+1.000^{+0.004}_{-0.004}C_{\substack{ll\\1221}} 
 		+0.355^{+0.012}_{-0.012}C_{HWB} 
 		-0.169^{+0.011}_{-0.011}C_{HD}
 		\nonumber\\ &
 		+0.000^{+0.058}_{-0.058}C_{\substack{lq\\3333}}^{(1)}
 		+0.000^{+0.055}_{-0.055}C_{\substack{lu\\3333}}
 		+0.000^{+0.049}_{-0.049}C_{\substack{lq\\3333}}^{(3)} 
 		+0.000^{+0.046}_{-0.046}C_{\substack{qe\\3333}}
 		\nonumber\\ &
 		+0.000^{+0.045}_{-0.045}C_{\substack{eu\\3333}}
 		+0.000^{+0.026}_{-0.026}\sum_{j=1,2} C_{\substack{lq\\jj33}}^{(3)}
 		+0.000^{+0.018}_{-0.018}C_{\substack{Hu\\33}}
 		+0.000^{+0.017}_{-0.017}C_{\substack{Hq\\33}}^{(1)}
 		\nonumber\\ &
 		+0.000^{+0.011}_{-0.011}C_{\substack{ll\\1122}}  
 		+0.000^{+0.008}_{-0.008}\sum_{j=1,2} C_{\substack{lq\\33jj}}^{(3)}
 		+0.000^{+0.006}_{-0.006}C_{\substack{uB\\33}}
 		\nonumber \\ &
 		+0.000^{+0.005}_{-0.005}C_{\substack{uW\\33}}
 		+\ldots
 		\bigg\}	\, ,
 	\label{eq:Ztautau_numres_alphaMu_LO}
 \end{align}
 \begin{align}
 		\Gamma^{\alpha_\mu}_{Z, \text{NLO}} =& \, 83.92^{+0.00}_{-0.00} \text{ MeV} + v_\mu^2 \Gamma^{\mu(4, 0)}_{Z\tau\tau} \bigg\{	
 		2.193^{+0.000}_{-0.003}C^{(1)}_{\substack{Hl\\33}} 
 		+2.181^{+0.000}_{-0.001}C^{(3)}_{\substack{Hl\\33}} 
 		-1.897^{+0.006}_{-0.002}C_{\substack{He\\33}} \nonumber\\			
 		& 
 		-1.029^{+0.001}_{-0.000}\sum_{j=1,2} C^{(3)}_{\substack{Hl\\jj}} 
 		+1.006^{+0.000}_{-0.000}C_{\substack{ll\\1221}} 
 		-0.289^{+0.009}_{-0.007}C_{HD} 
 		+0.258^{+0.003}_{-0.004}C_{HWB}
 		\nonumber\\	
 		&
 		-0.053^{+0.001}_{-0.001}C^{(3)}_{\substack{lq\\3333}} 
 		-0.049^{+0.003}_{-0.000}C_{\substack{lu\\3333}} 
 		+0.042^{+0.000}_{-0.002}C^{(1)}_{\substack{lq\\3333}} 
 		+0.040^{+0.002}_{-0.005}C_{\substack{eu\\3333}} 
 		\nonumber\\	
 		& 
 		-0.034^{+0.006}_{-0.002}C_{\substack{qe\\3333}} 
 		-0.020^{+0.016}_{-0.012}C^{(1)}_{\substack{Hq\\33}} 
 		+0.018^{+0.011}_{-0.016}C_{\substack{Hu\\33}} 
 		-0.017^{+0.000}_{-0.000}C_{\substack{ll\\1122}} 
 		\nonumber\\	
 		& 
 		+0.015^{+0.000}_{-0.001}\sum_{j=1,2} C^{(3)}_{\substack{lq\\jj33}}
 		+ ...	
 		\bigg\}	\, ,
 	\label{eq:Ztautau_numres_alphaMu}
 \end{align}
 %
 and in the LEP-scheme we find
 \begin{align}
 		\Gamma^{\text{LEP}}_{Z, \text{LO}} =& \, 
 		83.30^{+0.11}_{-0.11} \text{ MeV} 
 		+ v_\mu^2 \Gamma^{\text{LEP}(4,0)}_{Z\tau\tau} \bigg\{
 		2.121^{+0.035}_{-0.035} C^{(1)}_{\substack{Hl\\33}}
 		+2.121^{+0.012}_{-0.012} C^{(3)}_{\substack{Hl\\33}}
 		\nonumber\\ &
 		-1.863^{+0.069}_{-0.069} C_{\substack{He \\ 33}}
 		+1.173^{+0.031}_{-0.031} C_{\substack{ll \\ 1221}}
 		-1.173^{+0.008}_{-0.008} \sum_{j=1,2} C^{(3)}_{\substack{Hl\\jj}}
 		-0.587^{+0.026}_{-0.026}  C_{HD} 
 		\nonumber\\
 		&
 		-0.410^{+0.046}_{-0.046} C_{HWB} 
 		+0.000^{+0.061}_{-0.061} C^{(1)}_{\substack{Hq \\ 33}}
 		+0.000^{+0.060}_{-0.060} C_{\substack{Hu \\ 33}}
 		+0.000^{+0.056}_{-0.056} C^{(1)}_{\substack{lq \\ 3333}} 
 		\nonumber\\ 
 		&
 		+0.000^{+0.053}_{-0.053} C_{\substack{lu \\ 3333}}
 		+0.000^{+0.049}_{-0.049} C_{\substack{qe \\ 3333}}
 		+0.000^{+0.047}_{-0.047} C^{(3)}_{\substack{lq \\ 3333}}
 		+0.000^{+0.047}_{-0.047} C_{\substack{eu \\ 3333}} 
 		\nonumber\\
 		&
 		+0.000^{+0.030}_{-0.030} \sum_{j=1,2} C^{(3)}_{\substack{lq \\ jj33}} 
 		+0.000^{+0.013}_{-0.013} C_{\substack{ll \\ 1122}}
 		+0.000^{+0.008}_{-0.008} \sum_{j=1,2} C^{(3)}_{\substack{lq \\ 33jj}}  
 		\nonumber\\
 		&
 		+0.000^{+0.007}_{-0.007} C_{\substack{uB \\ 33}}
 		+0.000^{+0.006}_{-0.006} C_{\substack{uW \\ 33}}
 		+ ... \bigg\} \, ,
 	\label{eq:Ztautau_numres_LEP_LO}
 \end{align}
 \begin{align}
 		\Gamma^{\text{LEP}}_{Z, \text{NLO}} =& \, 84.12^{+0.00}_{-0.03} \text{ MeV}
 		+ v_\mu^2 \Gamma^{\text{LEP}(4,0)}_{Z\tau\tau} \bigg\{
 		2.219^{+0.003}_{-0.004} C^{(1)}_{\substack{Hl\\33}} 
 		+2.210^{+0.002}_{-0.001} C^{(3)}_{\substack{Hl\\33}}
 		\nonumber\\ &
 		-1.901^{+0.005}_{-0.000} C_{\substack{He \\ 33}}
 		-1.254^{+0.000}_{-0.004} \sum_{j=1,2} C^{(3)}_{\substack{Hl\\jj}}
 		+1.227^{+0.002}_{-0.000} C_{\substack{ll \\ 1221}}
 		-0.633^{+0.004}_{-0.003} C_{HD} 
 		\nonumber\\ 
 		&
 		-0.481^{+0.000}_{-0.012} C_{HWB} 
 		+0.055^{+0.002}_{-0.013} C_{\substack{Hu\\33}}
 		-0.052^{+0.011}_{-0.002} C^{(1)}_{\substack{Hq\\33}}
 		-0.051^{+0.000}_{-0.002} C^{(3)}_{\substack{lq\\3333}} 
 		\nonumber\\
 		&
 		-0.048^{+0.004}_{-0.002} C_{\substack{lu\\3333}}
 		+0.042^{+0.000}_{-0.004} C_{\substack{eu\\3333}}
 		+0.041^{+0.002}_{-0.003} C^{(1)}_{\substack{lq\\3333}}
 		-0.036^{+0.005}_{-0.000} C_{\substack{qe\\3333}}
 		\nonumber\\
 		&
 		+0.025^{+0.000}_{-0.005} C^{(3)}_{\substack{Hq\\33}}
 		-0.020^{+0.002}_{-0.000} C_{\substack{ll\\1122}}
 		+0.017^{+0.003}_{-0.001} \sum_{j=1,2} C^{(3)}_{\substack{lq\\jj33}}  
 		+ ...  \bigg\} \, .
 	\label{eq:Ztautau_numres_LEP}
 \end{align}

\section{Numerical results using universal corrections in SMEFT} 
\label{app:universal_corrections}

For completeness, we present here numerical results for the prefactors of the Wilson coefficients at different perturbative orders for $W$, $h$ and $Z$~decay which have not been shown in Section~\ref{sec:universal_corrections} yet. 
Table~\ref{tab:resummingAlphaMuScheme} shows the results for $W$ and $h$~decay in the $\alpha$~scheme.  For the LEP scheme, Table~\ref{tab:resummingLEPScheme} shows the results for $W$, and $Z$~decay, respectively. The $h$~decay results for the LEP scheme have been omitted since they only have very small (numerical) differences with respect to the numbers in the $\alpha_\mu$~scheme, which are presented in Table~\ref{tab:resummingAlphaMuScheme}.

For results in the $\alpha$~scheme and $Z$ decay in the $\alpha_\mu$ scheme, we refer to Tables~\ref{tab:resummingAlphaScheme} and \ref{tab:resummingZdecay_vmu} in Section~\ref{sec:universal_corrections}.

\setlength{\tabcolsep}{2pt}
\renewcommand{\arraystretch}{1.5}
\begin{table}[t!]
\centering
\small
\begin{tabular}{c||c|c|c}
$W\to \tau \nu$	&$C^{(3)}_{\substack{Hl\\jj}}$& $C^{(3)}_{\substack{lq\\jj33}}$ & $C_{\substack{ll\\1221}}$  \\[3mm]
	\hline \hline
	NLO
	&$-1.025^{+0.001}_{-0.000}$&$0.019^{+0.000}_{-0.001}$ &$0.998^{+0.000}_{-0.000}$		\\	
	$\text{NLO}_{t}$
	&$-1.019^{+0.001}_{-0.000}$&$0.019^{+0.004}_{-0.005}$  &$1.000^{+0.002}_{-0.002}$	\\
	\hline
	LO&$-1.000^{+0.015}_{-0.015}$&$0.000^{+0.026}_{-0.026}$ &$1.000^{+0.004}_{-0.004}$ \\
$\text{LO}_K$&$-1.019^{+0.011}_{-0.010}$&$0.019^{+0.000}_{-0.001}$  &$1.000^{+0.004}_{-0.004}$\\
\end{tabular}
\qquad
\begin{tabular}{c||c|c|c}
$h\to b \bar{b}$	&$C^{(3)}_{\substack{Hl\\jj}}$& $C^{(3)}_{\substack{lq\\jj33}}$ & $C_{\substack{ll\\1221}}$     \\[3mm]
	\hline \hline
	NLO
	&$-1.009^{+0.001}_{-0.000}$&$0.003^{+0.000}_{-0.000}$ & $0.992^{+0.000}_{-0.000}$ \\	
	$\text{NLO}_{t}$
	&$-1.009^{+0.002}_{-0.001}$&$0.003^{+0.003}_{-0.005}$ & $1.006^{+0.002}_{-0.002}$ 	\\
	\hline
	LO&$-1.000^{+0.014}_{-0.014}$&$0.000^{+0.026}_{-0.026}$  & $1.000^{+0.005}_{-0.005}$ 	\\
$\text{LO}_K$&$-1.003^{+0.013}_{-0.012}$ &$0.003^{+0.000}_{-0.001}$ & $1.000^{+0.005}_{-0.005}$\\
\end{tabular}

\caption{The numerical prefactors of the Wilson coefficients in the $\alpha_\mu$~scheme appearing in $K_W^{(6,1,\mu)}$ for various perturbative approximations. The tree-level decay rate as well as $v_\mu^2$ have been factored out and the results have been evaluated at the scale of the process. We show the results for $W$~decay (left) and Higgs decay (right).}
\label{tab:resummingAlphaMuScheme}
\end{table}
\renewcommand{\arraystretch}{1}


\setlength{\tabcolsep}{2pt}
\renewcommand{\arraystretch}{1.5}
\begin{table}[t!]
\centering
\small
\begin{tabular}{c||c|c|c|c|c|c|c}
$W\to \tau \nu$	&$C_{HD}$ & $C_{HWB}$ & $C^{(3)}_{\substack{Hq\\33}}$ & $C_{\substack{Hu\\33}}$  & $C^{(1)}_{\substack{Hq\\33}}$ & $C_{\substack{uB\\33}}$& $C_{\substack{uW\\33}}$  \\[3mm]
	\hline \hline
	NLO
	&$-1.165^{+0.012}_{-0.001}$&$-2.455^{+0.008}_{-0.000}$&$0.046^{+0.001}_{-0.010}$	&$0.116^{+0.002}_{-0.031}$	&$-0.103^{+0.029}_{-0.002}$	&$0.044^{+0.000}_{-0.008}$	&$0.019^{+0.000}_{-0.006}$	\\
	
	$\text{NLO}_{t}$
	&$-1.143^{+0.009}_{-0.002}$&$-2.434^{+0.024}_{-0.019}$&$0.040^{+0.002}_{-0.011}$	&$0.124^{+0.002}_{-0.028}$	&$-0.124^{+0.026}_{-0.002}$	&$0.045^{+0.000}_{-0.005}$	&$0.023^{+0.000}_{-0.004}$	\\
	\hline
	LO&$-1.078^{+0.073}_{-0.073}$&$-2.379^{+0.102}_{-0.102}$&$0.000^{+0.005}_{-0.005}$	&$0.000^{+0.109}_{-0.109}$	&$0.000^{+0.114}_{-0.114}$	&$0.000^{+0.040}_{-0.040}$	&$0.000^{+0.037}_{-0.037}$	\\

	$\text{LO}_{K}$ 
	&$-1.143^{+0.027}_{-0.018}$&$-2.434^{+0.045}_{-0.039}$&$0.040^{+0.000}_{-0.005}$	&$0.124^{+0.000}_{-0.025}$	&$-0.124^{+0.027}_{-0.004}$	&$0.045^{+0.000}_{-0.007}$	&$0.023^{+0.000}_{-0.005}$	\\
	\hline \hline
		&$C^{(3)}_{\substack{Hl\\jj}}$& $C^{(3)}_{\substack{lq\\jj33}}$ & $C_{\substack{ll\\1221}}$   & $C^{(3)}_{\substack{Hl\\33}}$  \\[3mm]
	\hline \hline
	NLO
	&$-1.742^{+0.002}_{-0.000}$&$0.032^{+0.001}_{-0.003}$&$1.697^{+0.000}_{-0.001}$ &$2.091^{+0.001}_{-0.001}$	\\	
	$\text{NLO}_{t}$
	&$-1.725^{+0.007}_{-0.005}$&$0.032^{+0.007}_{-0.010}$  &$1.693^{+0.003}_{-0.003}$ &$2.079^{+0.007}_{-0.009}$	\\
	\hline
	LO&$-1.173^{+0.008}_{-0.008}$&$0.000^{+0.030}_{-0.030}$ &$1.656^{+0.001}_{-0.001}$ &$2.000^{+0.019}_{-0.019}$	\\
$\text{LO}_{K}$ 
	&$-1.725^{+0.011}_{-0.009}$&$0.032^{+0.001}_{-0.003}$  &$1.693^{+0.001}_{-0.001}$  &$2.040^{+0.020}_{-0.020}$	\\
\multicolumn{4}{c}{}
\\[1mm]
$Z\to \tau \tau$	&$C_{HD}$ & $C_{HWB}$ & $C^{(3)}_{\substack{Hq\\33}}$ & $C_{\substack{Hu\\33}}$  & $C^{(1)}_{\substack{Hq\\33}}$ & $C_{\substack{uB\\33}}$& $C_{\substack{uW\\33}}$  \\[3mm]
	\hline \hline
	NLO
	&$-0.633^{+0.004}_{-0.003}$&$-0.481^{+0.000}_{-0.012}$&$0.025^{+0.000}_{-0.005}$	&$0.055^{+0.002}_{-0.013}$	&$-0.052^{+0.011}_{-0.002}$	&$0.007^{+0.002}_{-0.000}$	&$0.005^{+0.002}_{-0.000}$	\\
	
	$\text{NLO}_{t}$
	&$-0.631^{+0.012}_{-0.009}$&$-0.493^{+0.055}_{-0.057}$&$0.022^{+0.000}_{-0.005}$	&$0.056^{+0.001}_{-0.013}$	&$-0.056^{+0.012}_{-0.002}$	&$0.006^{+0.001}_{-0.000}$	&$0.006^{+0.001}_{-0.000}$	\\
	\hline
	LO&$-0.587^{+0.026}_{-0.026}$&$-0.410^{+0.046}_{-0.046}$&$0.000^{+0.001}_{-0.001}$	&$0.000^{+0.060}_{-0.060}$	&$0.000^{+0.061}_{-0.061}$	&$0.000^{+0.007}_{-0.007}$	&$0.000^{+0.006}_{-0.006}$	\\
	$\text{LO}_{K}$ &$-0.619^{+0.011}_{-0.013}$&$-0.496^{+0.056}_{-0.060}$&$0.022^{+0.000}_{-0.001}$	&$0.027^{+0.031}_{-0.034}$	&$-0.027^{+0.031}_{-0.034}$	&$0.010^{+0.004}_{-0.002}$	&$0.004^{+0.003}_{-0.002}$	\\
	\hline \hline
		&$C^{(3)}_{\substack{Hl\\jj}}$& $C^{(3)}_{\substack{lq\\jj33}}$ & $C_{\substack{ll\\1221}}$ & $C_{\substack{He\\33}}$ & $C^{(1)}_{\substack{Hl\\33}}$ & $C^{(3)}_{\substack{Hl\\33}}$  \\[3mm]
	\hline \hline
	NLO
	&$-1.254^{+0.000}_{-0.004}$&$0.017^{+0.003}_{-0.001}$ &$1.227^{+0.002}_{-0.000}$ &$-1.901^{+0.005}_{-0.000}$ &$2.219^{+0.003}_{-0.004}$ &$2.210^{+0.002}_{-0.001}$			\\	
	$\text{NLO}_{t}$
	&$-1.244^{+0.025}_{-0.024}$&$0.017^{+0.006}_{-0.006}$	&$1.227^{+0.027}_{-0.027}$	 &$-1.882^{+0.024}_{-0.019}$	 &$2.197^{+0.016}_{-0.020}$ &$2.197^{+0.017}_{-0.020}$	\\
	\hline
	LO&$-1.174^{+0.008}_{-0.008}$&$0.000^{+0.030}_{-0.030}$ &$1.174^{+0.031}_{-0.031}$ &$-1.863^{+0.069}_{-0.069}$	 &$2.121^{+0.035}_{-0.035}$ &$2.121^{+0.012}_{-0.012}$		\\
$\text{LO}_{K}$ &$-1.244^{+0.037}_{-0.037}$&$0.017^{+0.001}_{-0.001}$ 	&$1.227^{+0.030}_{-0.030}$	  &$-1.855^{+0.069}_{-0.069}$ &$2.166^{+0.037}_{-0.037}$ &$2.166^{+0.013}_{-0.013}$	 \\	
\end{tabular}
\caption{The numerical prefactors of the Wilson coefficients in the LEP~scheme appearing in $K_W^{(6,1,\mu)}$ and $\hat{\Delta}_{W,t}^{(6,1,\mu)}$ for various perturbative approximations. The tree-level decay rate as well as $v_\mu^2$ have been factored out and the results have been evaluated at the scale of the process. We show the results for $W$~decay (top) and $Z$~decay (bottom).
}
\label{tab:resummingLEPScheme}
\end{table}
\renewcommand{\arraystretch}{1}

\section{Comparison with previous literature}
\label{sec:DG_compare}

Electroweak precision observables at NLO in SMEFT have been calculated previously in~\cite{Dawson:2019clf,Dawson:2022bxd}.
In this section we compare the LEP-scheme results for $M_W$ and the $Z\to \ell\ell$ decay rate
with results given in that work.\footnote{The decay rate for the $W$~boson has not been compared since a leptonic partial branching fraction is not provided in the 
previous literature.}  In order to do so, we must take into account some differences in calculational set-ups.  

First, in contrast to the present paper, those works use $\alpha^{\rm O.S.}(0)$ as an input, so that large logarithms
of lepton masses and hadronic contributions appear in fixed order. We can convert our results to that renormalisation 
scheme by eliminating 
$\alpha(M_Z)$ through use of Eq.~(\ref{eq:ToOnShell}) and 
 \begin{align}
 	\label{eq:alphaMZ_to_alpha0}
	\alpha^{\rm O.S.}(M_Z) = \frac{\alpha^{\rm O.S.}(0)}{1-\Delta\alpha} \approx  \alpha^{\rm O.S.}(0)(1+\Delta\alpha) \, ,
\end{align}
where 
\begin{align}
	\Delta \alpha= \Delta \alpha_\text{lep} + \Delta \alpha^{(5)}_\text{had} = 0.03142 + 0.02764  \, . 
\end{align}
Expanding observables to linear order in  $\alpha^{\rm O.S.}(0)$ and $\Delta\alpha$ then yields SM predictions as given 
in \cite{Dawson:2019clf,Dawson:2022bxd}.  
After making this conversion and using a common set of input parameters also for heavy-particle masses, 
we can exactly reproduce the SM values for the $W$~mass at LO and NLO:
\begin{align}
	\label{eq:MWpredict_Dawson}
	M_W^\text{LO} = 80.939\text{ GeV},  \qquad M_W^\text{NLO} = 80.548\text{ GeV} \, .
\end{align}
The SMEFT results for the $W$-boson mass also agree, when the same set of flavour assumptions is made.

For the $Z\to \ell\ell$ decay rate, we agree with an analytic result in the $\alpha_\mu$~scheme provided to us 
by the authors of~\cite{Dawson:2019clf,Dawson:2022bxd} (after using the flavour assumptions of those papers).  
This forms the basis for LEP-scheme results. In our case these are obtained by using Eqs.~(\ref{eq:MW_prediction})~and~(\ref{eq:LEP_convert_X})
to express the result in terms of $\hat{M}_W$, while~\cite{Dawson:2019clf,Dawson:2022bxd} re-organise the 
SM part of the loop expansion in a way that is specified in the recent preprint \cite{Bellafronte:2023amz}.  
Taking into account these differences, as well as the renormalisation of $\alpha$ 
discussed above, we find numerical agreement with \cite{Dawson:2019clf,Dawson:2022bxd}.

\newpage
\begin{table}
\begin{center}
\small
\begin{minipage}[t]{4.4cm}
\renewcommand{\arraystretch}{1.5}
\begin{tabular}[t]{c|c}
\multicolumn{2}{c}{$1:X^3$} \\
\hline
$Q_G$                & $f^{ABC} G_\mu^{A\nu} G_\nu^{B\rho} G_\rho^{C\mu} $ \\
$Q_{\widetilde G}$          & $f^{ABC} \widetilde G_\mu^{A\nu} G_\nu^{B\rho} G_\rho^{C\mu} $ \\
$Q_W$                & $\epsilon^{IJK} W_\mu^{I\nu} W_\nu^{J\rho} W_\rho^{K\mu}$ \\ 
$Q_{\widetilde W}$          & $\epsilon^{IJK} \widetilde W_\mu^{I\nu} W_\nu^{J\rho} W_\rho^{K\mu}$ \\
\end{tabular}
\end{minipage}
%
\begin{minipage}[t]{2.5cm}
\renewcommand{\arraystretch}{1.5}
\begin{tabular}[t]{c|c}
\multicolumn{2}{c}{$2:H^6$} \\
\hline
$Q_H$       & $(H^\dag H)^3$ 
\end{tabular}
\end{minipage}
\begin{minipage}[t]{4.9cm}
\renewcommand{\arraystretch}{1.5}
\begin{tabular}[t]{c|c}
\multicolumn{2}{c}{$3:H^4 D^2$} \\
\hline
$Q_{H\Box}$ & $(H^\dag H)\Box(H^\dag H)$ \\
$Q_{H D}$   & $\ \left(H^\dag D_\mu H\right)^* \left(H^\dag D_\mu H\right)$ 
\end{tabular}
\end{minipage}
%
\begin{minipage}[t]{2.5cm}
\renewcommand{\arraystretch}{1.5}
\begin{tabular}[t]{c|c}
\multicolumn{2}{c}{$5: \psi^2H^3 + \hbox{h.c.}$} \\
\hline
$Q_{eH}$           & $(H^\dag H)(\bar l_p e_r H)$ \\
$Q_{uH}$          & $(H^\dag H)(\bar q_p u_r \widetilde H )$ \\
$Q_{dH}$           & $(H^\dag H)(\bar q_p d_r H)$\\
\end{tabular}
\end{minipage}

\begin{minipage}[t]{4.7cm}
\renewcommand{\arraystretch}{1.5}
\begin{tabular}[t]{c|c}
\multicolumn{2}{c}{$4:X^2H^2$} \\
\hline
$Q_{H G}$     & $H^\dag H\, G^A_{\mu\nu} G^{A\mu\nu}$ \\
$Q_{H\widetilde G}$         & $H^\dag H\, \widetilde G^A_{\mu\nu} G^{A\mu\nu}$ \\
$Q_{H W}$     & $H^\dag H\, W^I_{\mu\nu} W^{I\mu\nu}$ \\
$Q_{H\widetilde W}$         & $H^\dag H\, \widetilde W^I_{\mu\nu} W^{I\mu\nu}$ \\
$Q_{H B}$     & $ H^\dag H\, B_{\mu\nu} B^{\mu\nu}$ \\
$Q_{H\widetilde B}$         & $H^\dag H\, \widetilde B_{\mu\nu} B^{\mu\nu}$ \\
$Q_{H WB}$     & $ H^\dag \sigma^I H\, W^I_{\mu\nu} B^{\mu\nu}$ \\
$Q_{H\widetilde W B}$         & $H^\dag \sigma^I H\, \widetilde W^I_{\mu\nu} B^{\mu\nu}$ 
\end{tabular}
\end{minipage}
%
\begin{minipage}[t]{5.2cm}
\renewcommand{\arraystretch}{1.5}
\begin{tabular}[t]{c|c}
\multicolumn{2}{c}{$6:\psi^2 XH+\hbox{h.c.}$} \\
\hline
$Q_{eW}$      & $(\bar l_p \sigma^{\mu\nu} e_r) \sigma^I H W_{\mu\nu}^I$ \\
$Q_{eB}$        & $(\bar l_p \sigma^{\mu\nu} e_r) H B_{\mu\nu}$ \\
$Q_{uG}$        & $(\bar q_p \sigma^{\mu\nu} T^A u_r) \widetilde H \, G_{\mu\nu}^A$ \\
$Q_{uW}$        & $(\bar q_p \sigma^{\mu\nu} u_r) \sigma^I \widetilde H \, W_{\mu\nu}^I$ \\
$Q_{uB}$        & $(\bar q_p \sigma^{\mu\nu} u_r) \widetilde H \, B_{\mu\nu}$ \\
$Q_{dG}$        & $(\bar q_p \sigma^{\mu\nu} T^A d_r) H\, G_{\mu\nu}^A$ \\
$Q_{dW}$         & $(\bar q_p \sigma^{\mu\nu} d_r) \sigma^I H\, W_{\mu\nu}^I$ \\
$Q_{dB}$        & $(\bar q_p \sigma^{\mu\nu} d_r) H\, B_{\mu\nu}$ 
\end{tabular}
\end{minipage}
%
\begin{minipage}[t]{5cm}
\renewcommand{\arraystretch}{1.5}
\begin{tabular}[t]{c|c}
\multicolumn{2}{c}{$7:\psi^2H^2 D$} \\
\hline
$Q_{H l}^{(1)}$      & $(H^\dag i\overleftrightarrow{D}_\mu H)(\bar l_p \gamma^\mu l_r)$\\
$Q_{H l}^{(3)}$      & $(H^\dag i\overleftrightarrow{D}^I_\mu H)(\bar l_p \sigma^I \gamma^\mu l_r)$\\
$Q_{H e}$            & $(H^\dag i\overleftrightarrow{D}_\mu H)(\bar e_p \gamma^\mu e_r)$\\
$Q_{H q}^{(1)}$      & $(H^\dag i\overleftrightarrow{D}_\mu H)(\bar q_p \gamma^\mu q_r)$\\
$Q_{H q}^{(3)}$      & $(H^\dag i\overleftrightarrow{D}^I_\mu H)(\bar q_p \sigma^I \gamma^\mu q_r)$\\
$Q_{H u}$            & $(H^\dag i\overleftrightarrow{D}_\mu H)(\bar u_p \gamma^\mu u_r)$\\
$Q_{H d}$            & $(H^\dag i\overleftrightarrow{D}_\mu H)(\bar d_p \gamma^\mu d_r)$\\
$Q_{H u d}$ + h.c.   & $i(\widetilde H ^\dag D_\mu H)(\bar u_p \gamma^\mu d_r)$\\
\end{tabular}
\end{minipage}

\vspace{0.25cm}

\begin{minipage}[t]{4.75cm}
\renewcommand{\arraystretch}{1.5}
\begin{tabular}[t]{c|c}
\multicolumn{2}{c}{$8:(\bar LL)(\bar LL)$} \\
\hline
$Q_{ll}$        & $(\bar l_p \gamma_\mu l_r)(\bar l_s \gamma^\mu l_t)$ \\
$Q_{qq}^{(1)}$  & $(\bar q_p \gamma_\mu q_r)(\bar q_s \gamma^\mu q_t)$ \\
$Q_{qq}^{(3)}$  & $(\bar q_p \gamma_\mu \sigma^I q_r)(\bar q_s \gamma^\mu \sigma^I q_t)$ \\
$Q_{lq}^{(1)}$                & $(\bar l_p \gamma_\mu l_r)(\bar q_s \gamma^\mu q_t)$ \\
$Q_{lq}^{(3)}$                & $(\bar l_p \gamma_\mu \sigma^I l_r)(\bar q_s \gamma^\mu \sigma^I q_t)$ 
\end{tabular}
\end{minipage}
\begin{minipage}[t]{5.25cm}
\renewcommand{\arraystretch}{1.5}
\begin{tabular}[t]{c|c}
\multicolumn{2}{c}{$8:(\bar RR)(\bar RR)$} \\
\hline
$Q_{ee}$               & $(\bar e_p \gamma_\mu e_r)(\bar e_s \gamma^\mu e_t)$ \\
$Q_{uu}$        & $(\bar u_p \gamma_\mu u_r)(\bar u_s \gamma^\mu u_t)$ \\
$Q_{dd}$        & $(\bar d_p \gamma_\mu d_r)(\bar d_s \gamma^\mu d_t)$ \\
$Q_{eu}$                      & $(\bar e_p \gamma_\mu e_r)(\bar u_s \gamma^\mu u_t)$ \\
$Q_{ed}$                      & $(\bar e_p \gamma_\mu e_r)(\bar d_s\gamma^\mu d_t)$ \\
$Q_{ud}^{(1)}$                & $(\bar u_p \gamma_\mu u_r)(\bar d_s \gamma^\mu d_t)$ \\
$Q_{ud}^{(8)}$                & $(\bar u_p \gamma_\mu T^A u_r)(\bar d_s \gamma^\mu T^A d_t)$ \\
\end{tabular}
\end{minipage}
\begin{minipage}[t]{4.75cm}
\renewcommand{\arraystretch}{1.5}
\begin{tabular}[t]{c|c}
\multicolumn{2}{c}{$8:(\bar LL)(\bar RR)$} \\
\hline
$Q_{le}$               & $(\bar l_p \gamma_\mu l_r)(\bar e_s \gamma^\mu e_t)$ \\
$Q_{lu}$               & $(\bar l_p \gamma_\mu l_r)(\bar u_s \gamma^\mu u_t)$ \\
$Q_{ld}$               & $(\bar l_p \gamma_\mu l_r)(\bar d_s \gamma^\mu d_t)$ \\
$Q_{qe}$               & $(\bar q_p \gamma_\mu q_r)(\bar e_s \gamma^\mu e_t)$ \\
$Q_{qu}^{(1)}$         & $(\bar q_p \gamma_\mu q_r)(\bar u_s \gamma^\mu u_t)$ \\ 
$Q_{qu}^{(8)}$         & $(\bar q_p \gamma_\mu T^A q_r)(\bar u_s \gamma^\mu T^A u_t)$ \\ 
$Q_{qd}^{(1)}$ & $(\bar q_p \gamma_\mu q_r)(\bar d_s \gamma^\mu d_t)$ \\
$Q_{qd}^{(8)}$ & $(\bar q_p \gamma_\mu T^A q_r)(\bar d_s \gamma^\mu T^A d_t)$\\
\end{tabular}
\end{minipage}

\vspace{0.25cm}

\begin{minipage}[t]{3.75cm}
\renewcommand{\arraystretch}{1.5}
\begin{tabular}[t]{c|c}
\multicolumn{2}{c}{$8:(\bar LR)(\bar RL)+\hbox{h.c.}$} \\
\hline
$Q_{ledq}$ & $(\bar l_p^j e_r)(\bar d_s q_{tj})$ 
\end{tabular}
\end{minipage}
\begin{minipage}[t]{5.5cm}
\renewcommand{\arraystretch}{1.5}
\begin{tabular}[t]{c|c}
\multicolumn{2}{c}{$8:(\bar LR)(\bar L R)+\hbox{h.c.}$} \\
\hline
$Q_{quqd}^{(1)}$ & $(\bar q_p^j u_r) \epsilon_{jk} (\bar q_s^k d_t)$ \\
$Q_{quqd}^{(8)}$ & $(\bar q_p^j T^A u_r) \epsilon_{jk} (\bar q_s^k T^A d_t)$ \\
$Q_{lequ}^{(1)}$ & $(\bar l_p^j e_r) \epsilon_{jk} (\bar q_s^k u_t)$ \\
$Q_{lequ}^{(3)}$ & $(\bar l_p^j \sigma_{\mu\nu} e_r) \epsilon_{jk} (\bar q_s^k \sigma^{\mu\nu} u_t)$
\end{tabular}
\end{minipage}
\end{center}
\caption{\label{op59}
The 59 independent baryon number conserving dimension-six operators built from Standard Model fields, in 
the notation of \cite{Jenkins:2013zja}.  The subscripts $p,r,s,t$ are flavour indices, and $\sigma^I$ are Pauli
matrices.}
\end{table}
\FloatBarrier

\bibliography{literature}
\bibliographystyle{JHEP.bst}

\end{document}